%% file: MAIN.tex
\documentclass[sigconf]{acmart}

\AtBeginDocument{%
  }

\copyrightyear{2025}
\acmYear{2025}
\setcopyright{cc}
\setcctype{by}
\acmConference[CUI '25]{Proceedings of the 7th ACM Conference on Conversational User Interfaces}{July 8--10, 2025}{Waterloo, ON, Canada}
\acmBooktitle{Proceedings of the 7th ACM Conference on Conversational User Interfaces (CUI '25), July 8--10, 2025, Waterloo, ON, Canada}
\acmDOI{10.1145/3719160.3736639}
\acmISBN{979-8-4007-1527-3/2025/07}

\usepackage{rotating}
\usepackage{xspace}
\usepackage{soul}
\usepackage{colortbl}
\usepackage{listings}

\begin{document}

\input{macro}
\title{How Managers Perceive AI-Assisted Conversational Training for Workplace Communication}

\author{Lance T. Wilhelm}
\orcid{0009-0006-3741-2801}
\affiliation{%
  \institution{Virginia Tech}
  \city{Blacksburg}
  \state{Virginia}
  \country{USA}
}

\author{Xiaohan Ding}
\orcid{0009-0003-2679-3344}
\affiliation{%
  \institution{Virginia Tech}
  \city{Blacksburg}
  \state{Virginia}
  \country{USA}
}

\author{Kirk McInnis Knutsen}
\orcid{0009-0004-7404-7298}
\affiliation{%
  \institution{Virginia Tech}
  \city{Blacksburg}
  \state{Virginia}
  \country{USA}
}

\author{Buse Carik}
\orcid{0000-0002-4511-5827}
\affiliation{%
  \institution{Virginia Tech}
  \city{Blacksburg}
  \state{Virginia}
  \country{USA}
}

\author{Eugenia H. Rho}
\orcid{0000-0002-0961-4397}
\affiliation{%
  \institution{Virginia Tech}
  \city{Blacksburg}
  \state{Virginia}
  \country{USA}
}

\begin{abstract}
    Effective workplace communication is essential for managerial success, yet many managers lack access to tailored and sustained training. Although AI-assisted communication systems may offer scalable training solutions, little is known about how managers envision the role of AI in helping them improve their communication skills. To investigate this, we designed a conversational role-play system, \textsc{CommCoach}, as a functional probe to understand how managers anticipate using AI to practice their communication skills. Through semi-structured interviews, participants emphasized the value of adaptive, low-risk simulations for practicing difficult workplace conversations. They also highlighted opportunities, including human-AI teaming, transparent and context-aware feedback, and greater control over AI-generated personas. AI-assisted communication training should balance personalization, structured learning objectives, and adaptability to different user styles and contexts. However, achieving this requires carefully navigating tensions between adaptive and consistent AI feedback, realism and potential bias, and the open-ended nature of AI conversations versus structured workplace discourse.
\end{abstract}




\maketitle

\section{Introduction}
According to a recent survey conducted by The Harris Poll and Google, 86\% of new and aspiring managers believe that artificial intelligence (AI) could strengthen their leadership capabilities, with nearly half reporting that AI could help improve their communication skills and facilitate better workplace relationships \cite{google_NewAspiringLeaders_2024}. AI, particularly large language models (LLMs), are already embedded in professional workflows, from email auto-completions and smart replies to AI-generated reports and content. Previous research on LLMs in human-computer interaction (HCI) has focused primarily on how these tools augment user messages, helping with tone, structure, and efficiency \cite{pang_UnderstandingLLMificationCHI_2025, jakesch_AIMediatedCommunicationHow_2019}. However, polished text alone does not guarantee effective leadership communication \cite{charlier_EmergentLeadershipVirtual_2016}. Managers must be able to adapt their messages to different workplace contexts, navigate difficult conversations, and communicate with confidence, skills that go beyond AI-assisted message refinement. This raises a question: \textit{If AI were to be used as a training tool rather than just an assistant, how would leaders and managers envision its role?} This study explores how managers conceptualize AI as a tool for communication training, particularly in developing workplace communication skills through role-play and coaching.

Workplace communication plays an essential role in shaping organizational culture and success \cite{keyton_CommunicationOrganizationalCulture_2010}. Open and clear communication from managers not only fosters employee perceptions of organizational support \cite{neves_ManagementCommunicationEmployee_2012}, but also improves workplace performance \cite{kurtessis_PerceivedOrganizationalSupport_2017}. In contrast, poor communication practices from managers are associated with adverse outcomes, such as diminished employee mental health \cite{st-hilaire_WhatLeadersNeed_2019}, lower productivity, \cite{tepper_ConsequencesAbusiveSupervision_2000} and heightened perception of workplace aggression \cite{hershcovis_PredictingWorkplaceAggression_2007}. Although traditional leadership training programs and mentoring opportunities can help managers develop these skills \cite{scandura_MentoringTransformationalLeadership_2004}, many new managers are left to learn communication strategies through self-guided learning, broadly applicable self-help resources, or trial-and-error \cite{mccall_LeadershipDevelopmentExperience_2004, moldoveanu_FutureLeadershipDevelopment_2019}. Mentorship can provide valuable guidance, but access to experienced mentors is not always available, particularly for new or those from historically marginalized groups \cite{zambrana_DontLeaveUs_2015, scandura_MentorshipCareerMobility_1992, scandura_MentoringTransformationalLeadership_2004}. Without structured opportunities to practice handling complex workplace conversations, managers are often left to develop these skills in high-stakes, real-time environments \cite{mccall_LeadershipDevelopmentExperience_2004}.

Role-playing is an established method for the development of managerial skills, allowing managers to practice communication strategies in a controlled environment with real-time feedback from facilitators \cite{kark_GamesManagersPlay_2011, tabak_LearningDoingLeadership_2017}. These exercises help managers experiment with different conversational approaches and prepare for complicated workplace interactions \cite{ladousse_RolePlay_1987}. However, traditional role-playing programs require experienced facilitators and structured group sessions, making them costly and difficult to scale \cite{lacerenza_LeadershipTrainingDesign_2017, rashkin_EmpatheticOpendomainConversation_2019}. Many organizations, particularly those with limited training budgets or high learner-to-facilitator ratios, may be unable to provide personalized, ongoing communication training. LLMs present an opportunity to facilitate interactive workplace conversations, leading to increased interest in AI-driven role-play for training applications \cite{shaikh_RehearsalSimulatingConflict_2024, guyre_PromptEngineeringLLM_2024, hohn_UsingLargeLanguage_2024}. Researchers have explored LLM-based role-play for public speaking practice \cite{park_AudiLensConfigurableLLMGenerated_2023}, medical bedside manner training \cite{ng_RealTimeHybridLanguage_2023, borg_CreatingVirtualPatients_2024}, and conflict resolution simulations \cite{shaikh_RehearsalSimulatingConflict_2024}. However, while AI-driven role-play may open the door to more accessible training, its ability to effectively support workplace communication training remains underexplored.

Many professionals, including managers and supervisees, are already using AI tools such as ChatGPT for their workplace communication tasks (e.g., improving email composition) \cite{retkowsky_ManagingChatGPTempoweredWorkforce_2024, google_NewAspiringLeaders_2024, microsoft_2024WorkTrend_2024}. However, little is known about how managers conceptualize AI's role for workplace communication training and how they perceive its usefulness. This explores these perspectives by addressing the following research questions:

\begin{itemize}
    \item[\textbf{RQ 1}] What communication challenges do managers face in their workplace context and how do they currently address them? 
    \item[\textbf{RQ 2}] How do managers conceptualize the role of AI in communication training, and what expectations or concerns do they have about AI-driven training tools?
\end{itemize}

To answer these questions, we conducted semi-structured interviews with 17 managers with at least one year of supervisory experience. To facilitate discussions on AI-assisted communication training, we introduced participants to an LLM-powered functional probe, \textsc{CommCoach}, designed to simulate role-play scenarios. This study uses \textsc{CommCoach} to understand how managers engage with AI-driven training tools and what expectations, concerns, and needs they express.

Our findings reveal that managers perceive AI-assisted role-play as a tool to practice workplace communication, particularly when it offers customization, iterative feedback, and contextual adaptation. Participants envisioned AI feedback as most useful when it provided actionable guidance aligned with real-world supervisory challenges, helping them refine tone, balance empathy with assertiveness, and tailor messages to different audiences. Although they conceptualized AI as a means of facilitating low-risk repeatable training scenarios, they also saw its greatest potential to augment, rather than replace, human mentorship. In addition, they recognized tensions in AI-assisted communication training, such as balancing adaptability with consistency in feedback, realism with potential bias in AI-generated personas, and open-ended AI conversations with the structured nature of workplace discourse.

Our study contributes to HCI research by: (1) examining how managers conceptualize the role of AI-assisted communication training in the workplace; (2) identifying user expectations, challenges, and concerns that influence managers' engagement with AI-assisted training tools; (3) providing design implications for AI-assisted communication training systems, including the need to accommodate different communication styles, balance AI-generated feedback with human judgment, and address concerns about empathy and contextual appropriateness; and (4) offering \textsc{CommCoach}, as a functional probe to explore AI-mediated role-play in managerial training.

\section{Related Works}
\subsection{Leadership Theory}
\label{sec:leadership_theory}
Researchers and professionals have sought to define leadership since the beginning of the twentieth century \cite{rost_LeadershipTwentyFirstCentury_1993}. Initial research around leadership stemmed from Carlyle's "Great Man" theory \cite{carlyle_HeroesHeroworshipHeroic_1840} and focused on what traits make a person an effective leader. However, the traditional theory of the \textit{trait perspective of leadership} has garnered criticism due to its emphasis on innate or inborn characteristics, and some quantitative studies have shown only moderate correlations between an individual's traits and their leadership effectiveness \cite{hoffman_GreatManGreat_2011, judge_PersonalityLeadershipQualitative_2002}. In contrast, the \textit{leadership as a process} perspective views leadership as a phenomenon that occurs in the interactions between leaders and followers \cite{barker_HowCanWe_1997}. Northouse defines leadership as a "process whereby an individual influences a group of individuals to achieve a common goal" \cite{northouse_LeadershipTheoryPractice_2021}. Although we recognize the differences between management and leadership, we adopt the perspective that both leadership and management rely on the power of language and communication between the leader and the follower \cite{fairhurst_DiscursiveLeadershipConversation_2007, drath_MakingCommonSense_1994}.

The \textit{discourse} perspective of leadership diverges from traditional theories that focus on inherent traits or behaviors and offers a way to study leadership as a relational process through dialogue \cite{northouse_LeadershipTheoryPractice_2021}. According to Fairhurst and Uhl-Bien, discursive leadership is not just about the messages delivered by leaders, but also about how these messages are interpreted and acted upon by followers, thus shaping the social reality of an organization \cite{fairhurst_OrganizationalDiscourseAnalysis_2012}. Discursive leadership also acknowledges the importance of context, suggesting that the effectiveness of leadership discourse varies between different situations and organizational cultures \cite{mcdermott_SocialOrganizationBehavior_1978}. Aritz et al. extend the scholarship on discursive leadership by investigating the power of speech acts (e.g., greeting, apology, complaint, requests, refusal, etc.) within workplace interactions, demonstrating that the use of questions is a key resource in influencing and establishing leadership \cite{aritz_DiscourseLeadershipPower_2017}. Using the discursive perspective of leadership, this study examines how managers in the workplace envision the role of LLMs as a potential tool to help  them improve their managerial communication skills and needs. 

\subsection{Role of Technology in Communication Training}
Researchers in HCI have explored the application of various technologies to improve communication skills in different professional contexts. Chatbots, for example, have been used to simulate real-world interactions and practice communication skills \cite{augello_ModelSocialChatbot_2016, park_SocialSimulacraCreating_2022}. Researchers in these studies have also programmed chatbot agents to represent different personas and scenarios, allowing users to engage in realistic conversations and receive feedback on their communication style and technique \cite{shaikh_RehearsalSimulatingConflict_2024}. Similarly, researchers have also used virtual reality to simulate immersive training scenarios, enabling users to practice communication in realistic settings \cite{bryant_ReviewVirtualReality_2020, mcgovern_ApplicationVirtualReality_2020}. These virtual environments can be tailored to specific professional contexts, such as healthcare \cite{bracq_VirtualRealitySimulation_2019} or customer service \cite{metzger_HowMachinesAre_2017}, providing targeted training experiences.

In addition to chatbots and virtual reality, researchers have applied machine learning to analyze and provide feedback on users' communication practices. For example, Butow and Hoque used supervised and unsupervised machine learning algorithms to audit patterns in doctor-patient conversations associated with high versus low patient satisfaction \cite{butow_UsingArtificialIntelligence_2020}. These algorithms performed comparable to, and sometimes better than, human coding in identifying new communication variables linked with high patient satisfaction. Similarly, INWARD, a video reflection system for executive coaching, leverages machine learning to analyze coaching sessions and identify key points for reflection \cite{arakawa_INWARDComputerSupportedTool_2020}. By highlighting the video timeline with markers, INWARD facilitates meta-reflection and helps users think about their communication experiences.

More recently, researchers have explored the potential of using LLMs to improve communication skills. Park and Choi explored improving formal speech skills by leveraging LLMs to simulate virtual audience members with customized personas \cite{park_AudiLensConfigurableLLMGenerated_2023}. These LLM-powered agents would then provide feedback on the user's speech from multiple perspectives, giving the user real-time feedback on goal accomplishment. Similarly, studies by Ng et al. and Borg et al. demonstrated the effectiveness of using LLMs to create virtual patients for clinical training \cite{ng_RealTimeHybridLanguage_2023, borg_CreatingVirtualPatients_2024}. While these studies have created practical simulations for their respective use cases, they lack a coaching aspect that provides meaningful feedback to the trainee. Furthermore, there is a noticeable gap in research on the application of LLMs specifically for managerial communication training in workplace contexts. Although prior studies have explored LLMs for communication training in fields like healthcare and public speaking, their potential for workplace communication remains underexplored. This study aims to fill this gap by investigating how LLMs, paired with conversational coaching and feedback, could support reflective learning and skill development in this context.

\subsection{Reflective Learning}
Reflective learning, defined as the process through which individuals create knowledge from experience \cite{kolb_ExperientialLearningExperience_1984}, is integral to developing awareness of one's communication effectiveness \cite{karnieli-miller_ReflectivePracticeTeaching_2020, adams_ReflectionCriticalProficiency_2006}. In the context of communicative training systems, reflection allows users to internalize and refine their use of language and interaction strategies, which is crucial for areas such as interviewing \cite{lackner_HelpingUniversityStudents_2017}, public speaking \cite{zhou_VirtualRealityReflection_2021}, and healthcare \cite{pangh_EffectReflectionNursePatient_2019}. Prior research by Kolb suggests that learning is a dynamic process that involves concrete experience, reflective observation, abstract conceptualization, and active experimentation \cite{kolb_ExperientialLearningExperience_1984}. Kolb's conceptual model highlights the importance of reflection in enhancing communication skills through repeated practice and feedback. Schön and Boud et al. further support this concept, emphasizing iterative reflection as a critical component of effective learning \cite{schon_ReflectivePractitionerHow_1983, boud_ReflectionTurningExperience_1985}. Building on these theoretical foundations, HCI researchers have explored ways to integrate experiential and reflective learning into system design.

Drawing on Fleck and Fitzpatrick's framework \cite{fleck_ReflectingReflectionFraming_2010}, which categorizes levels of reflection within technology and suggests techniques to support reflection, HCI researchers have incorporated various approaches to facilitate reflective learning. For example, researchers have incorporated machine learning algorithms \cite{chen_FacilitatingCounselorReflective_2023, arakawa_INWARDComputerSupportedTool_2020, kimani_ConversationalAgentSupport_2019, ortiz_EnablingCriticalSelfReflection_2018} and various features of user interface design \cite{arakawa_INWARDComputerSupportedTool_2020, chang_ThreadCautionProactively_2022} to support various stages and types of reflection in conversational systems. Chen et al.'s counselor training system, \textit{Pin-MI}, for instance, allows counselors to annotate key moments in role-play sessions for later reflection \cite{chen_FacilitatingCounselorReflective_2023}, demonstrating how technology can facilitate reflective practice. Similarly, role-play-based systems like \textit{Chimeria:Grayscale} demonstrate the potential of interactive narratives to foster self-reflection, which is particularly relevant to address the complex interpersonal dynamics of managerial communication \cite{ortiz_EnablingCriticalSelfReflection_2018}. Reflective learning processes can be particularly useful in managerial communication, where iterative practice and feedback have been shown to help managers navigate complex interpersonal dynamics \cite{loo_JournalingLearningTool_2002}. It is under this premise that our research investigates managers' perspectives on how AI systems might support their communication skills, including the role of reflective practice in such training. 

\section{Formative Study}
Before developing our technical probe for AI-assisted communication training, we conducted a formative study with five managers from various professional backgrounds (academia, business, and military) and varying experiences overseeing teams of two to 100 supervisees. The goal was to obtain early insights into managerial communication training needs and initial reactions to an AI-assisted communication training concept before building the functional probe. This exploratory stage aimed to gather information about participants' prior experience with managerial communication training and their initial impressions of a mock AI-assisted communication training system.

\paragraph{Materials \& Procedure}
We engaged participants in semi-structured interviews in which they reflected on past experiences with managerial communication training and discussed their perspectives on AI's role in skill development. We presented them with static and video mockups illustrating the concept of an AI-assisted communication training system. The mockups, designed using established user experience (UX) principles \cite{yablonski_LawsUXUsing_2020, helander_HandbookHumanComputerInteraction_1997}, simulated a chat-based role-playing system featuring real-time feedback from an AI-driven coaching assistant. The mockup designs closely resembled the final function probe design (Appendix \ref{app:screenshots}), but varied by different features. We presented 17 different static mockups to participants with combinations of the following features:

\begin{table*}[ht]
  \centering
  \caption{Themes and key quotes from formative study participants.}
  \label{tab:formative_study}
  \begin{tabular}{%
      p{0.20\linewidth}|
      p{0.24\linewidth}
      p{0.24\linewidth}
      p{0.24\linewidth}
    }
    \hline
    \textbf{Theme}
      & \multicolumn{3}{c}{\textbf{Participant Quotes}} \\
    \hline
    \themea{\textbf{Customizable and realistic scenarios are crucial for effective training}}
      & \textbf{F1:} \textit{If you could \themeaunder{give a specific scenario that you're struggling with} at work, then maybe you can role-play with this chatbot a little bit, maybe get some meaningful insights into your approach.}
      & \textbf{F4:} \textit{\themeaunder{If I can put in my own scenario} of like, “Hey, this is exactly what's going on,” then, maybe it will help me find the best fit model to address the scenario that I'm struggling with. [...] That would definitely be crucial for me.}
      & \textbf{F3:} \textit{\themeaunder{Making the scenarios as applicable as possible is key}. I just think you need \themeaunder{very, very real life examples}, not something that is just made up and just standard [...] I think the scenario needs to be contextual from the very get-go.} \\
    \hline
    \themeb{\textbf{Immediate feedback is preferred over post-session feedback}}
      & \textbf{F1:} \textit{I feel like there's more value in the \themebunder{immediate feedback [...] immediate feedback} allows the user to address it or change [their approach] on the fly.}
      & \textbf{F4:} \textit{I really like \themebunder{seeing my problems as I'm hitting them} instead of just at the end. I don't want that.}
      & \textbf{F3:} \textit{I never, ever attend to the reflections when it's at the very end of the module or the session [...] It's best to have that \themebunder{embedded into these sessions} than at the very end.} \\
    \hline
    \themec{\textbf{Simulating diverse conversational outcomes for practice and learning}}
      & \textbf{F5:} \textit{Having the ability, maybe after they have had a difficult conversation with a person, and it didn't go so well, and they'd like \themecunder{to practice it again}, or get feedback on what they did. I could certainly see that being useful.}
      & \textbf{F3:} \textit{If I were proposed this [system] as a part of my orientation to have to go through, \themecunder{I'd be spending time on this}. And if someone told me that this is based on real live scenarios that have actually happened with our customers, I'd be like, I'm all in.}
      & \textbf{F5:} \textit{I liked that you could \themecunder{go back and forth between the two versions} of what you had said. So you could see the first feedback and the second feedback.} \\
    \hline
    \themed{\textbf{Clear and direct feedback with opportunities for further exploration}}
      & \textbf{F5:} \textit{I almost wonder if after you get the first feedback, if the person doesn't understand the feedback, they could click a question mark or something that would then \themedunder{allow them to interact with the coach}. They'd be like, “I don't know why you're telling me this” or could send whatever message.}
      & \textbf{F1:} \textit{I'm thinking that the \themedunder{ability for the user to engage with the coach, on a specific feedback} adds something of value [...] You can maybe say, “okay I understand [the feedback] that was presented, but I don't understand how I responded to this situation would work well with this situation.” I think that it'd be \themedunder{useful to have a side dialogue [with the coach]} separate from the main dialogue.}
      & 
      \\
    \hline
  \end{tabular}
\Description{The table titled "Themes and key quotes from formative study participants" presents five key themes identified from user feedback on a communication training tool. The first theme highlights the importance of customizable and realistic scenarios, with participants emphasizing the value of using specific, real-life examples to enhance training relevance. The second theme focuses on the preference for immediate feedback over post-session reflections, allowing users to adjust their approach in real time. The third theme underscores the benefit of simulating diverse conversational outcomes, where participants can practice difficult conversations and receive feedback for improvement. The fourth theme stresses the importance of clear and direct feedback with opportunities for further exploration, with users noting the value of engaging with a coach to clarify feedback and explore specific situations. Overall, the table reveals that users appreciate a training tool that offers realism, timely feedback, and opportunities for re-engagement and deeper reflection.}
\end{table*}

\begin{itemize}
    \item Immediate feedback
    \item Post role-play feedback
    \item Chat session management
    \item AI highlighting of portions of user dialogue that need attention
    \item Learning objectives alongside the scenario
    \item Immediate reflective dialogue or journaling
    \item Post role-play reflective dialogue or journaling
    \item Hints and co-writing with AI
\end{itemize}

Particpants were encouraged to draw on or visually annotate the mockups, providing direct input on system features, potential usability challenges, and desired functionality.

\paragraph{Key Takeaways}
The first author conducted and transcribed the interviews using Zoom's transcription service. They then used open coding \cite{charmaz_ConstructingGroundedTheory_2006} to identify common themes across the five interviews. After deliberation, the research team distiled four major themes from the interviews (Table \ref{tab:formative_study}): (1) customization and realism in scenarios, (2) a preference for immediate feedback, (3) the ability to explore multiple conversational outcomes, and (4) clear, interactive feedback mechanisms. 

\begin{figure*}[ht!]
  \centering
  \includegraphics[width=\textwidth]{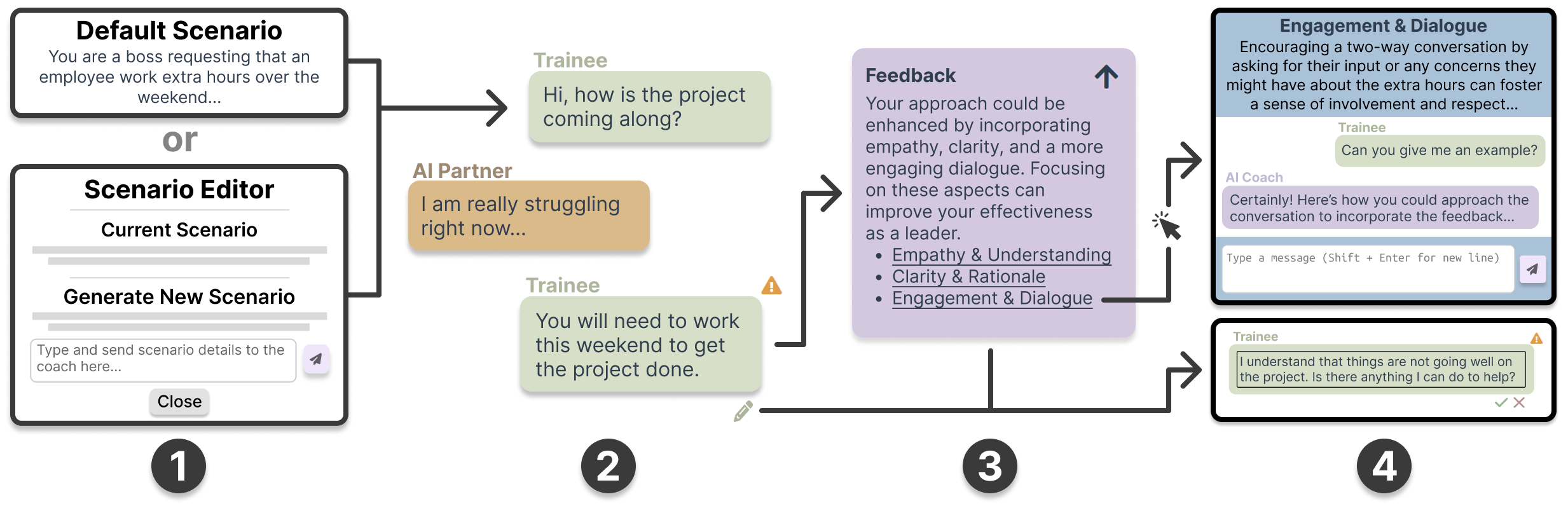}
  \caption{User-experience flow for \textsc{CommCoach}, the functional probe used during user study interviews (\textit{Note: text has been truncated in some cases to accommodate space, indicated with elipses}). \textbf{(1)} Users can either start with the default scenario or design custom scenarios through an interactive process with the coach agent. Once a user is satisfied with a scenario, they can accept it and \textbf{(2)} start a new conversation with the conversational partner agent. \textbf{(3)} If the coach agent deems it necessary, it will intervene and provide immediate context-aware feedback. \textbf{(4)} Users then have the ability to explore a specific feedback item further through a dialogue with the coach, or can retry messages to simulate different conversational outcomes.}
  \label{fig:system_flow}
  \Description{The figure illustrates the user-experience flow for CommCoach, an AI-driven conversational coaching system, structured into four steps. First, users select a scenario, either choosing a Default Scenario, such as a boss requesting an employee to work extra hours, or creating a custom scenario using the Scenario Editor, where they can enter scenario details manually. Once a scenario is selected, users engage in a conversation with an AI Partner simulating a human counterpart. In the example shown, the user (Trainee) initiates the dialogue by asking about project progress, to which the AI Partner responds that they are struggling. The Trainee then instructs the AI Partner to work over the weekend, triggering a warning icon, indicating a potential issue. If the system detects room for improvement, it provides real-time Feedback in a panel suggesting enhancements in Empathy & Understanding, Clarity & Rationale, and Engagement & Dialogue. These categories are hyperlinked, suggesting that users can explore them further. To refine their response, users can engage with an AI Coach, who provides suggestions for improved phrasing. In the example, the Trainee requests an example from the AI Coach, who then offers a revised response: "I understand that things are not going well on the project. Is there anything I can do to help?" The figure is visually structured with arrows connecting steps, and different colors distinguish elements such as trainee responses, AI Partner dialogue, feedback, and coaching interactions.}
\end{figure*}

\section{\textsc{CommCoach} System Design}
\label{sec:SystemDesign}
To explore user perceptions of AI-assisted communication training systems, we developed \textsc{CommCoach}, a functional probe informed by themes from the formative study and previous research (Figure \ref{fig:system_flow}). \textsc{CommCoach} is an AI-assisted communication training system designed to help users practice managerial communication skills based on role-play simulation. \textsc{CommCoach} uses a role-play sandbox approach \cite{othlinghaus-wulhorst_TechnicalConceptualFramework_2020}, allowing users to engage with custom scenarios and receive feedback from an AI coach agent. Informed by insights from our formative study, we integrated features such as conversation re-tries, flagging, and revisiting discussions in \textsc{CommCoach} to facilitate reflective learning \cite{schon_ReflectivePractitionerHow_1983, boyd_ReflectiveLearningKey_1983, boud_ReflectionTurningExperience_1985, fleck_ReflectingReflectionFraming_2010} as part of our training system.

\textsc{CommCoach} resembles a chatbot or messaging application, enabling users to chat with a conversational partner agent (playing the role of a supervisee in a role-play scenario), while a coach agent provides feedback in a parallel pane based on the participant's conversations with the conversational partner. Our team built \textsc{CommCoach} as a web-based application using a client-server architecture. The system consists of a Vue.js front end, a Python and MongoDB back end served through FastAPI, and two LLM agents powered by the GPT-4 API. We kept the system prompts for each agent and their tasks as minimal as possible, since \textsc{CommCoach} was designed to serve as an exploratory tool rather than an ideal solution. More details on the implementation, including prompts, can be found in Appendix \ref{app:system_design}. Screenshots of \textsc{CommCoach}'s UI are provided in Appendix \ref{app:screenshots} as well. 

Below, we describe the primary design features of \textsc{CommCoach} in more detail. 

\subsection{Customizable and realistic scenario-based role-playing}
Existing research highlights the effectiveness of role-play in simulating and practicing challenging conversations \cite{agboolasogunro_EfficacyRolePlaying_2004, chen_FacilitatingCounselorReflective_2023}. Participants in our formative study who had prior communication training also emphasized the value of role-playing as a training method. However, they noted that prescribed role-play scenarios can sometimes feel sterile and lack realism. To maximize the effectiveness of role-play training, participants expressed a desire for realistic and complex scenarios that closely mimic real-world situations. Traditionally, generating and practicing such role-playing scenarios and evaluating performance against objectives has required human peers or expert resources.

\textsc{CommCoach} incorporates a scenario editor that enables users to create customized scenarios based on their specific management challenges (Figure \ref{fig:system_flow}, Part 1). The system facilitates the generation and refinement of customized scenarios through an iterative process involving back-and-forth interaction between the user (trainee) and the coach agent, ensuring relevant and practical role-playing experiences that optimize learning outcomes.

\subsection{Timely context-aware in-situ feedback for immediate reflection}
Feedback plays a crucial role in promoting learning and fostering change among users, with timely delivery being particularly valued by participants in our formative study. HCI research on computer tutoring systems shows that feedback is most effective when tailored to align with the user's interactions and conversation history \cite{deeva_ReviewAutomatedFeedback_2021}. However, despite recent advances in real-time feedback facilitation \cite{chen_FacilitatingCounselorReflective_2023, arakawa_INWARDComputerSupportedTool_2020, shaikh_RehearsalSimulatingConflict_2024}, there remains a gap in the literature on immediate feedback specifically tailored to managerial communication.

We developed a prompt framework that builds on prior studies in prompt chaining \cite{wu_AIChainsTransparent_2022, ding_LeveragingPromptBasedLarge_2024, park_GenerativeAgentsInteractive_2023a, wu_LargeLanguageModels_2023a}. Our framework analyzes the user's input, conversation history (Figure \ref{fig:system_flow}, Part 2), and predefined scenario to generate a prompt for the coach agent, which then provides tailored feedback, identifying areas for improvement, proposing alternative strategies, and offering actionable insights. The timely delivery of context-aware feedback is designed to support immediate reflection, aligning with Fleck and Fitzpatrick's conceptual framework on reflection \cite{fleck_ReflectingReflectionFraming_2010}. \textsc{CommCoach} displays a warning triangle next to messages that trigger coach interventions, encouraging users to revisit and reflect on their communication approach in real-time (Figure \ref{fig:system_flow}, Part 3). The intervention threshold is set through a simple prompt to the coach agent. This immediate feedback is aimed at promoting reflective description \cite{fleck_ReflectingReflectionFraming_2010}, prompting users to consider the impact of their messages and explore alternative strategies.

\subsection{Facilitating dialogic reflection through user-coach feedback discussions}
To foster reflective learning, \textsc{CommCoach} allows users to engage in focused discussions with the coach agent about specific feedback items (Figure \ref{fig:system_flow}, Part 4, top panel). When users receive feedback on how they communicate with the conversational partner, they can click on individual feedback items to open a dedicated modal window and initiate a conversation with the coach agent. This feature allows users to ask questions, seek clarification, and explore the feedback in greater depth, promoting a deeper understanding of the concepts and their application to the user's context. By facilitating these targeted discussions, \textsc{CommCoach} aims to encourage users to move beyond surface-level reflection and engage in dialogic reflection \cite{fleck_ReflectingReflectionFraming_2010}, where they actively construct meaning through dialogue with the coach agent.

\subsection{Applying feedback and exploring alternative approaches through chat branching}
\textsc{CommCoach} additionally supports reflective learning \cite{fleck_ReflectingReflectionFraming_2010} through chat branching, which allows users to diverge from the original conversation flow and create alternative paths by revising their messages based on the coach agent's feedback or personal curiosity. Users can access an input field by clicking on a message they want to modify, where they can edit their original message, incorporating the insights and suggestions provided by the coach agent (Figure \ref{fig:system_flow}, Part 4, bottom panel). When the user sends the revised message, \textsc{CommCoach} generates a new branch in the conversation, allowing the user to explore how the revised approach affects the outcome of the interaction. Users can toggle between different conversation branches, comparing the feedback and outcomes associated with each path. We also added this feature based on participant feedback and studies indicating that breaks and checkpoints facilitate user progression without significantly hindering skill development \cite{johanson_HelpingPlayersProgress_2023,iacovides_PlayerStrategiesBreakthroughs_2014}. The ability to pause or retry scenarios after reaching unsatisfactory results further encourages users to engage in dialogic reflection \cite{fleck_ReflectingReflectionFraming_2010}. The chat branching feature is also designed to engage users in experiential learning \cite{kolb_ExperientialLearningExperience_1984}, where participants can actively experiment with different communication approaches and learn from the ensuing responses from the conversational partner in a low-stakes environment.

\section{User Study with \textsc{CommCoach}}
\label{sec:user_study}
\subsection{Participants}
For our study, we recruited 17 managers (11 males, 6 females) between 20 and 65 years old ($\bar{x}=41.9$, $s=14.3$) from diverse industries, including academia, research, military, human resources, sales, engineering, and healthcare. Participants had varying levels of managerial experience, overseeing between 2 to 84 ($\bar{x}=21.8$, $s=21.6$) supervisees. Particpants were required to have at least one year of supervisory experience. Table \ref{tab:participants} provides a detailed breakdown of participant demographics. As this study was reviewed and approved by our Institutional Review Board (IRB), all participants acknowledged an informed consent document before agreeing to participate in the study. Before the interview, they were given another opportunity to review the informed consent document before continuing.

Participants were recruited using online methods including email announcements at the authors' institution, social media platforms such as Facebook, LinkedIn and X, and word-of-mouth referrals. Word-of-mouth recruitment was particularly effective, yielding the majority of participants. As noted in the Limitations section (\S \ref{sec:discussion_limitations}), this approach comes with associated limitations regarding potential biases or representation. Despite this, recruitment methods ensured diversity in managerial backgrounds, industry sectors, and levels of experience.

\begin{table*}[]
\caption{User Study Participant Demographic Data.}
\begin{tabular}{llllll}
\hline
\textbf{P\#} & \textbf{Age} & \textbf{Gender} & \textbf{Ed. Level}  & \textbf{\begin{tabular}[c]{@{}l@{}}Report \#\\ (employees)\end{tabular}} & \textbf{\begin{tabular}[c]{@{}l@{}}Field of Supervisory \\ Experience\end{tabular}} \\ \hline
1            & 40           & Male            & Professional degree & 50                                                                       & Military                                                                            \\
2            & 41           & Female          & Doctorate           & 15                                                                       & Academia; Research                                                                  \\
3            & 26           & Male            & 4-year degree       & 30                                                                       & Non-profit organization                                                             \\
4            & 43           & Male            & Professional degree & 84                                                                       & Military; HR; Sales                                                                 \\
5            & 20           & Male            & Some college        & 13                                                                       & Math tutoring                                                                       \\
6            & 34           & Female          & 4-year degree       & 10                                                                       & Military; Clinical Laboratory                                                       \\
7            & 56           & Female          & Doctorate           & 14                                                                       & Academia                                                                            \\
8            & 63           & Female          & 4-year degree       & 10                                                                       & Contractor                                                                          \\
9            & 33           & Male            & 4-year degree       & 35                                                                       & Military; Finance                                                                   \\
10           & 60           & Male            & Doctorate           & 40                                                                       & Human Resources; Training                                                           \\
11           & 65           & Male            & 4-year degree       & 3                                                                        & Operations                                                                          \\
12           & 64           & Female          & 2-year degree       & 3                                                                        & Medical                                                                             \\
13           & 35           & Male            & 4-year degree       & 30                                                                       & IT and HR                                                                           \\
14           & 35           & Male            & 4-year degree       & 7                                                                        & Engineering                                                                         \\
15           & 37           & Male            & 4-year degree       & 22                                                                       & Maritime Manufacturing                                                              \\
16           & 29           & Male            & 2-year degree       & 2                                                                        & Military; Personnel                                                                 \\
17           & 32           & Female          & Doctorate           & 3                                                                        & UX Research                                                                         \\ \hline
\end{tabular}
\label{tab:participants}
\Description{The table titled "User Study Participant Demographic Data" provides details on 17 participants, organized into six columns: participant number (P#), age, gender, education level, the number of employees they supervise (Report #), and their field of supervisory experience. Participants vary in age from 20 to 65 years old, and include both males (10) and females (7). Their educational backgrounds range from "Some college" to "Doctorate" and "Professional degree." The number of employees they supervise spans from 2 to 84. Participants have experience in various fields, including military, academia, research, non-profit organizations, human resources, finance, medical, engineering, and more specialized areas like UX research, maritime manufacturing, and clinical laboratories. This demographic table provides a diverse range of supervisory experiences across different industries and educational levels.}
\end{table*}

\subsection{Procedure}
This study aimed to explore how managers perceive and engage with AI-assisted communication training, building on the themes identified in the formative study. The web-based function probe, \textsc{CommCoach}, was used as a research probe to facilitate discussions and observations about user interactions with AI-driven coaching tools. The user study consisted of three phases:

\begin{enumerate}
    \item \textbf{Management and Training Background:} Participants first underwent a semi-structured interview session based on their indicated supervisory and training experience gathered from a pre-survey with the objective of gathering background information about their experience with managerial communication training. Specific questions were designed to understand their past experiences with traditional and technology-based training methods and their perceptions of the effectiveness of these methods. This initial step set the context for their subsequent interaction with the probe.
    \item \textbf{User Interaction with Functional Probe:} Participants were then introduced to our functional probe, \textsc{CommCoach} (Figure \ref{fig:system_flow}) and engaged in both a predetermined scenario and custom scenarios based on challenges from their personal experiences. Participants engaged with the probe while verbalizing their thoughts, providing insight into how such tools might support managerial communication training.
    \item \textbf{Post-Interaction Reflection:} Following the interaction with the probe, participants were asked to reflect on the alignment of AI-generated response with their expectations, preferences regarding feedback received from such systems, and how they envision AI's role in managerial communication training. This phase provided deeper insights into user preferences and highlighted potential focus areas for AI-assisted communication training systems.
\end{enumerate}

The protocol for the semi-structured interviews can be found in Appendix \ref{app:user_study_interview_protocol}.

\subsection{Data Collection and Analysis}
The collected data consisted of transcribed recordings of the sessions. These recordings captured participants' responses about their supervisory background and perceptions and interactions with \textsc{CommCoach}. The first author used a grounded theory approach \cite{charmaz_ConstructingGroundedTheory_2006, corbin_BasicsQualitativeResearch_2015, muller_GroundedTheoryMethod_2012} for qualitative analysis. The interview transcripts were first analyzed through an open-coding process, resulting in 346 initial codes that represent every significant utterance in the data. These initial open codes were then organized through axial coding, identifying connections and relationships between them, which resulted in 69 axial codes. Next, these axial codes were synthesized into 15 selective codes, providing overarching thematic categories, though in four cases the axial codes were directly adopted as selective codes due to their specificity and clarity. Finally, the research team collaboratively reviewed and refined the resulting themes to ensure consistency and accuracy.

This hierarchical coding structure was closely aligned with the semi-structured interview design, where the first half addressed the backgrounds of the participants and the previous training experiences in managerial communication, and the second half focused on their interactions with \textsc{CommCoach}. When interpreting the findings, the first author referred to the resulting code table to identify and extract participant quotes relevant to each selective and axial code. This structured coding facilitated a detailed exploration of the two research questions, as the axial codes directly corresponded to answering RQ1 and RQ2.


\section{Results}
\subsection{Challenges in Management Communication Practices and Prior Training Experiences}
\label{sec:findings_background}
\subsubsection{Management Communication Challenges in the Workplace}
All participants emphasized the importance of effective managerial communication for their roles (P1-P17), highlighting adapting their communication styles as managers to different situations and individual supervisees at their workplace (P4, P10, P15, P17). Participants used a diverse range of discursive strategies, such as storytelling (P1), humor (P13), socratic questioning (P1), directness (P1), avoiding escalating language (P8), signaling empathy (P7), and encouraging competence and reliability (P17). However, these strategies were often hit or miss and were adapted through trial and error over time. In addition, some found that their communication strategies as managers were misaligned with workplace demands and sometimes counter to their individual personalities (P6, P12). For example, P6, who is a clinical laboratory scientist, mentioned the challenges of balancing empathy with a dislike of conflict, stating \textit{"I don't like conflict a lot, so I think [my words] can come off as passive-aggressive,"}, even though she intends to be more empathetic. Similarly, P12 often doubted her ability to handle conflict as a supervisor, stating \textit{"I don't think I'm a really good supervisor because I don't like confrontation."} Participants also highlighted various other communication barriers, including emotional reactions (P1, P2), power dynamics (P2), fear of being misunderstood (P6), and differences in seniority (P17).

\subsubsection{Lack of Sustained and Tailored Training}
Twelve participants experienced some type of prior management communication training, although most described it as one-off sessions, typically early in their careers. This limited their ability to consistently apply the lessons learned from their training programs. Many relied on self-guided learning through the Internet, literature, and mentors (P2, P7, P9, P10, P17). For example, P2 described turning to Google for guidance before a challenging conversation: \textit{"I was looking for training and maybe I would have appreciated just know what the norms are and what people think the best approaches are."} Participants highlighted the need for a supportive and nonjudgmental environment to facilitate their development in communication strategies, expressing concerns about being judged or making mistakes while handling delicate situations with supervisees (P5, P13, P14). Some, like P13, who is a miltary officer, found personal mentors helpful in understanding how to improve their communication in such contexts: "I appreciated [mentorship] more because it felt safe...They would tell me hard things without making me feel stupid." However, not all participants had access to such mentors, let alone training experiences through which they could learn how to handle such delicate situations (P2).

\subsubsection{Challenges in Applying Prior Training to Real-World Situations}
Participants faced challenges in applying their previous training to real-world situations. Some participants questioned the relative importance of training versus experience for managerial effectiveness (P1, P8). P1 claimed that training merely \textit{"makes up 30 to 45\% of a manager's effectiveness,"} suggesting that practical experience plays a significant role in the development of management skills. Despite this, participants recognized some advantages of management communication training, such as exposing people to what they do not know (P1), particularly benefiting introverted individuals through role-playing in person (P4), and providing tools to understand others' behavior and thoughts (P5). However, they also expressed various preferences for training methods, including learning through experience (P1, P3), hands-on learning or exercises (P1, P4, P5, P11, P14), expert mentors (P7, P13), and critical thinking and self-reflection (P4). Some participants, like P7, reported learning through trial and error on the job, while others, like P10, relied on negative examples of management as \textit{"anti-examples"} to inform their communication strategies.

\subsection{Setup and Customization of AI Systems}
\label{sec:findings_setup}
\subsubsection{Customizable Scenarios}
Participants envisioned AI-assisted communication training as more effective when it could \textbf{personalize scenarios based on specific workplace situations or current experiences with their supervisees}. They noted that AI systems such as \textsc{CommCoach} should reflect their specific workplace challenges and communication needs (P1, P2, P4, P12).  Unlike generic training programs, participants felt that AI has the potential to create bespoke scenarios that resonate with users' specific roles, industries, and communication challenges. Participants actively sought the ability to create and refine custom scenarios using the scenario editor. Participants felt that this allowed them to focus on communication challenges directly applicable to their roles, creating the perception of more effective training (P1-P7, P9, P12-P17). Several participants highlighted the importance of tailoring AI scenarios to reflect specific communication needs and challenges, particularly those they face in their actual work environments (P1, P5-P7, P10, P13, P15-P17). More specifically, they valued the ability to create custom scenarios to simulate challenging interaction\textbf{s}, such as managing conflicts, navigating sensitive topics, and practicing constructive feedback as a supervisor. A list of participant-generated scenarios is given in Appendix \ref{app:participant_scenarios}

\subsubsection{Adjustable Conversational Partner Behavior}
Furthermore, participants sought more features beyond simply customizing the scenarios themselves. Several desired the ability to \textbf{adjust the behavior of the conversation partner to create a more challenging and realistic training environment} (P3, P5, P7, P10-P12, P16, P17). For example, P5 suggested including predefined personality traits for users to choose from when creating a scenario, while P16 recommended adding agent demographics such as educational background or age. P11 also highlighted the benefit of simulating difficult conversational partners, suggesting that a more \textit{"recalcitrant"} or \textit{"reserved"} AI character might be beneficial, stating, \textit{"It would be interesting to see if they were difficult, how I'd have to react."} Similarly, P12 desired a more \textit{"confrontational"} AI partner to learn how to \textit{"use words to deflect the situation."} P5 further advocated for a feature that allows the AI to adopt varying levels of difficulty, explaining that \textit{"having an AI character that’s not always cooperative would make the learning more valuable, as it mirrors real-life situations where not everyone is agreeable."}

\subsubsection{Training Objectives and Structured Tasks}
Beyond scenario and interlocutor customization, participants expressed a strong interest in having the system set \textbf{specific mangerial communication training objectives to provide structure and milestone directions} for continued training and improvement over time with AI. These objectives were considered beneficial to help users focus their efforts, track their learning progress, and stay motivated, especially for those new to managerial or communication roles (P10, P13, P17). This structured approach to goal setting was considered a key advantage, offering a more customized and engaging training experience than traditional methods. P13 suggested adding reminders to revisit scenarios before important conversations, explaining, \textit{"If I could set a reminder...I'd review what I learned before I talk to [the person]."}

\subsection{Feedback}
\label{sec:findings_feedback}
\subsubsection{What Feedback Should the AI Provide to the User?} 
\label{sec:what_feedback_should}
Participants' expectations for AI-coach's feedback were shaped by their existing experiences as managers, with many anticipating feedback that aligned with real-world supervisory challenges (P1-P4, P6, P9-P13, P15-P17). They preferred \textbf{actionable feedback tailored to specific challenges they faced as a supervisor at work}. Rather than receiving generic advice, which is commonly found in internet-based training resources, participants wanted feedback tailored to their specific role and situation they often faced in their supervisory position, offering concrete strategies they could implement immediately with their current supervisees (P2, P3, P5, P6, P9, P10).

Participants also emphasized the \textbf{importance of AI helping them to understand how their tone impacts potential conversation outcomes}. Many expressed a strong desire for the AI coach to provide nuanced feedback on their tone of communication when interacting with the practice supervisee played by the AI agent. For instance, several desired explicit insight on why their word choices could be perceived negatively or why the coach's alternative would lead to more constructive dialogue outcomes (P5, P6, P9). Participants envisioned this feature would be especially useful in rehearsing delicate or emotionally charged situations with their supervisees. Furthermore, some expressed that they were not always aware of how their tone is perceived by colleagues and supervisees, stating that a system such as \textsc{CommCoach} that could detect potential blind spots in this area would be useful: \textit{"it picked up on some stuff that I might have blind spots to...could be doing better with balancing that firmness, sensitivity, etc."} (P4).

Participants often sought guidance on balancing seemingly contradictory tones, such as being both empathetic and direct, to ensure clarity without sounding harsh or detached (P6, P9, P10). They anticipated that AI coach's feedback would account for individual communication styles and intent, expecting it to adapt to their unique characteristics over time. Some envisioned AI systems that could refine their suggestions based on ongoing interactions, aligning more closely with personal communication approaches.

Participants considered the \textbf{integration of established leadership principles into AI feedback} as a potential way to improve managerial communication training. Some, like P13, expected AI-driven coaching to align with recognized leadership frameworks, such as those by John Maxwell \cite{maxwell_21IndispensableQualities_2007, maxwell_21IrrefutableLaws_2007, maxwell_DevelopingLeaderYou_1993}, to provide structured guidance. They viewed this as a means of connecting AI coach's feedback with widely accepted managerial strategies, making the training more relevant and practically applicable to their professional contexts (P13). However, P10 cautioned that feedback must also align with organizational competencies and standards and not just individual preferences. He emphasized that the AI system's guidance should be consistent with the organization’s goals and values, which could prevent potential conflicts and improve the effectiveness of the training (P10).

In addition to corrective feedback, some participants sought \textbf{positive reinforcement from the AI coach} after creating a well-constructed message, as it validated their efforts and increased their confidence in their communication skills, encouraging them to continue applying effective strategies (P3, P4, P6, P9, P14). P9 likened this to classical conditioning and suggested that \textit{"When I had the good message, the feedback could have still been helpful if they said your message was good because of X, Y, Z."} Some interpreted lack of feedback during use of the functional probe as a positive sign that they were improving (P6).

\subsubsection{How Should the Feedback Be Provided to the User?}
Most participants preferred immediate feedback from the system. They felt that \textbf{receiving in-situ feedback on their communication style as they progressed through a conversation with the AI} helped them correct errors in real time, preventing repeated errors, and strengthened their learning in the moment (P1, P2, P4, P5, P6, P9, P10, P17). This feature was particularly valued in the role-playing context of the system, where participants could immediately adjust their strategies based on AI coach's feedback (P4, P13, P14). As P13 explained, \textit{"I don't want to be wrong the entire time before I take the feedback and learn a critical point. I want to find out now that there's a place that I could pivot to get better."} However, not all participants felt that the timing of immediate feedback from the AI coach was ideal. P1 and P10 expressed concerns that the coach's feedback arrived too early in a conversation. Despite these concerns, participants often clarified that they planned to apply the specific feedback in later messages, suggesting that the AI coach's feedback was still appropriate but mistimed. Furthermore, some participants saw potential advantages in delayed feedback, which could facilitate deeper reflection after completing a scenario (P2, P5, P16). As P3 and P12 suggested, delayed feedback from the AI coach could provide a broader perspective on performance, allowing participants to reflect on their overall approach.

Beyond the timing of the feedback, participants expressed a preference for \textbf{feedback as an ongoing dialogue with the AI coach as part of the training process}, rather than a one-time response (P5, P9). They saw value in the ability to revisit and refine their responses, using iterative practice to internalize communication strategies (P13). The opportunity to pause, reflect, and retry different approaches, especially when presented with multiple response options, was viewed as a key factor in improving their ability to navigate workplace conversations effectively.

Participants reported that receiving feedback prompted immediate reflection on their messages (P2, P4-P6, P9-P16). In some cases, participants could even anticipate the feedback they would receive, indicating a greater awareness of their communication patterns. P15's comment illustrates this: \textit{"I knew as soon as I sent it and as soon as I saw that there was a triangle that something was being generated. So I bet it's because I'm not being direct enough in this moment."}

\subsection{Practicing}
\label{sec:findings_practice}
\subsubsection{Human-AI Teaming}
Participants highlighted the potential for \textbf{human-AI teaming} as a way to improve their managerial communication training experience. Although acknowledging the capabilities of AI systems such as \textsc{CommCoach}, participants highlighted the limitations of relying solely on AI for sensitive and context-specific training. This sentiment was summarized by P10, who leads professional development training for a large orgnaization:

\begin{quote}
\textbf{P10:} \textit{"It doesn't know me. I like to think I mean well. We all have our own blind spots, but I like to think I am empathetic to people's situations. Now, if [the coach] has worked with me for a while, they may even know that I'm overly empathetic."}
\end{quote}

P10's statements imply a larger concern that AI systems lack the contextual understanding that human experts provide. Several participants suggested that integrating human expertise into the training process could address this gap (P9, P13, P15, P17). For example, participants imagined scenarios where experienced managers or leadership coaches contribute by designing realistic scenarios, contextualizing feedback, and even reviewing AI-mediated interactions to provide deeper insights. 

Additionally, P15 described a vision of human-AI collaboration where \textit{he}, as a coach, would work alongside the AI system rather than being replaced by it. As a CEO, he explained how this partnership would function:

\begin{quote}
\textbf{P15:} \textit{"I can educate my teammates...I've got a project manager that doesn't have the empathy side yet...So okay, put them in the seat, run the scenario. See how they respond. It'll tell me a lot about that project manager...And I'm like, Oh, yeah, you didn't do that right. Let me show you why, let me show you what this response to their question does."}
\end{quote}

In this model, P15 would leverage the AI system to create revealing simulations while maintaining his essential role as a coach who interprets results and provides contextualized guidance. This approach illustrates how managers envision AI as a diagnostic and demonstration tool within human-led coaching relationships rather than as a standalone training solution.

Likewise, P13 highlighted the possibility of using the system in collaborative peer-learning contexts,envisioning team-based applications beyond individual training. When asked how he would use such a system, he explained: 

\begin{quote}
\textbf{P13:} \textit{"One way that I would use it is to let my team use it to see how they communicate and see how many coach feedback reflection similarities there are. You know, like, wow, the entire team... we all had clarity and communication as one of our feedback options."}
\end{quote}

P13's perspective reveals how managers see potential for AI systems to identify collective communication patterns and team-wide improvement areas. By allowing team members to engage with the system individually while aggregating insights at the group level, P13 envisioned a data-informed approach leveraging AI systems like \textsc{CommCoach} to address recurring communication barriers within his team. Such peer-learning experiences are viewed as a means of improving both indvidual and group communication skills, using AI-generated feedback to stimulate meaningful conversations about shared communication challenges.

\subsubsection{Developing Procedural Communication Memory While Retaining Agency}
Participants described the process of developing communication skills as akin to \textbf{developing procedural communication memory}, or "muscle memory". P1 compared this to strength training, noting, \textit{"For a newer leader...maybe this will help them get some reps and sets in."} Participants recognized that repeated practice helped internalize these strategies, making them easier to apply instinctively when needed. The opportunity to refine communication skills through repeated attempts was particularly valued, as it allowed participants to build confidence and competence over time (P3, P13, P16).

Despite the majority preference for feedback-driven practice, participants also took note of \textbf{maintaining a balance between relying on AI guidance and trusting their own instincts}. Although most participants highly valued the immediate feedback provided by the coach agent, P16 raised a counterpoint, stating, \textit{"It might be for the best that I try to handle the task just from my point of view first, and then see what the coach recommends."} P16's thought suggests a potential benefit from systems that encourage users to first form their responses independently before seeking AI guidance, as reflected in the statement. This focus on maintaining personal agency is consistent with the varied communication styles participants reported using effectively in their previous experiences (\S\ref{sec:findings_background}).

In addition, underscoring the importance of personal agency, participants highlighted the need for systems to \textbf{adapt to managers at different career stages}, as training requirements often vary depending on experience and responsibilities (P1, P8). Early career managers may benefit more from foundational communication strategies, while experienced leaders may require nuanced approaches tailored to advanced managerial contexts. As P1 explained, \textit{"every leader is at a different stage in their career, based on their experience, based on their education, based on their rank...a leader that's been leading for ten years, is going to probably need something different than somebody who's only been leading for two."} This adaptability suggests that systems should empower users by addressing their individual growth trajectories rather than adopting a one-size-fits-all approach. P12 and P16 echoed this sentiment, emphasizing that training effectiveness depends on aligning resources with the specific needs and challenges faced by leaders at different stages of their careers.

\subsubsection{Experimenting Conversational Paths and Outcomes in a Safe Manner}
Participants emphasized the importance of iterative practice in refining their communication skills. They valued opportunities to revisit scenarios, adjust their communication approaches, and reflect on feedback. P16 described the benefit of exploring different conversational paths using the chat branching feature, stating, \textit{"You can see how the other person might react if you say something differently or take a different approach."} They felt that this process was particularly valuable for mastering complex communication skills, such as balancing empathy with assertiveness or tailoring messages to different audiences (P2, P7, P11, P16).

Beyond the benefits of iterative practice, participants noted the importance of \textbf{having a safe, non-judgmental environment to practice and experiment} with different communication techniques (P1, P3, P4, P10, P11, P13, P14, P16, P17). They felt they could push their limits and try new strategies that they otherwise might hesitate to use in workplace situations (P5, P12, P17). However, P9, who has experience leading teams in both the military and finance industry, cautioned that while systems like \textsc{CommCoach} provide an effective environment for practice, users still need human-to-human practice:

\begin{quote}
\textbf{P9:} \textit{I think it can teach you how to think about how you're gonna give the talk, but to actually get better at doing it verbally, you have to stand up and look someone in the eye and actually do it. That could be like a higher pressure [situation].}
\end{quote}

\section{Discussion}
\begin{figure*}[h]
    \centering
    \includegraphics[width=\textwidth]{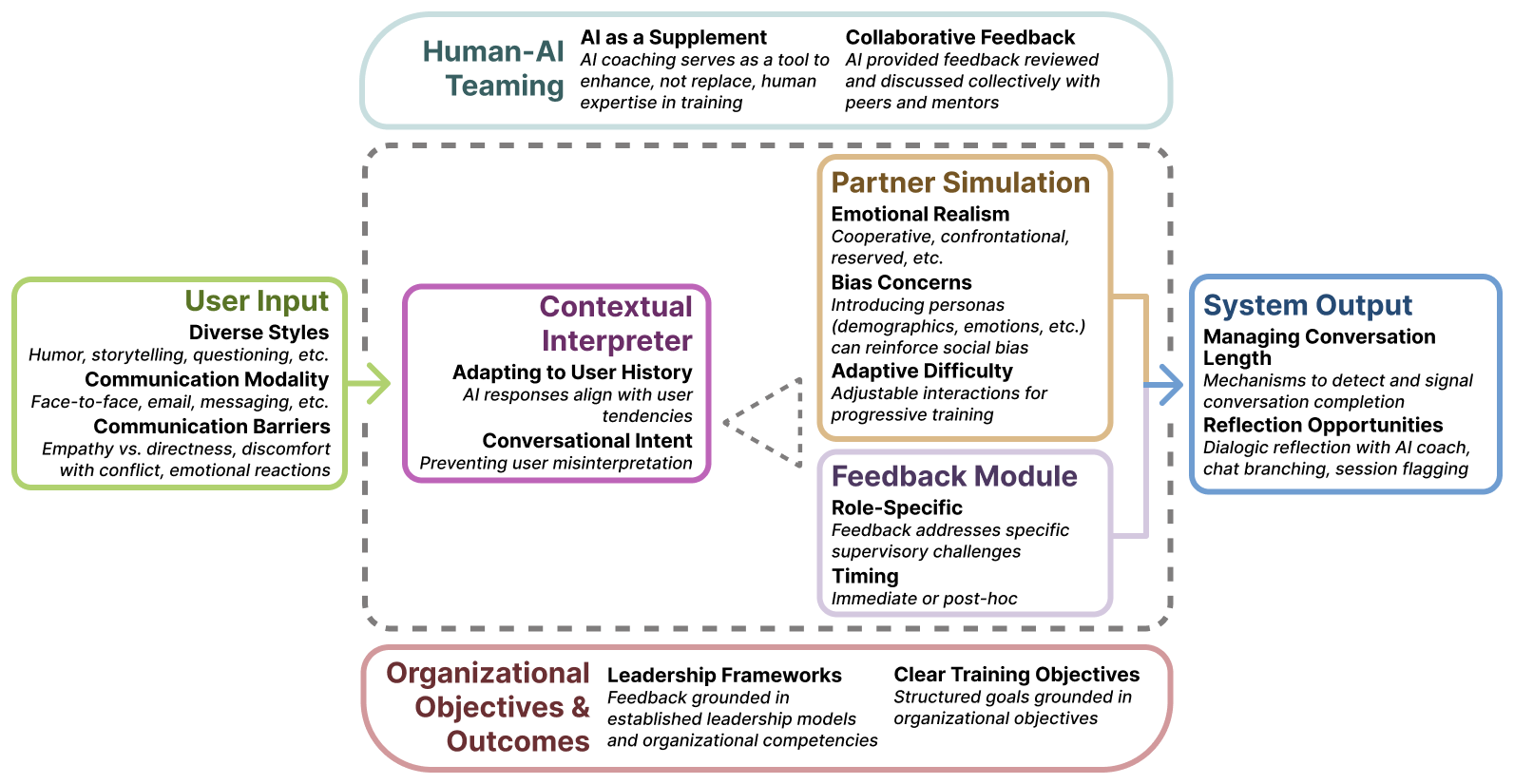}
    \caption{Framework for AI-assisted managerial communication training. \textbf{User Input} introduces diverse communication styles, approaches, and barriers that the system must interpret and adapt to in order to provide effective training. The three core components within the dotted box, the \textbf{Contextual Interpreter}, \textbf{Partner Simulation}, and \textbf{Feedback Module}, work together to adapt AI-driven role-play and feedback to user communication styles, ensuring relevant and effective training. These components interact dynamically to refine AI responses, balance realism in conversational partners, and provide tailored feedback, ultimately shaping the \textbf{System Output} in the form of structured reflection and controlled conversation flow. To maintain alignment with real-world managerial training needs, the system is guardrailed by \textbf{Organizational Objectives \& Outcomes} and \textbf{Human-AI Teaming}. Organizational Objectives \& Outcomes ensure AI responses remain consistent with workplace objectives and structured leadership frameworks, while Human-AI Teaming positions AI as a supplemental tool that supports, rather than replaces, human mentorship and collaborative learning.}
    \label{fig:system-concept}
    \Description{The figure presents a framework for AI-assisted managerial communication training, structured around key components that interact to provide effective coaching. User Input represents diverse communication styles, modalities, and barriers, such as humor, directness, or discomfort with conflict, which the AI system must interpret and adapt to. At the core of the framework, three interconnected modules—Contextual Interpreter, Partner Simulation, and Feedback Module—work together to refine AI-driven role-play and feedback. The Contextual Interpreter personalizes AI responses based on user history and conversational intent, preventing misinterpretations. Partner Simulation ensures realistic interactions by adjusting emotional realism, addressing bias concerns, and modifying difficulty levels for progressive training. The Feedback Module delivers role-specific guidance and timely interventions, either immediately or post-hoc, to enhance communication strategies. These elements shape the System Output, which manages conversation flow through mechanisms for detecting conversation completion and facilitating reflection opportunities, such as AI-coached dialogic reviews and session flagging. The framework is further supported by Organizational Objectives & Outcomes, ensuring AI-driven coaching aligns with structured leadership frameworks and training goals. Finally, Human-AI Teaming positions AI as a collaborative supplement to human expertise, emphasizing AI's role in enhancing, rather than replacing, mentorship and peer learning.}
\end{figure*}

Building on our findings, this section examines the theoretical and design tensions that emerge when conceptualizing AI-assisted communication training for workplace contexts. We present a framework (Figure \ref{fig:system-concept}) that illustrates the interconnected components of such systems: \GreenHL{\textbf{User Input}}, \PinkHL{\textbf{Contextual Interpreter}}, \PeachHL{\textbf{Partner Simulation}}, \PurpleHL{\textbf{Feedback Module}}, and \BlueHL{\textbf{System}} \\ \BlueHL{\textbf{Output}}, all bounded by \RedHL{\textbf{Organizational Objectives}} and \\ \LightBlueHL{\textbf{Human-AI Teaming}} considerations. 

Our discussion unpacks how managers' diverse communication styles and expectations create challenges for AI system design, highlights the delicate balance between adaptability and consistency in AI feedback, and explores the implications of human-AI teaming in managerial communication training. These insights extend beyond simple implementation considerations to address fundamental questions about how AI might meaningfully participate in the development of managerial communication skills in workplace contexts.

\subsection{\GreenHL{User Input} \& \PinkHL{Contextual Interpretation}: Navigating Different Discursive Styles Between System Users}
\label{sec:discussion_discursive_leadership_styles}
Discourse theory of leadership holds that speech acts are essential to successful leader-follower communication (\S \ref{sec:leadership_theory}). This point of view also recognizes the importance of context, which requires customized approaches depending on the situation or organizational culture \cite{iivari_RelationshipOrganizationalCulture_2007}. We found that participants came with a wide variety of previous training experiences in addition to a \ul{wide range of preferred discursive management styles}, such as humor, storytelling, and signaling empathy (\textbf{RQ 1}).

Participants also adapted their communication modality based on the interaction context, aligning with how professionals change their communication strategies in different workplace environments (\textbf{RQ 1}). P7 and P17 initially approached their system use as if they were composing an email or a written message, while others engaged in a more conversational, face-to-face style. When asked to retry the conversation in a simulated spoken interaction, we observed the conversational partner agent's responses change to match their style and the coach's interventions decline. Ultimately, \ul{the differences in communication modalities} shaped how participants interacted with the conversational partner agent, which in turn affected the AI partner's responses and the coach's feedback. 

To improve system adaptability, AI-assisted communication training systems should not only interpret the nuances of workplace conversations, but also recognize common \ul{communication barriers} experienced by users. These challenges include balancing empathy with directness, discomfort with conflict resolution, and difficulty in adapting tone and structure depending on the modality (e.g., email vs. face-to-face) (\textbf{RQ 1}). Reflecting past HCI research, Johnston et al. showed that adaptive algorithms can learn from user interactions, dynamically adjusting system responses to better align with user experiences \cite{johnston2019framework}. Similarly, Wintersberger et al. emphasize the importance of continuous adaptation in human-AI systems, particularly in contexts like AI-assisted text entry, where traditional methods may be inadequate \cite{wintersberger2022designing}. 

Theese findings suggest that AI-assisted communication training systems may need to consider ways to balance adaptability, personalization, and consistency. This presents an interesting challenge for future researchers and system designers: How might a system potentially adapt to diverse styles while maintaining consistent performance? Future work could explore how to recognize, interpret, and respond to the preferred modalities of various user styles while aiming to provide a reliable user experience.

\subsection{\PeachHL{Partner Simulation}: Balancing Emotional Realism of LLM Agents with Potential to Reinforce Social Biases}
\label{sec:discussion-biases}
Participants in our study expressed a strong desire to \ul{control the tone or emotions of the conversational partner agent} (\S \ref{sec:findings_setup}). Some participants sought to shape the personality of the conversational partner with demographic details such as age and education level, while others used names that were clearly gender-specific when creating custom scenarios (\textbf{RQ 2}). Although LLMs are emerging as a potential means of understanding human emotion \cite{rashkin_EmpatheticOpendomainConversation_2019, wang_EmotionalIntelligenceLarge_2023, cuadra_IllusionEmpathyNotes_2024} and simulating realistic human behavior \cite{park_GenerativeAgentsInteractive_2023a, kovacevic_ChatbotsAttitudeEnhancing_2024}, accurately simulating human emotions remains a challenge. For instance, NLP researchers have identified biases in language models' detection and interpretation of linguistic characteristics associated with certain demographic groups, such as African-Americans and women. For example, African-American vernacular English (AAVE) is often tagged as more toxic \cite{sap_RiskRacialBias_2019} and abusive \cite{davidson_RacialBiasHate_2019} compared to standard American English. Women's speech is also often classified as more emotional, aligning with societal stereotypes rather than the actual emotional experiences of women \cite{plaza-del-arco_AngryMenSad_2024}. 

If participants want to bring more details to the conversational partner agent through customizable personas, \ul{LLM agents may not only reflect their inherent biases} based on their model of these specific social demographic groups but also \ul{train their users to communicate with groups based on these biases} \cite{sun_BuildingBetterAI_2024}. Finding the right balance between emotional authenticity in the conversational partner and training effectiveness requires careful design considerations. When designing our system, we specifically masked names to reduce gender bias. However, controlling for such issues may become more difficult if system designers provide users with the option to customize LLM agent personas and demographic traits. System designers should pay careful attention to biases in both users and the system in an effort to maintain system consistency and promote fairness.

\subsection{\BlueHL{System Output} \& \RedHL{Organizational Guidance}: When Does Training Stop?}
\label{sec:discussion_whendoestrainingstop}
When asked about the importance of communication in their roles as supervisors, all our participants mentioned its role in achieving specific workplace objectives, such as task delegation, briefing, or conflict resolution (\textbf{RQ 1}). Similarly, when using \textsc{CommCoach}, participants were implicitly assigned conversational objectives in default and custom scenarios. For example, persuading an employee to work late over the weekend or navigating a difficult feedback session. Most, if not all, workplace discourses have natural stopping points via conversational goals \cite{koester_InvestigatingWorkplaceDiscourse_2006}, be that of some acknowledgment of one conversational partner or simply the exhaustion of a topic. This begs the question of \ul{when and how the system should stop a conversation based on the objective}.

Open-domain language models acting as chatbots, such as ChatGPT, are designed to carry out a conversation indefinitely \cite{serban_BuildingEndToEndDialogue_2016, gao_NeuralApproachesConversational_2018}, in contrast to the natural tendency of workplace discourse. Although we did not design when the conversation should end, system designers should consider appropriate end points of conversations. In our study, some participants indicated their desire or the option to receive feedback at the end of a conversation. However, this cannot happen without the design for such instances. Furthermore, as discussed in \S \ref{sec:discussion_discursive_leadership_styles}, users have different styles and perceptions of discursive leadership. This may also mean that they follow different conversational arcs or have different thoughts about what is considered the end of a conversation. Designers should then consider appropriate frameworks and thresholds that match organizational objectives or allow the customization of objectives by a system administrator \cite{wen_NetworkbasedEndtoEndTrainable_2017, liu_DialogueLearningHuman_2018}. One way this can be achieved is through the integration of external knowledge sources related to \ul{organizational training objectives} \cite{guyre_PromptEngineeringLLM_2024}. Advancements in LLM techniques, such as retrieval-augmented generation (RAG) \cite{lewis_RetrievalAugmentedGenerationKnowledgeIntensive_2021} and fine-tuning \cite{raffel_ExploringLimitsTransfer_2019}, have the potential to enhance the system's ability to adapt to different conversational dynamics and ensure that they align with organizational competencies.

\subsection{\LightBlueHL{Human-AI Teaming}}
Human-AI teaming \cite{horvitz_PrinciplesMixedinitiativeUser_1999, shneiderman_HumanCenteredArtificialIntelligence_2020, zhang_IdealHumanExpectations_2021} represents a potential avenue to improve AI-assisted communication training systems. We use the term "human-AI teaming" to emphasize the role of artificial intelligence in facilitating collaborative processes, leveraging its advanced capabilities in tandem with human expertise. Participants in our study recognized the limitations of relying solely on AI for communication training, particularly in complex and context-sensitive scenarios (\S\ref{sec:findings_practice}) (\textbf{RQ 2}). Effective human-AI collaboration relies on aligning AI capabilities with human expertise to maximize system effectiveness and ensure that the system's outputs are relevant and actionable \cite{inkpen_AdvancingHumanAIComplementarity_2023, mahmud_StudyHumanAI_2023}. This partnership utilizes the complementary strengths of humans, such as contextual understanding and judgment, alongside AI features such as scalability and pattern recognition, leading to results that neither could achieve independently \cite{licklider_ManComputerSymbiosis_1960}. Several participants envisioned hybrid systems where human experts and AI systems collaboratively support training processes (\textbf{RQ 2}). For example, human coaches could design realistic training scenarios and provide tailored feedback to address the specific needs of users. This is consistent with broader findings in human-AI interaction, which emphasize the importance of balancing autonomous AI actions with human oversight, particularly in high-stakes domains where errors carry significant consequences \cite{horvitz_PrinciplesMixedinitiativeUser_1999}. 

One participant envisioned using the system in team meetings, where colleagues could \ul{collaboratively discuss AI-provided feedback on communication scenarios} (\textbf{RQ 2}). This aligns with research that emphasizes the effectiveness of AI in improving collaboration and learning outcomes in peer-to-peer environments. For example, Magnisalis et al. reviewed adaptive and intelligent systems supporting collaborative learning, finding that well-designed AI systems can improve domain knowledge and collaboration skills through unobtrusive interventions and pre-task support \cite{magnisalis_AdaptiveIntelligentSystems_2011}. McLaren et al. demonstrated how AI techniques can identify key contributions and patterns in group discussions, helping facilitators guide meaningful debates and improve group learning outcomes \cite{mclaren_SupportingCollaborativeLearning_2010}. These examples highlight the potential for AI to improve personal learning and encourage deeper team interactions by providing consistent, context-aware guidance, and practical feedback. However, Mena-Guacas et al. emphasized that while AI holds significant promise for collaborative learning, its success depends on addressing design challenges and ensuring that AI augments rather than replaces traditional mentorship frameworks \cite{mena-guacas_CollaborativeLearningSkill_2023}.

AI-assisted communication training systems may sometimes serve better as \ul{components of broader training environments rather than standalone solutions} \cite{arakawa_CoachingCopilotBlended_2024}. Careful design is required to merge effective human-AI teaming within AI-mediated training contexts to complement AI's scalability with the nuanced expertise of human mentors. This aligns with design principles advocating AI systems that enable granular feedback, transparency, and adaptable interfaces to support dynamic collaboration \cite{barmer_HumanCenteredAI_2021}. Furthermore, Amershi et al.'s guidelines for human-AI interaction stress the necessity of making AI capabilities and limitations transparent (G1 and G2) to facilitate this collaboration effectively \cite{amershi_GuidelinesHumanAIInteraction_2019}. Future research may investigate how best to divide responsibilities between AI systems and human experts, considering scalability, cost-effectiveness, and training efficacy. Comparative studies examining AI-facilitated peer learning models against traditional mentoring methods could provide valuable information on optimizing human-AI collaboration in training contexts. Finally, further exploration of the socio-emotional dynamics in human-AI interaction, such as fostering trust and maintaining engagement, could inform the design of more robust and user-centered training systems \cite{kolomaznik_RoleSocioemotionalAttributes_2024}.

\subsection{Ethical Considerations}
While AI-assisted training systems such as \textsc{CommCoach} offer promising opportunities to enhance managerial communication skills, they also come with potential ethical concerns. In our findings, P15 mentioned using a system like \textsc{CommCoach} to help him find areas for improvement in his employees. Although his intentions may have been purely for professional development, there is a risk of misuse in such instances, especially if organizations deploy these tools not just for developmental training, but as instruments of employee performance evaluation or surveillance. This usage could compromise employee trust, privacy, and sense of autonomy \cite{ball_ElectronicMonitoringSurveillance_2021, hickok_PolicyPrimerRoadmap_2023}. Workplace surveillance has been found to undermine morale and psychological safety, leading to reduced engagement and a breakdown in the employer–employee trust relationship \cite{thiel2022monitoring}.

In addition, reliance on the approval provided by AI systems may lead managers to prioritize performance within the training system over genuine building of interpersonal relationships. This could prove harmful if systems are not adequately aligned with organizational competencies and agreed upon leadership principles, potentially reducing managerial effectiveness and harming workplace culture \cite{raisch_ArtificialIntelligenceManagement_2021, hickok_PolicyPrimerRoadmap_2023}. Algorithmic feedback loops may incentivize managers to focus on what is easily measurable or rewarded by the system, rather than on meaningful, trust-building interactions with their teams \cite{kochling_DiscriminatedAlgorithmSystematic_2020}. Future research and practical implementations should explore the boundaries of AI use within workplace communication training contexts, focusing on transparency, consent, and maintaining balance between qualitative human judgment and AI-driven insights.  This balance is essential to ensure that AI augments, rather than undermines, human-centered leadership development.

\subsection{Limitations and Future Research Directions}
\label{sec:discussion_limitations}
This study was conducted as an exploratory investigation of the potential of AI systems to support managerial communication training. Although this approach allowed us to uncover initial user perceptions and highlight key design considerations, it does not provide definitive evidence of the system's efficacy in real-world applications. We also attempted to sample a diverse set of managers with respect to managerial experience, but the number of participants ($n=17$) may not fully represent the wide range of managerial styles, organizational contexts, or other cultural factors that influence communication practices. Consequently, the findings may not be broadly generalizable in all managerial contexts.

The think-aloud protocol used during the semi-structured interview could cause users to provide biased answers if they provide what they believe we desire \cite{zhang_ThinkaloudProtocols_2019} and their personalities may affect their comfort in verbalizing their thoughts during the use of the system \cite{saliba_PersonalityParticipationWho_2014}. We tried to ask questions and appropriately investigate users when it appeared that they were thinking, but this does not guarantee transparency or authenticity. These interviews may not have effectively captured all the potential user needs for AI-assisted communication training systems. Additionally, the study was conducted in a controlled environment (e.g., in a Zoom meeting with hypothetical role-play scenarios), which does not reflect the potential use cases of the system in everyday training settings. Finally, while grounded theory was used to analyze the qualitative data and generate exploratory insights, this approach does not provide quantitative evidence of the system's efficacy. The design of this study does not allow conclusions to be drawn regarding the comparison of the system with traditional training methods or its quantifiable effects on communication skills in leadership. These limitations should inform future research that seeks to assess AI systems in wider and more diverse contexts.

\subsubsection{Future Work}
The long-term effects of systems like \textsc{CommCoach} on lasting behavioral change and, consequently, organizational effectiveness should be the subject of future research. Researchers may be able to gain a more comprehensive understanding of the complex efficacy of these systems through longitudinal studies, including specific samples of managers from different sectors. This research may also yield valuable information on how to incorporate AI-assisted communication training systems into existing managerial development programs. Extending research of such systems within the existing managerial training field may also reveal critical managerial skills from many industry domains and cultures that should be integrated into these kinds of system. Finally, future studies should investigate how AI-assisted communication training systems, especially in underserved communities, can expand access to expert-level resources.

\section{Conclusion}
This work provides an exploratory investigation into the potential for AI-assisted systems to serve as tools for the training of managerial communication. By investigating the user perceptions of 17 managers of an AI-based role-play functional probe, we sought to understand their experiences with workplace communication training (\textbf{RQ 1}) and gathered insights into their perceptions of AI-assisted communication training potential (\textbf{RQ 2}). We developed a functional probe, \textsc{CommCoach}, to facilitate this exploration. \textsc{CommCoach} was developed as a role-play sandbox for managers to practice and improve their workplace communication skills with the help of an AI partner and coach. Participants engaged with an AI conversational partner agent according to a predefined or customized scenario, while an AI coach agent observed and intervened to provide feedback. By engaging in such scenarios, users could practice and experiment with different styles and techniques of conversation at work. 

With regard to \textbf{RQ 1}, we found that managers face various challenges in balancing communication styles, emotional dynamics, and situational contexts, often addressing them through ad hoc methods with varying effectiveness. Regarding \textbf{RQ 2}, managers perceived AI tools as a potential means to provide personalized, iterative, and safe environments to practice communication skills, although they strongly emphasized the need to integrate human oversight to ensure ethical and practical effectiveness. Specifically, participants highlighted the value of realism in simulated conversations and reported finding the coach agent's feedback meaningful, which supported reflection on their communication approaches. However, users sought features that they felt would enhance the customization and realism of their simulated conversations, such as the persona development of AI partner and speech capabilities. Participants noted the limitations of standalone AI systems, particularly in addressing context-sensitive or emotionally complex scenarios, and expressed a preference for hybrid approaches that integrate human oversight and expertise. This aligns with the concept of human-AI teaming, in which AI systems offer scalable, consistent support while humans provide the nuanced, contextual understanding necessary for more complex situations. 

Our findings also point to critical considerations when designing AI-assisted communication training systems. These systems must balance personalization and adaptability with fairness and consistency, addressing challenges such as various discursive styles among users and the risk of perpetuating biases inherent in LLMs. Encouraging open and interactive exchanges, AI-assisted communication training systems can more effectively adjust to the requirements of various user groups, while ensuring that their outputs conform to both organizational guidelines and ethical standards.


\bibliographystyle{ACM-Reference-Format}
\bibliography{references}

\clearpage
\begin{onecolumn}
\appendix
\section{\textsc{CommCoach} System Design}
\label{app:system_design}

\subsection{Front End}
The front end of \textsc{CommCoach} is designed with a minimal aesthetic similar to popular chatbots and messaging systems such as ChatGPT, Messenger, and WhatsApp. This familiar mental model helps users quickly understand and navigate the interface, reducing the learning curve and enabling them to focus on the communication training experience. The user interface is organized into three main areas: the chat pane, where messages to and from the interlocutor agent are displayed; the scenario pane, which presents the current role-play scenario; and the feedback pane, where feedback messages from the coach agent are shown alongside the associated chat messages.

\subsection{Back End}
The \textsc{CommCoach} back end handles core functionality, including managing WebSocket connections, processing user input, and generating agent responses. The back-end architecture consists of the following components:

\begin{itemize}
    \item WebSocket Server: Implemented using FastAPI, the WebSocket server receives messages from the front end and routes them to the appropriate functions for processing.
    \item LLM Agents: Two GPT-4-powered agents, the interlocutor agent and the coach agent, are responsible for generating content and providing feedback based on the user's input and the current scenario. The agents are implemented using a custom Python framework and the GPT-4 API (\verb|gpt-4-0125-preview|). To minimize the effect of gender bias in the agent responses \cite{brown_LanguageModelsAre_2020}, user and interlocutor agent names are masked using regex replace methods before being sent to the agents.
    \item Database: MongoDB is used to store all messages from the user and agents, along with keys associating messages to conversation sessions and users.
\end{itemize}

The back end handles two main tasks: intervention assessment and feedback response.

\subsubsection{Intervention Assessment}
The purpose of the intervention assessment is to determine whether the coach agent should provide feedback based on the conversation history and scenario. The process involves the following steps:

\begin{enumerate}
    \item The conversation history, scenario, and intervention instruction are combined into a prompt and sent to the coach agent.
    \item The coach agent analyzes the prompt and responds with a boolean value indicating whether it believes an intervention is necessary.
    \item \begin{itemize}
        \item[(a)] If the value is \verb|false|, the interlocutor agent is prompted to provide the next response in the conversation.
        \item[(b)] If the value is \verb|true|, the coach agent is prompted to provide specific feedback on the user's latest response.
    \end{itemize}
\end{enumerate}

\subsubsection{Feedback Response}
When the intervention assessment determines that feedback is required, the back end generates a feedback response using the following process:

\begin{enumerate}
    \item The conversation history, scenario, and feedback instruction are combined into a prompt and sent to the coach agent.
    \item The coach agent analyzes the prompt and provides specific feedback on what needs to be addressed in the user's latest response.
    \item The feedback is formatted as JSON for easier parsing and display on the front end.
\end{enumerate}
 
\subsubsection{Scenario Editing}
Users can create custom scenarios through the \textsc{CommCoach} scenario editor by sending initial requests to the coach agent. The coach agent generates a scenario based on the user's input, which the user can accept, refine, or clear. If the user chooses to refine the scenario, they provide additional suggestions that are combined with the previous iteration and sent back to the coach agent for an update. This iterative process continues until the user is satisfied with the generated scenario.

\subsubsection{Dialogic Reflection with the Coach Agent}
To facilitate deeper reflection and understanding, \textsc{CommCoach} enables users to engage in focused discussions with the coach agent about specific feedback items. When a user clicks on a feedback item, a modal window appears containing the feedback text and a chat box. The user can send messages directly to the coach agent, asking questions or seeking clarification about the feedback. The coach agent's responses are displayed within the modal, distinguished from the interlocutor agent's messages by color.

\subsubsection{Chat Branching}
\textsc{CommCoach} supports conversational experimentation (reattempts) through chat branching, allowing users to explore alternative paths and observe the impact of their communication choices. Chat messages are organized into a tree structure, with each message linked to its parent and child messages. When a user edits and saves a message, a new branch is created, preserving the original conversation flow. Users can switch between different branches to compare feedback and outcomes, promoting a deeper understanding of the consequences of their communication strategies. The front end handles the display of the active branch, while the back end serves all messages in a session to minimize latency when switching branches. Only the messages in the active branch are considered during the intervention assessment, interlocutor response, and feedback response processes.

\subsection{Prompts and Parameters}
\textbf{Actor System Prompt}:
\begin{promptbox}
You are an actor who is playing the role of an employee, <actor_name>. You are participating in a role-play dialogue with a leadership trainee human whose name is <user_name>. The scenario that was provided to the trainee was the following:

Scenario: {scenario}

Your specific job is to provide a response to the latest communication from the user based on your role as <actor_name>. Keep it short and give room for the trainee to give you direction and practice their leadership skills. Do not provide any suggestions or feedback to the trainee. Respond in a colloquial manner. Sometimes have emotional fluctuations in your responses.
\end{promptbox}

\textbf{Coach System Prompt}:
\begin{promptbox}
You are an expert leadership coach with lots of knowledge of leadership theory. You observe a role-play dialogue between a human trainee and an AI actor.

The human user is <user_name>, and the actor is <actor_name>. The scenario that was provided to the trainee is the following:

Scenario: {scenario}
\end{promptbox}

\textbf{Feedback Task}:
\begin{promptbox}
Your specific job is the following:
Evaluate the last communication from <user_name>. Provide feedback addressing <user_name> based on leadership theory and approaches that could help them improve their communication to become more effective. Only critique <user_name>'s communication and not <actor_name>'s.

The current dialogue is given between the triple backticks below.
```{dialogue}```
\end{promptbox}

\textbf{Feedback Task Output Format}:
\begin{promptbox}
The output should be formatted as a JSON instance that conforms to the JSON schema below.

As an example, for the schema {"properties": {"foo": {"title": "Foo", "description": "a list of strings", "type": "array", "items": {"type": "string"}}}, "required": ["foo"]}
the object {"foo": ["bar", "baz"]} is a well-formatted instance of the schema. The object {"properties": {"foo": ["bar", "baz"]}} is not well-formatted.

Here is the output schema:
```
{"properties": {"bluf": {"title": "Bluf", "description": "Summary of feedback items directed to the human trainee of what they can improve. Be empathetic and avoid should statements", "type": "string"}, "item_titles": {"title": "Item Titles", "description": "list of titles for detailed items to improve", "type": "array", "items": {"type": "string"}}, "items": {"title": "Items", "description": "list of detailed items to improve", "type": "array", "items": {"type": "string"}}}, "required": ["bluf", "item_titles", "items"]}
```
\end{promptbox}

\textbf{Intervention Task}:
\begin{promptbox}
Your specific job is the following:
Evaluate the last communication chat from <user_name> and judge whether or not you should intervene.

The current dialogue is given between the triple backticks below.
```{dialogue}```
\end{promptbox}

\textbf{Intervention Task Output Format}:
\begin{promptbox}
The output should be formatted as a JSON instance that conforms to the following JSON schema

{'intervention': true/false}
\end{promptbox}

\textbf{Actor LLM Model}:
gpt-4-0125-preview

\textbf{Coach LLM Model}:
gpt-4-0125-preview

\textbf{Actor Temperature}:
0

\textbf{Coach Temperature}:
0

\textbf{Scenario Generation Task}:
\begin{promptbox}
You are an expert leadership coach with lots of knowledge of leadership theory. You observe a role-play dialogue between a human trainee and an AI actor.

The human user is <user_name>, and the actor is <actor_name>. Your specific job is to take the inputs the user provides and generate a scenario prompt for the user. The scenario prompt should be directed to the user using `you` language. If the user provides the name of an actor, replace the actor's name with the string '<actor_name>'. If the user does not provide a name, use the string 'Taylor'. You will also provide a short explanation as to why you created the scenario in the way that you did. If there is a previous scenario edit provided, then you should refine that based on the user input. In the scenario do not ask questions, give hints or tips, or objectives or tasks for the user. Keep in mind that this scenario will be used for a role-play exercise that the user will take part in.
\end{promptbox}

\textbf{Scenario Task Output Format}:
\begin{promptbox}
The output should be formatted as a JSON instance that conforms to the following JSON schema 

{'scenario': 'string', 'actor_name': 'string', 'scenario_explanation': 'string'}
\end{promptbox}

\textbf{Discussion Task}:
\begin{promptbox}
Your specific job is to respond to the latest communication from the user regarding the following feedback that you have provided to them on the following dialogue.

Dialogue:
{dialogue}
Feedback:
{feedback}
\end{promptbox}

\section{User Study}
\subsection{Interview Protocol}
\label{app:user_study_interview_protocol}
\subsubsection{Leadership Experience}
\begin{itemize}
  \item \textbf{General Experience:} \newline
  How important do you believe effective communication was/is in your experience as a supervisor?
\end{itemize}

\subsubsection{Communication Training Experience}
\begin{itemize}
  \item \textbf{Communication Training}
  \begin{itemize}
    \item Have you ever received any training for effective workplace communication as a supervisor? If so, how much (hours, if possible) can you describe your experience?
    \item Can you describe a particularly effective or ineffective past communication training session you have experienced? What made it so?
    \item What methods were used in your previous communication training, and how did they impact your learning?
    \item What feedback mechanisms were present in your past training, and how did they contribute to your growth as a leader or manager?
    \item Do you have any previous experience with roleplaying for leadership training? If yes, can you describe your experience?
    \item Reflecting on your past training, what elements were missing that could have improved the experience? (limitations)
    \item Was there anything in the training that helped you reflect on what you learned? Did it add to the training?
    \item How have previous communication training experiences shaped your approach to leadership and management?
  \end{itemize}
  \item \textbf{Technology in Training}
  \begin{itemize}
    \item Was there any technology used for the training process in past training? What was it like?
    \item How familiar are you with using AI chatbots?
  \end{itemize}
\end{itemize}

\subsubsection{CommCoach System Use}
Participants will now use the system for about 20 minutes. Ask the participants to think aloud while they are using the system. Take note of any interesting actions or thoughts.

\subsubsection{Roleplay Training}
\begin{itemize}
  \item \textbf{Chatbot Roleplaying}
  \begin{itemize}
    \item How was performing roleplaying via a chatbot?
    \item \textbf{Follow-up:} Do you feel any advantages or disadvantages to this type of system for roleplaying?
    \item How did the immediacy of chatbot roleplay impact your learning process compared to previous training methods?
  \end{itemize}
  
  \item \textbf{Actor}
  \begin{itemize}
    \item Were the training partner's responses realistic and natural? Why or why not?
    \item How did the chatbot's tone or language style affect the believability of the scenarios?
  \end{itemize}
  
  \item \textbf{Scenario}
  \begin{itemize}
    \item Was the scenario realistic or relevant? What made it so or not?
    \item Were you able to explore the scenario and try different responses as desired?
    \item Did the roleplay scenarios challenge your existing communication skills? In what ways?
  \end{itemize}
\end{itemize}

\subsubsection{Coach Feedback}
\begin{itemize}
  \item \textbf{Quality}
  \begin{itemize}
    \item Was the feedback relevant, actionable, and easy to understand? Why or why not?
    \item How personalized did the feedback feel to your specific responses?
  \end{itemize}
  \item \textbf{Quantity}
  \begin{itemize}
    \item Was the amount and length of feedback appropriate, or did you feel overwhelmed/underwhelmed?
  \end{itemize}
  \item \textbf{Timing}
  \begin{itemize}
    \item How did you feel about receiving real-time (in-situ or in-action) feedback?
    \item Was there a moment when you expected feedback but didn’t receive any?
  \end{itemize}
  \item \textbf{Format}
  \begin{itemize}
    \item How did you feel about the format of the feedback (e.g., BLUF summary, hover details)?
  \end{itemize}
  \item \textbf{Trust}
  \begin{itemize}
    \item Do you trust that the feedback is accurate/reliable? Why or why not?
    \begin{itemize}
      \item What information would help make the system feedback more reliable?
      \item What could enhance the credibility of the feedback for you?
    \end{itemize}
  \end{itemize}
  \item \textbf{Effect}
  \begin{itemize}
    \item How did the feedback influence your future interactions within the roleplay?
  \end{itemize}
\end{itemize}

\subsubsection{Reflective Learning}
\begin{itemize}
  \item \textbf{Perception}
  \begin{itemize}
    \item Do you feel that the feedback provided by the coach spurred you to reflect on what you said?
    \item Was your ability to reflect on your conversation boosted or limited in any way?
  \end{itemize}
  \item \textbf{Timing}
  \begin{itemize}
    \item Do you feel that performing reflection immediately (in-situ) was/is beneficial?
  \end{itemize}
  \item \textbf{Perspective-taking}
  \begin{itemize}
    \item Did the feedback that the coach provided broaden your perspective of your roleplaying partner? That is, do you feel as though you have a broader appreciation of their burden?
    \item How did perspective-taking influence your approach to communication challenges?
  \end{itemize}
\end{itemize}

\subsubsection{Overall System Design}
\begin{itemize}
  \item \textbf{Navigation}
  \begin{itemize}
    \item Did you have any trouble navigating around the system and performing actions?
    \item What improvements could make the system more user-friendly?
  \end{itemize}
  \item \textbf{Layout}
  \begin{itemize}
    \item Did the layout feel intuitive? Why or why not?
    \item How could the layout be altered to facilitate learning better?
  \end{itemize}
  \item \textbf{Purposefulness}
  \begin{itemize}
    \item Were there system features that you found particularly beneficial or unnecessary?
  \end{itemize}
\end{itemize}

\subsubsection{Overarching Questions}
\begin{itemize}
  \item \textbf{Human-AI Interaction}
  \begin{itemize}
    \item How was working with AI in this roleplaying setting?
    \item What preconceptions about AI were challenged during this experience?
  \end{itemize}
  \item \textbf{Workplace Communication}
  \begin{itemize}
    \item Have you experienced moments like the one in the scenario?
    \item How often do you feel that problematic communication scenarios arise in the workplace?
    \item Do you feel this tool would have been handy in your past experiences?
  \end{itemize}
  \item \textbf{CommCoach Usage}
  \begin{itemize}
    \item To what extent would you use a system like CommCoach for workplace communication training?
    \item How would you see yourself using this tool?
    \item How does this system fit into your current or desired workflow for communication training?
  \end{itemize}
  \item \textbf{Feedback for Us}
  \begin{itemize}
    \item Are there any specific concerns you feel we, as human-computer interaction researchers, should focus on when researching these communications systems?
  \end{itemize}
\end{itemize}

\subsection{Default and Participant-Generated Scenarios}
\subsubsection{Default} 
You are a boss requesting that an employee Taylor, work extra hours over the weekend to complete the XYZ project before a critical deadline. They are the only person that can do the job, and their progress has been substandard. 

This scenario can also be found in Figure \ref{fig:dg1} in Appendix \ref{app:screenshots}.

\subsubsection{Participant-Generated Scenarios}
\label{app:participant_scenarios}
The following scenarios were generated by participants using \textsc{CommCoach}

\begin{itemize}
    \item You are in a challenging situation where one of your employees, Sharla, is experiencing conflict with their direct supervisor. Sharla has expressed a desire to transfer to a different work center, but the options are limited. The only available position is in a smaller, three-person work center where the workload and the importance of the mission are significantly reduced. Additionally, there are no other vacancies to accommodate Sharla's transfer. You must address Sharla's concerns, manage the dynamics within the current work center, and explore potential solutions to this interpersonal conflict without compromising the team's effectiveness or Sharla's career development.
    \item You are gearing up to assign a crucial task to Taylor, a diligent member of your team. Recognizing your own concerns about the clarity of your communication, you decide to implement a new strategy. After explaining the task, you plan to engage in a reflective conversation with Taylor, where you will encourage them to ask questions and express any uncertainties. This method aims to foster an open dialogue, ensuring that both of you have a clear and shared understanding of the task's requirements and expectations.
    \item You are in the midst of a challenging situation at your company, where you've been tasked with the responsibility of handling a sensitive issue. Taylor, a member of the leadership team, was initially being let go due to performance issues. However, during the process of Taylor's departure, you've uncovered evidence of serious misconduct. Now, you find yourself needing to transition the reason for Taylor's dismissal from performance-related to disciplinary. Today, you are preparing to meet with Taylor to discuss this change. You aim to communicate the situation with a balance of firmness and sensitivity, ensuring that Taylor understands the seriousness of the misconduct while also respecting their dignity during this difficult conversation. 
    \item You stride confidently into the tutoring lab, noticing Taylor, your sole tutor for MATH 1234, rushing in 20 minutes late. A group of students, clearly frustrated, are checking their watches and whispering among themselves. As Taylor hurriedly sets up, you prepare to approach them. You understand the critical role Taylor plays in supporting the students' learning during this hour, and you're ready to discuss the importance of punctuality and the impact of their tardiness on the students' experience. Taylor is defensive because they don't have a valid justification for this act. 
    \item You are the manager at a small company where Morgan has been frequently calling in to work and often leaves early, citing the need to pick up her child from daycare. You've noticed that this pattern has started to affect the team's productivity and morale. Today, you've decided to address the situation. You schedule a private meeting with Morgan in your office. The room is set up to be welcoming, with two chairs facing each other to encourage open communication. As Morgan enters, you greet her warmly, acknowledging her dedication to her family while expressing your concern about how her current schedule is impacting the team and the work environment. You emphasize the importance of finding a balance that respects both her responsibilities as a parent and her commitments to her role within the company.
    \item You are a manager at a company where your employee, Taylor, has expressed dissatisfaction with having to travel on Sunday to prepare for and conduct a training session on Monday. Taylor feels that this arrangement encroaches on their personal time and has come to you seeking a resolution. You value Taylor's contribution to the team and understand the importance of work-life balance. You are now faced with the challenge of addressing Taylor's concerns while ensuring the training session's success.
    \item You are the manager of a small team at a tech company, and you've been facing ongoing performance and attitude issues with an employee named Bob. Despite multiple warnings and attempts at corrective action, Bob's behavior and work quality have not improved. Today, you have scheduled a meeting with Bob to discuss his employment status. As you prepare for this difficult conversation, you review the company's termination procedures, gather documentation of Bob's performance issues, and plan how to communicate the decision compassionately yet firmly.
    \item You are in a leadership position at your organization, and you've been tasked with implementing a new, agile video-based training program to meet the urgent demands of senior leadership. However, your senior subordinate, Taylor, is resistant to this new approach. Taylor believes that by not following the full instructional design models, this initiative cuts corners and compromises the quality of the training. Despite Taylor's resistance, you need to navigate this situation carefully, balancing the urgency from senior leadership with the concerns raised by Taylor, ensuring the project moves forward effectively.
    \item You sit down in your office, the weight of the upcoming deadline for the XYZ project pressing heavily on your shoulders. You know that the team has been working hard, but there's still a significant amount of work to be done. You call Taylor into your office, aware of his tendency to be a bit recalcitrant and reserved. Despite this, you've always found his work to be generally acceptable, if not commendable at times. You explain the critical nature of the deadline and the importance of the project to the company. You acknowledge Taylor's contributions so far but stress the need for extra effort from everyone, including him, to ensure the project's success. You request that he works extra hours over the weekend, emphasizing how crucial his role is in the completion of the XYZ project.
    \item You are leading a critical project at your company, and you've assigned a task to a staff member, Taylor, that is crucial for the project's success. However, Taylor refuses to carry out the task, stating that you are not their direct supervisor and therefore, they do not have to follow your instructions. This situation occurs in the middle of a team meeting, with all eyes on you, waiting to see how you will handle this unexpected challenge.
    \item In your role at X corporation, you've observed that Tim, who often works solo, feels isolated from the rest of the team. His work, however, is vital for the company's success across various departments. Your challenge is to make Tim recognize his significant contributions and feel more connected to the team, by pointing out specific examples of how his efforts have led to team victories.
    \item You are a supervisor at a bustling tech company, known for its fast-paced environment and cutting-edge projects. Among your team members is Taylor, an individual who is notably introverted and has always shied away from the spotlight. Despite their preference for staying out of the limelight, Taylor has been an exceptional performer, consistently meeting deadlines well ahead of schedule and managing projects significantly below cost. Recognizing Taylor's contributions is important to you, but you are also mindful of their discomfort with public acknowledgment. You find yourself contemplating the best approach to honor Taylor's achievements in a way that respects their personality and preferences.
    \item You are the manager of a team where Taylor, a key employee, has been consistently arriving late to work, accumulating 10 late arrivals in the last 30 days. Despite previous meetings where the attendance policy was clearly explained, Taylor's behavior has not changed. This policy allows for no more than 3 late arrivals in a 30-day period. Taylor's role is critical to the success of the business, making the situation delicate and requiring a nuanced approach. You need to address this issue once again, balancing the importance of Taylor's contributions with the necessity of adhering to company policies and maintaining team morale.
    \item You are the supervisor of a new airman, Taylor, who has recently joined your office. Over the weekend, Taylor received a DUI and has called you late at night to inform you of the situation. Despite the late hour and the shock of the news, you recognize the gravity of the situation and the impact it could have on Tavior's career and wellbeing You understand the impodance of handling the situation with care ensunno Tavior is safe and aware of the serousness of the incident while also prenaring to navinate the administrative and disciolinay processes that will follow.
    \item You sit down in your office, preparing yourself to have a constructive conversation with Taylor. Over the past few weeks, you've noticed a decline in the quality of Taylor's work. It's been sloppy, and it seems like he's not paying attention to the details that are crucial in your line of work. You've gathered specific examples of where his work has fallen short, aiming to provide clear, actionable feedback. Your goal is to help Taylor understand the impact of his current performance and to encourage him to ask for help or resources if he's struggling. You believe in Taylor's potential and want to support him in getting back on track, ensuring he knows that you're there to guide him through this process.
\end{itemize}

\section{Functional Probe UI Screenshots}
\label{app:screenshots}

\begin{figure}[!h]
    \centering
    \includegraphics[width=0.90\textwidth]{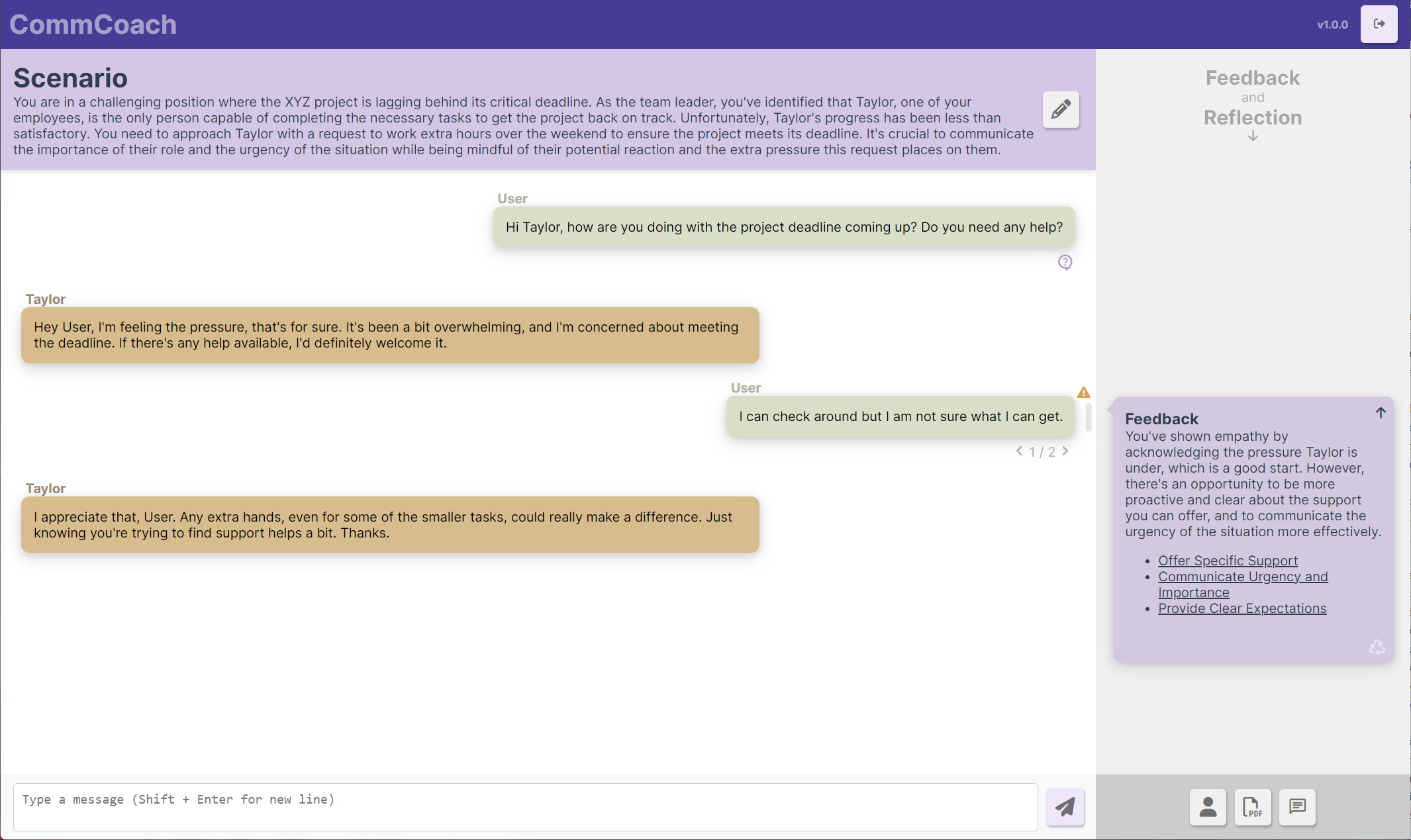}
    \caption{UI screenshot of the functional probe, \textsc{CommCoach}}
    \label{fig:system-screenshot}
    \Description{The CommCoach user interface is designed for AI-assisted communication training in managerial contexts, featuring a structured scenario where the user, acting as a team leader, must navigate a conversation with an employee, Taylor, about working extra hours due to a project deadline. The Scenario Panel at the top outlines the situation, emphasizing the importance of balancing urgency with empathy. In the main conversation area, the User and Taylor exchange messages, with Taylor expressing stress and a need for support, while the User acknowledges the pressure but remains uncertain about available assistance. Messages are color-coded, with User responses in green and Taylor’s messages in brown, and a warning icon appears next to a User response, signaling a potential area for improvement. On the right-hand side, the Feedback Panel provides AI-generated suggestions, recognizing the User’s empathy but advising improvements in offering specific support, communicating urgency, and setting clear expectations, with interactive links for further guidance. A text input box at the bottom allows the User to compose responses, and additional UI controls in the lower right corner provide options for downloading conversation history as a PDF or accessing settings. The interface integrates real-time AI coaching into workplace simulations, helping users refine communication strategies in a structured, interactive environment.}
\end{figure}

\begin{figure}[!h]
    \centering
    \includegraphics[width=0.90\textwidth]{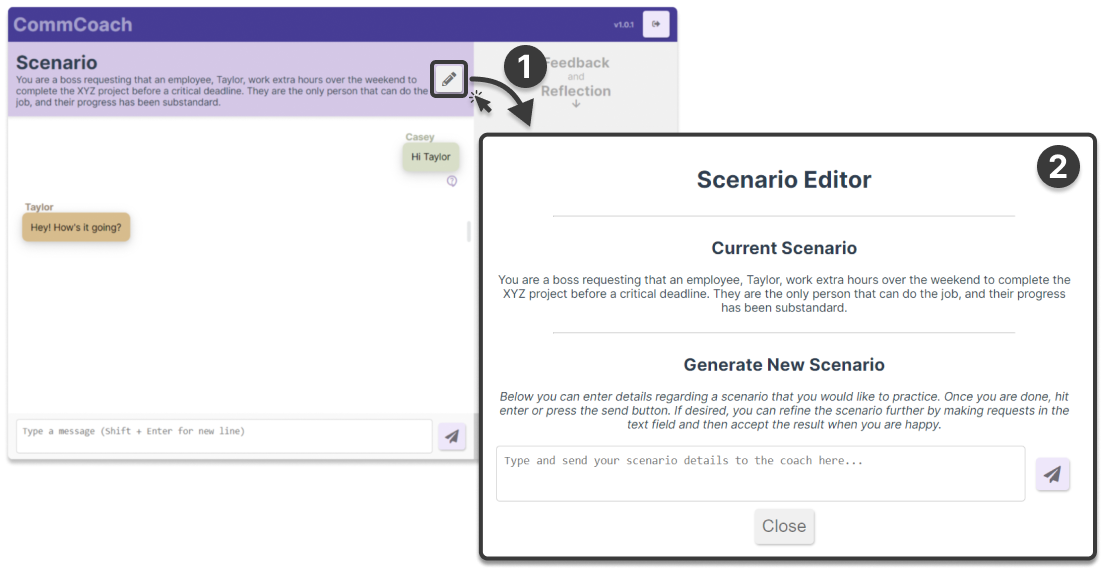}
    \caption{Users can create custom scenarios through the scenario editor.}
    \label{fig:dg1}
    \Description{The CommCoach interface allows users to create custom scenarios using the Scenario Editor, providing flexibility in training. In the main interface, users begin with a predefined scenario, but by selecting the edit icon (1), they can open the Scenario Editor (2), where they can modify the existing scenario or generate a new one. The editor presents the current scenario and includes a text input box for users to describe a different situation they would like to practice. Users can refine their scenario iteratively by interacting with the AI coach before finalizing it. This feature enables personalized training experiences tailored to different workplace communication challenges.}
\end{figure}

\begin{figure}[!h]
    \centering
    \includegraphics[width=0.90\textwidth]{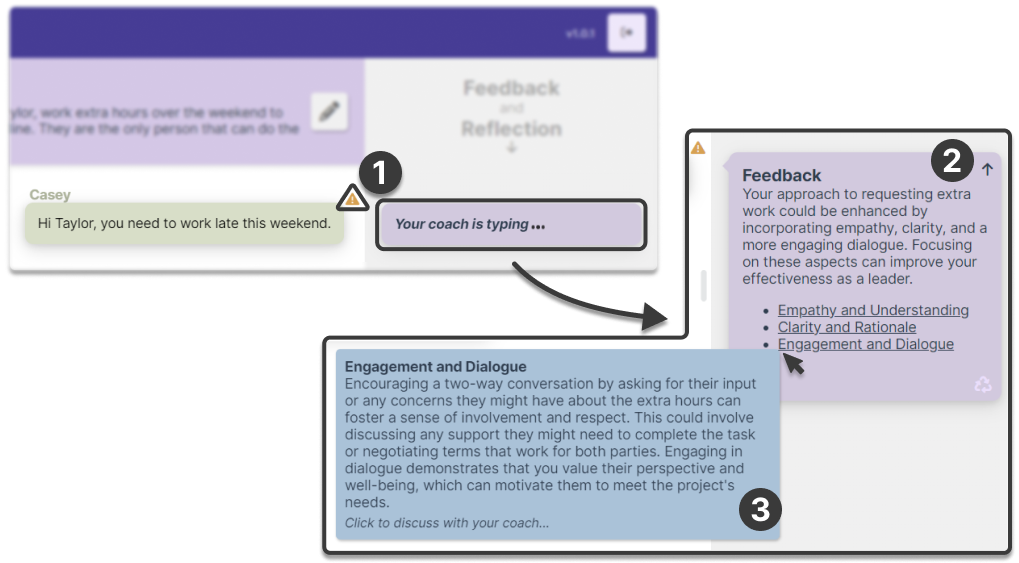}
    \caption{Immediate feedback is provided to users via the side panel. Hovering over specific feedback items will show more information in a popup.}
    \label{fig:dg2}
    \Description{The CommCoach system provides immediate feedback to users through a side panel, helping them refine their communication strategies in real time. When a user submits a message, the system detects potential areas for improvement and signals this with a warning icon (1), followed by a notification that the AI coach is generating feedback. The Feedback Panel (2) then appears on the right, suggesting ways to improve the message by incorporating elements like empathy, clarity, and engagement. Each feedback item is an interactive link, allowing users to explore further guidance. Hovering over a specific feedback item, such as Engagement and Dialogue (3), reveals a detailed popup window explaining how two-way conversations can foster involvement, respect, and motivation. This dynamic feedback mechanism ensures that users receive targeted, actionable suggestions that they can immediately apply to improve their managerial communication.}
\end{figure}

\begin{figure}[!h]
    \centering
    \includegraphics[width=0.90\textwidth]{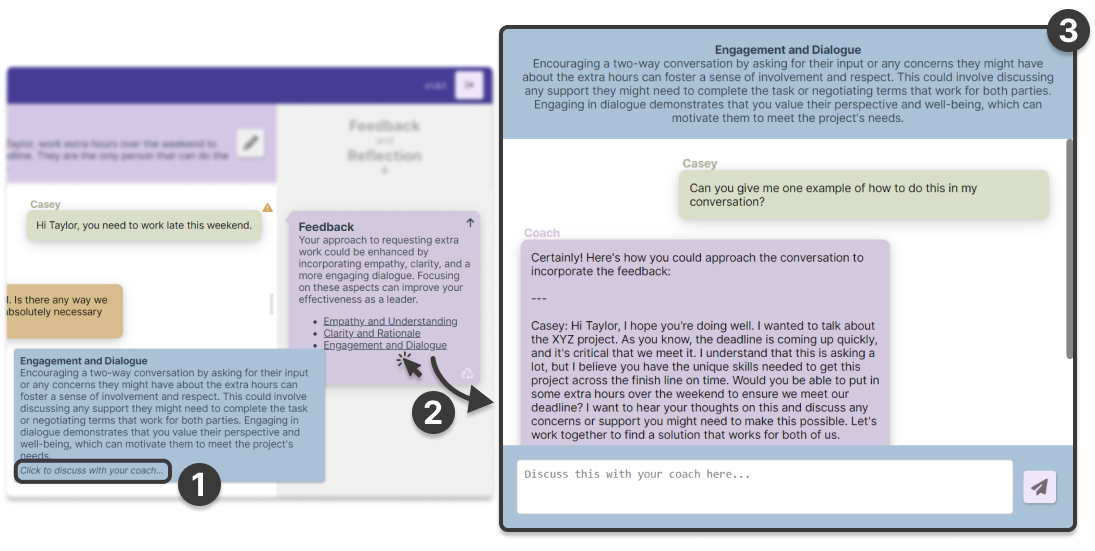}
    \caption{Users can discuss specific feedback items with the coach via an additional popup window.}
    \label{fig:dg3}
    \Description{The CommCoach system allows users to engage in deeper discussions about specific feedback items through an additional popup window. When a user receives feedback, they can select an area for improvement, such as Engagement and Dialogue, which expands into a detailed explanation box (1) outlining strategies for fostering a more interactive conversation. Users can then click a discussion button to further explore the feedback with the AI coach. Upon selecting a feedback item (2), a popup window (3) appears, providing a refined coaching experience where users can request concrete examples of how to apply the feedback in their conversations. In this example, the user asks, "Can you give me an example of how to do this in my conversation?" and the AI coach responds with a structured message that demonstrates an improved way to frame the request for extra work, incorporating empathy, clarity, and collaboration. This interactive mechanism allows users to iteratively refine their communication approach with real-time guidance.}
\end{figure}
\end{onecolumn}
\end{document}

%% file: macro.tex
\definecolor{buse_skyblue}{rgb}{0.1098, 0.5686, 0.7411}
\definecolor{buse_orange}{rgb}{0.9450, 0.6117, 0}
\definecolor{buse_plum}{rgb}{0.7, 0.3, 0.5}
\definecolor{buse_green}{rgb}{0.0274, 0.4274, 0.2352}
\definecolor{buse_mustard}{rgb}{0.8, 0.7, 0.2} 
\definecolor{buse_red}{rgb}{0.8392, 0.3921, 0.2666}
\definecolor{buse_purple}{rgb}{0.6, 0.4, 0.7} 
\definecolor{buse_seafoam}{rgb}{0.5, 0.75, 0.6}
\definecolor{buse_blue}{rgb}{0.3333, 0.3764, 0.6627}

\definecolor{lance_green}{RGB}{176,209,11}
\definecolor{lance_pink}{RGB}{189,100,184}
\definecolor{lance_peach}{RGB}{219,163,121}
\definecolor{lance_purple}{RGB}{212,200,223}
\definecolor{lance_blue}{RGB}{110,157,209}
\definecolor{lance_light_blue}{RGB}{199,222,223}
\definecolor{lance_red}{RGB}{210,154,156}

\newcommand{\textft}[1]{#1}
\newcommand{\customul}[2][black]{\setulcolor{#1}\ul{#2}\setulcolor{black}}

\newcommand{\themea}[1]{\textcolor{buse_skyblue}{#1}}
\newcommand{\themeb}[1]{\textcolor{buse_purple}{#1}}
\newcommand{\themec}[1]{\textcolor{buse_green}{#1}}
\newcommand{\themed}[1]{\textcolor{buse_orange}{#1}}
\newcommand{\themee}[1]{\textcolor{buse_plum}{#1}}
\newcommand{\themef}[1]{\textcolor{buse_red}{#1}}
\newcommand{\themeg}[1]{\textcolor{buse_seafoam}{#1}}
\newcommand{\themeh}[1]{\textcolor{buse_mustard}{#1}}
\newcommand{\themei}[1]{\textcolor{buse_blue}{#1}}

\newcommand{\themeaunder}[1]{\customul[buse_skyblue]{#1}}
\newcommand{\themebunder}[1]{\customul[buse_purple]{#1}}
\newcommand{\themecunder}[1]{\customul[buse_green]{#1}}
\newcommand{\themedunder}[1]{\customul[buse_orange]{#1}}
\newcommand{\themeeunder}[1]{\customul[buse_plum]{#1}}
\newcommand{\themefunder}[1]{\customul[buse_red]{#1}}
\newcommand{\themegunder}[1]{\customul[buse_seafoam]{#1}}
\newcommand{\themehunder}[1]{\customul[buse_mustard]{#1}}
\newcommand{\themeiunder}[1]{\customul[buse_blue]{#1}}

\newcommand{\themegreenunder}[1]{\customul[lance_green]{#1}}
\newcommand{\themepinkunder}[1]{\customul[lance_pink]{#1}}
\newcommand{\themepeachunder}[1]{\customul[lance_peach]{#1}}
\newcommand{\themepurpleunder}[1]{\customul[lance_purple]{#1}}
\newcommand{\themeblueunder}[1]{\customul[lance_blue]{#1}}
\newcommand{\themelightblueunder}[1]{\customul[lance_light_blue]{#1}}
\newcommand{\themeredunder}[1]{\customul[lance_red]{#1}}

\definecolor{lance_green}{RGB}{176,209,11}
\definecolor{lance_pink}{RGB}{189,100,184}
\definecolor{lance_peach}{RGB}{219,163,121}
\definecolor{lance_purple}{RGB}{212,200,223}
\definecolor{lance_blue}{RGB}{110,157,209}
\definecolor{lance_light_blue}{RGB}{199,222,223}
\definecolor{lance_red}{RGB}{210,154,156}

\definecolor{buse_skyblue}{rgb}{0.1098,0.5686,0.7411}
\definecolor{buse_orange}{rgb}{0.9450,0.6117,0}
\definecolor{buse_plum}{rgb}{0.7,0.3,0.5}
\definecolor{buse_green}{rgb}{0.0274,0.4274,0.2352}
\definecolor{buse_mustard}{rgb}{0.8,0.7,0.2}
\definecolor{buse_red}{rgb}{0.8392,0.3921,0.2666}
\definecolor{buse_purple}{rgb}{0.6,0.4,0.7}
\definecolor{buse_seafoam}{rgb}{0.5,0.75,0.6}
\definecolor{buse_blue}{rgb}{0.3333,0.3764,0.6627}

\newlength\HLpad 
\setlength\HLpad{2pt}

\newcommand{\GreenHL}[1]{%
  {\setlength{\fboxsep}{\HLpad}%
   \colorbox{lance_green!55!white}{#1}}%
}
\newcommand{\PinkHL}[1]{%
  {\setlength{\fboxsep}{\HLpad}%
   \colorbox{lance_pink!55!white}{#1}}%
}
\newcommand{\PeachHL}[1]{%
  {\setlength{\fboxsep}{\HLpad}%
   \colorbox{lance_peach!65!white}{#1}}%
}
\newcommand{\PurpleHL}[1]{%
  {\setlength{\fboxsep}{\HLpad}%
   \colorbox{lance_purple!65!white}{#1}}%
}
\newcommand{\BlueHL}[1]{%
  {\setlength{\fboxsep}{\HLpad}%
   \colorbox{lance_blue!55!white}{#1}}%
}
\newcommand{\LightBlueHL}[1]{%
  {\setlength{\fboxsep}{\HLpad}%
   \colorbox{lance_light_blue!65!white}{#1}}%
}
\newcommand{\RedHL}[1]{%
  {\setlength{\fboxsep}{\HLpad}%
   \colorbox{lance_red!55!white}{#1}}%
}

\lstdefinestyle{promptstyle}{
  basicstyle=\ttfamily\footnotesize,
  frame=single,                          
  rulecolor=\color{black},               
  backgroundcolor=\color{white},         
  breaklines=true,                       
  breakindent=0pt,                       
  showspaces=false,
  showstringspaces=false,
  xleftmargin=2pt, xrightmargin=2pt,
  aboveskip=1ex, belowskip=1ex
}
\lstnewenvironment{promptbox}{%
  \lstset{style=promptstyle}%
}{}

%% file: MAIN.bbl

\begin{thebibliography}{117}


\ifx \showCODEN    \undefined \def \showCODEN     #1{\unskip}     \fi
\ifx \showDOI      \undefined \def \showDOI       #1{#1}\fi
\ifx \showISBNx    \undefined \def \showISBNx     #1{\unskip}     \fi
\ifx \showISBNxiii \undefined \def \showISBNxiii  #1{\unskip}     \fi
\ifx \showISSN     \undefined \def \showISSN      #1{\unskip}     \fi
\ifx \showLCCN     \undefined \def \showLCCN      #1{\unskip}     \fi
\ifx \shownote     \undefined \def \shownote      #1{#1}          \fi
\ifx \showarticletitle \undefined \def \showarticletitle #1{#1}   \fi
\ifx \showURL      \undefined \def \showURL       {\relax}        \fi
\providecommand\bibfield[2]{#2}
\providecommand\bibinfo[2]{#2}
\providecommand\natexlab[1]{#1}
\providecommand\showeprint[2][]{arXiv:#2}

\bibitem[mic(2024)]%
        {microsoft_2024WorkTrend_2024}
 \bibinfo{year}{2024}\natexlab{}.
\newblock \bibinfo{booktitle}{\emph{2024 {Work} {Trend} {Index} {Annual} {Report}}}.
\newblock \bibinfo{type}{{T}echnical {R}eport}. \bibinfo{institution}{Microsoft}.
\newblock


\bibitem[goo(2024)]%
        {google_NewAspiringLeaders_2024}
 \bibinfo{year}{2024}\natexlab{}.
\newblock \bibinfo{title}{New and aspiring leaders use {AI} to increase their {Impact} at work}.
\newblock
\newblock
\urldef\tempurl%
\url{https://workspace.google.com/blog/ai-and-machine-learning/rising-leaders-embrace-ai-new-research-google-workspace-and-harris-poll}
\showURL{%
\tempurl}


\bibitem[Adams et~al\mbox{.}(2006)]%
        {adams_ReflectionCriticalProficiency_2006}
\bibfield{author}{\bibinfo{person}{Cindy~L. Adams}, \bibinfo{person}{Debra Nestel}, {and} \bibinfo{person}{Peter Wolf}.} \bibinfo{year}{2006}\natexlab{}.
\newblock \showarticletitle{Reflection: {A} {Critical} {Proficiency} {Essential} to the {Effective} {Development} of a {High} {Competence} in {Communication}}.
\newblock \bibinfo{journal}{\emph{Journal of Veterinary Medical Education}} \bibinfo{volume}{33}, \bibinfo{number}{1} (\bibinfo{date}{March} \bibinfo{year}{2006}), \bibinfo{pages}{58--64}.
\newblock
\showISSN{0748-321X}
\urldef\tempurl%
\url{https://doi.org/10.3138/jvme.33.1.58}
\showDOI{\tempurl}
\newblock
\shownote{Publisher: University of Toronto Press}.


\bibitem[Agboola~Sogunro(2004)]%
        {agboolasogunro_EfficacyRolePlaying_2004}
\bibfield{author}{\bibinfo{person}{Olusegun Agboola~Sogunro}.} \bibinfo{year}{2004}\natexlab{}.
\newblock \showarticletitle{Efficacy of role‐playing pedagogy in training leaders: some reflections}.
\newblock \bibinfo{journal}{\emph{Journal of Management Development}} \bibinfo{volume}{23}, \bibinfo{number}{4} (\bibinfo{date}{Jan.} \bibinfo{year}{2004}), \bibinfo{pages}{355--371}.
\newblock
\showISSN{0262-1711}
\urldef\tempurl%
\url{https://doi.org/10.1108/02621710410529802}
\showDOI{\tempurl}
\newblock
\shownote{Publisher: Emerald Group Publishing Limited}.


\bibitem[Amershi et~al\mbox{.}(2019)]%
        {amershi_GuidelinesHumanAIInteraction_2019}
\bibfield{author}{\bibinfo{person}{Saleema Amershi}, \bibinfo{person}{Dan Weld}, \bibinfo{person}{Mihaela Vorvoreanu}, \bibinfo{person}{Adam Fourney}, \bibinfo{person}{Besmira Nushi}, \bibinfo{person}{Penny Collisson}, \bibinfo{person}{Jina Suh}, \bibinfo{person}{Shamsi Iqbal}, \bibinfo{person}{Paul~N. Bennett}, \bibinfo{person}{Kori Inkpen}, \bibinfo{person}{Jaime Teevan}, \bibinfo{person}{Ruth Kikin-Gil}, {and} \bibinfo{person}{Eric Horvitz}.} \bibinfo{year}{2019}\natexlab{}.
\newblock \showarticletitle{Guidelines for {Human}-{AI} {Interaction}}. In \bibinfo{booktitle}{\emph{Proceedings of the 2019 {CHI} {Conference} on {Human} {Factors} in {Computing} {Systems}}} \emph{(\bibinfo{series}{{CHI} '19})}. \bibinfo{publisher}{Association for Computing Machinery}, \bibinfo{address}{New York, NY, USA}, \bibinfo{pages}{1--13}.
\newblock
\showISBNx{978-1-4503-5970-2}
\urldef\tempurl%
\url{https://doi.org/10.1145/3290605.3300233}
\showDOI{\tempurl}


\bibitem[Arakawa and Yakura(2020)]%
        {arakawa_INWARDComputerSupportedTool_2020}
\bibfield{author}{\bibinfo{person}{Riku Arakawa} {and} \bibinfo{person}{Hiromu Yakura}.} \bibinfo{year}{2020}\natexlab{}.
\newblock \showarticletitle{{INWARD}: {A} {Computer}-{Supported} {Tool} for {Video}-{Reflection} {Improves} {Efficiency} and {Effectiveness} in {Executive} {Coaching}}. In \bibinfo{booktitle}{\emph{Proceedings of the 2020 {CHI} {Conference} on {Human} {Factors} in {Computing} {Systems}}} \emph{(\bibinfo{series}{{CHI} '20})}. \bibinfo{publisher}{Association for Computing Machinery}, \bibinfo{address}{New York, NY, USA}, \bibinfo{pages}{1--13}.
\newblock
\showISBNx{978-1-4503-6708-0}
\urldef\tempurl%
\url{https://doi.org/10.1145/3313831.3376703}
\showDOI{\tempurl}


\bibitem[Arakawa and Yakura(2024)]%
        {arakawa_CoachingCopilotBlended_2024}
\bibfield{author}{\bibinfo{person}{Riku Arakawa} {and} \bibinfo{person}{Hiromu Yakura}.} \bibinfo{year}{2024}\natexlab{}.
\newblock \showarticletitle{Coaching {Copilot}: {Blended} {Form} of an {LLM}-{Powered} {Chatbot} and a {Human} {Coach} to {Effectively} {Support} {Self}-{Reflection} for {Leadership} {Growth}}. In \bibinfo{booktitle}{\emph{Proceedings of the 6th {ACM} {Conference} on {Conversational} {User} {Interfaces}}} \emph{(\bibinfo{series}{{CUI} '24})}. \bibinfo{publisher}{Association for Computing Machinery}, \bibinfo{address}{New York, NY, USA}, \bibinfo{pages}{1--14}.
\newblock
\showISBNx{9798400705113}
\urldef\tempurl%
\url{https://doi.org/10.1145/3640794.3665549}
\showDOI{\tempurl}


\bibitem[Aritz et~al\mbox{.}(2017)]%
        {aritz_DiscourseLeadershipPower_2017}
\bibfield{author}{\bibinfo{person}{Jolanta Aritz}, \bibinfo{person}{Robyn Walker}, \bibinfo{person}{Peter Cardon}, {and} \bibinfo{person}{Zhang Li}.} \bibinfo{year}{2017}\natexlab{}.
\newblock \showarticletitle{Discourse of {Leadership}: {The} {Power} of {Questions} in {Organizational} {Decision} {Making}}.
\newblock \bibinfo{journal}{\emph{International Journal of Business Communication}}  \bibinfo{volume}{0} (\bibinfo{date}{Jan.} \bibinfo{year}{2017}), \bibinfo{pages}{232948841668705}.
\newblock
\urldef\tempurl%
\url{https://doi.org/10.1177/2329488416687054}
\showDOI{\tempurl}


\bibitem[Augello et~al\mbox{.}(2016)]%
        {augello_ModelSocialChatbot_2016}
\bibfield{author}{\bibinfo{person}{Agnese Augello}, \bibinfo{person}{Manuel Gentile}, \bibinfo{person}{Lucas Weideveld}, {and} \bibinfo{person}{Frank Dignum}.} \bibinfo{year}{2016}\natexlab{}.
\newblock \showarticletitle{A {Model} of a {Social} {Chatbot}}. In \bibinfo{booktitle}{\emph{Intelligent {Interactive} {Multimedia} {Systems} and {Services} 2016}}, \bibfield{editor}{\bibinfo{person}{Giuseppe~De Pietro}, \bibinfo{person}{Luigi Gallo}, \bibinfo{person}{Robert~J. Howlett}, {and} \bibinfo{person}{Lakhmi~C. Jain}} (Eds.). \bibinfo{publisher}{Springer International Publishing}, \bibinfo{address}{Cham}, \bibinfo{pages}{637--647}.
\newblock
\showISBNx{978-3-319-39345-2}
\urldef\tempurl%
\url{https://doi.org/10.1007/978-3-319-39345-2_57}
\showDOI{\tempurl}


\bibitem[Ball(2021)]%
        {ball_ElectronicMonitoringSurveillance_2021}
\bibfield{author}{\bibinfo{person}{Kirstie Ball}.} \bibinfo{year}{2021}\natexlab{}.
\newblock \bibinfo{booktitle}{\emph{Electronic monitoring and surveillance in the workplace: literature review and policy recommendations}}.
\newblock \bibinfo{publisher}{Publications Office of the European Union}, \bibinfo{address}{Luxembourg}.
\newblock
\showISBNx{978-92-76-43340-8}
\newblock
\shownote{OCLC: 1289305461}.


\bibitem[Barker(1997)]%
        {barker_HowCanWe_1997}
\bibfield{author}{\bibinfo{person}{Richard~A. Barker}.} \bibinfo{year}{1997}\natexlab{}.
\newblock \showarticletitle{How {Can} {We} {Train} {Leaders} if {We} {Do} {Not} {Know} {What} {Leadership} {Is}?}
\newblock \bibinfo{journal}{\emph{Human Relations}} \bibinfo{volume}{50}, \bibinfo{number}{4} (\bibinfo{date}{April} \bibinfo{year}{1997}), \bibinfo{pages}{343--362}.
\newblock
\showISSN{0018-7267}
\urldef\tempurl%
\url{https://doi.org/10.1177/001872679705000402}
\showDOI{\tempurl}
\newblock
\shownote{Publisher: SAGE Publications Ltd}.


\bibitem[Barmer et~al\mbox{.}(2021)]%
        {barmer_HumanCenteredAI_2021}
\bibfield{author}{\bibinfo{person}{Hollen Barmer}, \bibinfo{person}{Rachel Dzombak}, \bibinfo{person}{Matthew Gaston}, \bibinfo{person}{Vijaykumar Palat}, \bibinfo{person}{Frank Redner}, \bibinfo{person}{Carol Smith}, {and} \bibinfo{person}{Tanisha Smith}.} \bibinfo{year}{2021}\natexlab{}.
\newblock \bibinfo{booktitle}{\emph{Human-{Centered} {AI}}}.
\newblock \bibinfo{type}{report}. \bibinfo{institution}{Carnegie Mellon University}.
\newblock
\urldef\tempurl%
\url{https://doi.org/10.1184/R1/16560183.v1}
\showDOI{\tempurl}


\bibitem[Borg et~al\mbox{.}(2024)]%
        {borg_CreatingVirtualPatients_2024}
\bibfield{author}{\bibinfo{person}{Alexander Borg}, \bibinfo{person}{Ioannis Parodis}, {and} \bibinfo{person}{Gabriel Skantze}.} \bibinfo{year}{2024}\natexlab{}.
\newblock \showarticletitle{Creating {Virtual} {Patients} using {Robots} and {Large} {Language} {Models}: {A} {Preliminary} {Study} with {Medical} {Students}}. In \bibinfo{booktitle}{\emph{Companion of the 2024 {ACM}/{IEEE} {International} {Conference} on {Human}-{Robot} {Interaction}}} \emph{(\bibinfo{series}{{HRI} '24})}. \bibinfo{publisher}{Association for Computing Machinery}, \bibinfo{address}{New York, NY, USA}, \bibinfo{pages}{273--277}.
\newblock
\showISBNx{9798400703232}
\urldef\tempurl%
\url{https://doi.org/10.1145/3610978.3640592}
\showDOI{\tempurl}


\bibitem[Boud et~al\mbox{.}(1985)]%
        {boud_ReflectionTurningExperience_1985}
\bibfield{author}{\bibinfo{person}{David Boud}, \bibinfo{person}{Rosemary Keogh}, {and} \bibinfo{person}{David Walker}.} \bibinfo{year}{1985}\natexlab{}.
\newblock \bibinfo{booktitle}{\emph{Reflection: {Turning} {Experience} {Into} {Learning}}}.
\newblock \bibinfo{publisher}{Kogan Page}.
\newblock
\showISBNx{978-0-85038-864-0}
\newblock
\shownote{Google-Books-ID: xBshIryFdr0C}.


\bibitem[Boyd and Fales(1983)]%
        {boyd_ReflectiveLearningKey_1983}
\bibfield{author}{\bibinfo{person}{Evelyn~M. Boyd} {and} \bibinfo{person}{Ann~W. Fales}.} \bibinfo{year}{1983}\natexlab{}.
\newblock \showarticletitle{Reflective {Learning}: {Key} to {Learning} from {Experience}}.
\newblock \bibinfo{journal}{\emph{Journal of Humanistic Psychology}} \bibinfo{volume}{23}, \bibinfo{number}{2} (\bibinfo{date}{April} \bibinfo{year}{1983}), \bibinfo{pages}{99--117}.
\newblock
\showISSN{0022-1678}
\urldef\tempurl%
\url{https://doi.org/10.1177/0022167883232011}
\showDOI{\tempurl}
\newblock
\shownote{Publisher: SAGE Publications Inc}.


\bibitem[Bracq et~al\mbox{.}(2019)]%
        {bracq_VirtualRealitySimulation_2019}
\bibfield{author}{\bibinfo{person}{Marie-Stéphanie Bracq}, \bibinfo{person}{Estelle Michinov}, {and} \bibinfo{person}{Pierre Jannin}.} \bibinfo{year}{2019}\natexlab{}.
\newblock \showarticletitle{Virtual {{Reality Simulation}} in {{Nontechnical Skills Training}} for {{Healthcare Professionals}}: {{A Systematic Review}}}.
\newblock \bibinfo{journal}{\emph{Simulation in Healthcare}} \bibinfo{volume}{14}, \bibinfo{number}{3} (\bibinfo{year}{2019}), \bibinfo{pages}{188}.
\newblock
\showISSN{1559-2332}
\urldef\tempurl%
\url{https://doi.org/10.1097/SIH.0000000000000347}
\showDOI{\tempurl}


\bibitem[Brown et~al\mbox{.}(2020)]%
        {brown_LanguageModelsAre_2020}
\bibfield{author}{\bibinfo{person}{Tom~B. Brown}, \bibinfo{person}{Benjamin Mann}, \bibinfo{person}{Nick Ryder}, \bibinfo{person}{Melanie Subbiah}, \bibinfo{person}{Jared Kaplan}, \bibinfo{person}{Prafulla Dhariwal}, \bibinfo{person}{Arvind Neelakantan}, \bibinfo{person}{Pranav Shyam}, \bibinfo{person}{Girish Sastry}, \bibinfo{person}{Amanda Askell}, \bibinfo{person}{Sandhini Agarwal}, \bibinfo{person}{Ariel Herbert-Voss}, \bibinfo{person}{Gretchen Krueger}, \bibinfo{person}{Tom Henighan}, \bibinfo{person}{Rewon Child}, \bibinfo{person}{Aditya Ramesh}, \bibinfo{person}{Daniel~M. Ziegler}, \bibinfo{person}{Jeffrey Wu}, \bibinfo{person}{Clemens Winter}, \bibinfo{person}{Christopher Hesse}, \bibinfo{person}{Mark Chen}, \bibinfo{person}{Eric Sigler}, \bibinfo{person}{Mateusz Litwin}, \bibinfo{person}{Scott Gray}, \bibinfo{person}{Benjamin Chess}, \bibinfo{person}{Jack Clark}, \bibinfo{person}{Christopher Berner}, \bibinfo{person}{Sam McCandlish}, \bibinfo{person}{Alec Radford}, \bibinfo{person}{Ilya Sutskever},
  {and} \bibinfo{person}{Dario Amodei}.} \bibinfo{year}{2020}\natexlab{}.
\newblock \bibinfo{title}{Language {Models} are {Few}-{Shot} {Learners}}.
\newblock
\newblock
\urldef\tempurl%
\url{https://doi.org/10.48550/arXiv.2005.14165}
\showDOI{\tempurl}
\newblock
\shownote{arXiv:2005.14165 [cs]}.


\bibitem[Bryant et~al\mbox{.}(2020)]%
        {bryant_ReviewVirtualReality_2020}
\bibfield{author}{\bibinfo{person}{Lucy Bryant}, \bibinfo{person}{Melissa Brunner}, {and} \bibinfo{person}{Bronwyn Hemsley}.} \bibinfo{year}{2020}\natexlab{}.
\newblock \showarticletitle{A review of virtual reality technologies in the field of communication disability: implications for practice and research}.
\newblock \bibinfo{journal}{\emph{Disability and Rehabilitation: Assistive Technology}} \bibinfo{volume}{15}, \bibinfo{number}{4} (\bibinfo{date}{May} \bibinfo{year}{2020}), \bibinfo{pages}{365--372}.
\newblock
\showISSN{1748-3107}
\urldef\tempurl%
\url{https://doi.org/10.1080/17483107.2018.1549276}
\showDOI{\tempurl}
\newblock
\shownote{Publisher: Taylor \& Francis \_eprint: https://doi.org/10.1080/17483107.2018.1549276}.


\bibitem[Butow and Hoque(2020)]%
        {butow_UsingArtificialIntelligence_2020}
\bibfield{author}{\bibinfo{person}{Phyllis Butow} {and} \bibinfo{person}{Ehsan Hoque}.} \bibinfo{year}{2020}\natexlab{}.
\newblock \showarticletitle{Using artificial intelligence to analyse and teach communication in healthcare}.
\newblock \bibinfo{journal}{\emph{The Breast}}  \bibinfo{volume}{50} (\bibinfo{date}{April} \bibinfo{year}{2020}), \bibinfo{pages}{49--55}.
\newblock
\showISSN{0960-9776, 1532-3080}
\urldef\tempurl%
\url{https://doi.org/10.1016/j.breast.2020.01.008}
\showDOI{\tempurl}
\newblock
\shownote{Publisher: Elsevier}.


\bibitem[Carlyle(1840)]%
        {carlyle_HeroesHeroworshipHeroic_1840}
\bibfield{author}{\bibinfo{person}{Thomas Carlyle}.} \bibinfo{year}{1840}\natexlab{}.
\newblock \bibinfo{booktitle}{\emph{On {Heroes}, {Hero}-worship and the {Heroic} in {History}}}.
\newblock \bibinfo{publisher}{Chapman and Hall}.
\newblock
\showISBNx{978-1-57179-831-2}


\bibitem[Chang et~al\mbox{.}(2022)]%
        {chang_ThreadCautionProactively_2022}
\bibfield{author}{\bibinfo{person}{Jonathan~P. Chang}, \bibinfo{person}{Charlotte Schluger}, {and} \bibinfo{person}{Cristian Danescu-Niculescu-Mizil}.} \bibinfo{year}{2022}\natexlab{}.
\newblock \showarticletitle{Thread {With} {Caution}: {Proactively} {Helping} {Users} {Assess} and {Deescalate} {Tension} in {Their} {Online} {Discussions}}.
\newblock \bibinfo{journal}{\emph{Proceedings of the ACM on Human-Computer Interaction}} \bibinfo{volume}{6}, \bibinfo{number}{CSCW2} (\bibinfo{date}{Nov.} \bibinfo{year}{2022}), \bibinfo{pages}{1--37}.
\newblock
\showISSN{2573-0142}
\urldef\tempurl%
\url{https://doi.org/10.1145/3555603}
\showDOI{\tempurl}


\bibitem[Charlier et~al\mbox{.}(2016)]%
        {charlier_EmergentLeadershipVirtual_2016}
\bibfield{author}{\bibinfo{person}{Steven~D. Charlier}, \bibinfo{person}{Greg~L. Stewart}, \bibinfo{person}{Lindsey~M. Greco}, {and} \bibinfo{person}{Cody~J. Reeves}.} \bibinfo{year}{2016}\natexlab{}.
\newblock \showarticletitle{Emergent leadership in virtual teams: {A} multilevel investigation of individual communication and team dispersion antecedents}.
\newblock \bibinfo{journal}{\emph{The Leadership Quarterly}} \bibinfo{volume}{27}, \bibinfo{number}{5} (\bibinfo{date}{Oct.} \bibinfo{year}{2016}), \bibinfo{pages}{745--764}.
\newblock
\showISSN{1048-9843}
\urldef\tempurl%
\url{https://doi.org/10.1016/j.leaqua.2016.05.002}
\showDOI{\tempurl}


\bibitem[Charmaz(2006)]%
        {charmaz_ConstructingGroundedTheory_2006}
\bibfield{author}{\bibinfo{person}{Kathy Charmaz}.} \bibinfo{year}{2006}\natexlab{}.
\newblock \bibinfo{booktitle}{\emph{Constructing {Grounded} {Theory}: {A} {Practical} {Guide} {Through} {Qualitative} {Analysis}}}.
\newblock \bibinfo{publisher}{SAGE}.
\newblock
\showISBNx{978-0-7619-7353-9}
\newblock
\shownote{Google-Books-ID: v1qP1KbXz1AC}.


\bibitem[Chen et~al\mbox{.}(2023)]%
        {chen_FacilitatingCounselorReflective_2023}
\bibfield{author}{\bibinfo{person}{Tianying Chen}, \bibinfo{person}{Michael~Xieyang Liu}, \bibinfo{person}{Emily Ding}, \bibinfo{person}{Emma O'Neil}, \bibinfo{person}{Mansi Agarwal}, \bibinfo{person}{Robert~E Kraut}, {and} \bibinfo{person}{Laura Dabbish}.} \bibinfo{year}{2023}\natexlab{}.
\newblock \showarticletitle{Facilitating {Counselor} {Reflective} {Learning} with a {Real}-time {Annotation} tool}. In \bibinfo{booktitle}{\emph{Proceedings of the 2023 {CHI} {Conference} on {Human} {Factors} in {Computing} {Systems}}} \emph{(\bibinfo{series}{{CHI} '23})}. \bibinfo{publisher}{Association for Computing Machinery}, \bibinfo{address}{New York, NY, USA}, \bibinfo{pages}{1--17}.
\newblock
\showISBNx{978-1-4503-9421-5}
\urldef\tempurl%
\url{https://doi.org/10.1145/3544548.3581551}
\showDOI{\tempurl}


\bibitem[Corbin and Strauss(2015)]%
        {corbin_BasicsQualitativeResearch_2015}
\bibfield{author}{\bibinfo{person}{Juliet Corbin} {and} \bibinfo{person}{Anselm Strauss}.} \bibinfo{year}{2015}\natexlab{}.
\newblock \bibinfo{booktitle}{\emph{Basics of {Qualitative} {Research}}}.
\newblock \bibinfo{publisher}{SAGE}.
\newblock
\showISBNx{978-1-4129-9746-1}
\newblock
\shownote{Google-Books-ID: Dc45DQAAQBAJ}.


\bibitem[Cuadra et~al\mbox{.}({[n.\,d.]})]%
        {cuadra_IllusionEmpathyNotes_2024}
\bibfield{author}{\bibinfo{person}{Andrea Cuadra}, \bibinfo{person}{Maria Wang}, \bibinfo{person}{Lynn~Andrea Stein}, \bibinfo{person}{Malte~F. Jung}, \bibinfo{person}{Nicola Dell}, \bibinfo{person}{Deborah Estrin}, {and} \bibinfo{person}{James~A. Landay}.} \bibinfo{year}{[n.\,d.]}\natexlab{}.
\newblock \showarticletitle{The {{Illusion}} of {{Empathy}}? {{Notes}} on {{Displays}} of {{Emotion}} in {{Human-Computer Interaction}}}. In \bibinfo{booktitle}{\emph{Proceedings of the {{CHI Conference}} on {{Human Factors}} in {{Computing Systems}}}} (New York, NY, USA, 2024-05-11) \emph{(\bibinfo{series}{{{CHI}} '24})}. \bibinfo{publisher}{Association for Computing Machinery}, \bibinfo{pages}{1--18}.
\newblock
\showISBNx{9798400703300}
\urldef\tempurl%
\url{https://doi.org/10.1145/3613904.3642336}
\showDOI{\tempurl}


\bibitem[Davidson et~al\mbox{.}(2019)]%
        {davidson_RacialBiasHate_2019}
\bibfield{author}{\bibinfo{person}{Thomas Davidson}, \bibinfo{person}{Debasmita Bhattacharya}, {and} \bibinfo{person}{Ingmar Weber}.} \bibinfo{year}{2019}\natexlab{}.
\newblock \showarticletitle{Racial {Bias} in {Hate} {Speech} and {Abusive} {Language} {Detection} {Datasets}}. In \bibinfo{booktitle}{\emph{Proceedings of the {Third} {Workshop} on {Abusive} {Language} {Online}}}, \bibfield{editor}{\bibinfo{person}{Sarah~T. Roberts}, \bibinfo{person}{Joel Tetreault}, \bibinfo{person}{Vinodkumar Prabhakaran}, {and} \bibinfo{person}{Zeerak Waseem}} (Eds.). \bibinfo{publisher}{Association for Computational Linguistics}, \bibinfo{address}{Florence, Italy}, \bibinfo{pages}{25--35}.
\newblock
\urldef\tempurl%
\url{https://doi.org/10.18653/v1/W19-3504}
\showDOI{\tempurl}


\bibitem[Deeva et~al\mbox{.}(2021)]%
        {deeva_ReviewAutomatedFeedback_2021}
\bibfield{author}{\bibinfo{person}{Galina Deeva}, \bibinfo{person}{Daria Bogdanova}, \bibinfo{person}{Estefanía Serral}, \bibinfo{person}{Monique Snoeck}, {and} \bibinfo{person}{Jochen De~Weerdt}.} \bibinfo{year}{2021}\natexlab{}.
\newblock \showarticletitle{A Review of Automated Feedback Systems for Learners: {{Classification}} Framework, Challenges and Opportunities}.
\newblock \bibinfo{journal}{\emph{Computers \& Education}}  \bibinfo{volume}{162} (\bibinfo{year}{2021}), \bibinfo{pages}{104094}.
\newblock
\showISSN{03601315}
\urldef\tempurl%
\url{https://doi.org/10.1016/j.compedu.2020.104094}
\showDOI{\tempurl}


\bibitem[Ding et~al\mbox{.}(2024)]%
        {ding_LeveragingPromptBasedLarge_2024}
\bibfield{author}{\bibinfo{person}{Xiaohan Ding}, \bibinfo{person}{Buse Carik}, \bibinfo{person}{Uma~Sushmitha Gunturi}, \bibinfo{person}{Valerie Reyna}, {and} \bibinfo{person}{Eugenia~H. Rho}.} \bibinfo{year}{2024}\natexlab{}.
\newblock \bibinfo{title}{Leveraging {{Prompt-Based Large Language Models}}: {{Predicting Pandemic Health Decisions}} and {{Outcomes Through Social Media Language}}}.
\newblock
\newblock
\urldef\tempurl%
\url{https://doi.org/10.1145/3613904.3642117}
\showDOI{\tempurl}
\showeprint[arxiv]{2403.00994}~[cs]


\bibitem[Drath and Palus(1994)]%
        {drath_MakingCommonSense_1994}
\bibfield{author}{\bibinfo{person}{Wilfred~H. Drath} {and} \bibinfo{person}{Charles~J. Palus}.} \bibinfo{year}{1994}\natexlab{}.
\newblock \bibinfo{booktitle}{\emph{Making {Common} {Sense}: {Leadership} as {Meaning}-making in a {Community} of {Practice}}}.
\newblock \bibinfo{publisher}{Center for Creative Leadership}.
\newblock
\showISBNx{978-1-932973-51-8}
\newblock
\shownote{Google-Books-ID: mLcqAgAAQBAJ}.


\bibitem[Fairhurst(2007)]%
        {fairhurst_DiscursiveLeadershipConversation_2007}
\bibfield{author}{\bibinfo{person}{Gail Fairhurst}.} \bibinfo{year}{2007}\natexlab{}.
\newblock \bibinfo{booktitle}{\emph{Discursive {Leadership}: {In} {Conversation} with {Leadership} {Psychology}}}.
\newblock \bibinfo{publisher}{SAGE}.
\newblock
\showISBNx{978-1-4522-7899-5}


\bibitem[Fairhurst and Uhl-Bien(2012)]%
        {fairhurst_OrganizationalDiscourseAnalysis_2012}
\bibfield{author}{\bibinfo{person}{Gail~T. Fairhurst} {and} \bibinfo{person}{Mary Uhl-Bien}.} \bibinfo{year}{2012}\natexlab{}.
\newblock \showarticletitle{Organizational discourse analysis ({ODA}): {Examining} leadership as a relational process}.
\newblock \bibinfo{journal}{\emph{The Leadership Quarterly}} \bibinfo{volume}{23}, \bibinfo{number}{6} (\bibinfo{date}{Dec.} \bibinfo{year}{2012}), \bibinfo{pages}{1043--1062}.
\newblock
\showISSN{1048-9843}
\urldef\tempurl%
\url{https://doi.org/10.1016/j.leaqua.2012.10.005}
\showDOI{\tempurl}


\bibitem[Fleck and Fitzpatrick(2010)]%
        {fleck_ReflectingReflectionFraming_2010}
\bibfield{author}{\bibinfo{person}{Rowanne Fleck} {and} \bibinfo{person}{Geraldine Fitzpatrick}.} \bibinfo{year}{2010}\natexlab{}.
\newblock \showarticletitle{Reflecting on Reflection: Framing a Design Landscape}. In \bibinfo{booktitle}{\emph{Proceedings of the 22nd {{Conference}} of the {{Computer-Human Interaction Special Interest Group}} of {{Australia}} on {{Computer-Human Interaction}}}} ({New York, NY, USA}, 2010-11-22) \emph{(\bibinfo{series}{{{OZCHI}} '10})}. \bibinfo{publisher}{{Association for Computing Machinery}}, \bibinfo{pages}{216--223}.
\newblock
\showISBNx{978-1-4503-0502-0}
\urldef\tempurl%
\url{https://doi.org/10.1145/1952222.1952269}
\showDOI{\tempurl}


\bibitem[Gao et~al\mbox{.}(2018)]%
        {gao_NeuralApproachesConversational_2018}
\bibfield{author}{\bibinfo{person}{Jianfeng Gao}, \bibinfo{person}{Michel Galley}, {and} \bibinfo{person}{Lihong Li}.} \bibinfo{year}{2018}\natexlab{}.
\newblock \showarticletitle{Neural {{Approaches}} to {{Conversational AI}}}. In \bibinfo{booktitle}{\emph{The 41st {{International ACM SIGIR Conference}} on {{Research}} \& {{Development}} in {{Information Retrieval}}}} (New York, NY, USA, 2018-06-27) \emph{(\bibinfo{series}{{{SIGIR}} '18})}. \bibinfo{publisher}{Association for Computing Machinery}, \bibinfo{pages}{1371--1374}.
\newblock
\showISBNx{978-1-4503-5657-2}
\urldef\tempurl%
\url{https://doi.org/10.1145/3209978.3210183}
\showDOI{\tempurl}


\bibitem[Guyre et~al\mbox{.}(2024)]%
        {guyre_PromptEngineeringLLM_2024}
\bibfield{author}{\bibinfo{person}{Melissa Guyre}, \bibinfo{person}{Liz Holland}, \bibinfo{person}{Nirva Shah}, {and} \bibinfo{person}{Rahul~R. Divekar}.} \bibinfo{year}{2024}\natexlab{}.
\newblock \showarticletitle{Prompt {Engineering} an {LLM} into {Roleplaying} a {Management} {Coach}: a {Short} {Guide} by and for {Non}-{NLP} {Experts}}. In \bibinfo{booktitle}{\emph{Proceedings of the 6th {ACM} {Conference} on {Conversational} {User} {Interfaces}}} \emph{(\bibinfo{series}{{CUI} '24})}. \bibinfo{publisher}{Association for Computing Machinery}, \bibinfo{address}{New York, NY, USA}, \bibinfo{pages}{1--10}.
\newblock
\showISBNx{9798400705113}
\urldef\tempurl%
\url{https://doi.org/10.1145/3640794.3665570}
\showDOI{\tempurl}


\bibitem[Helander(1997)]%
        {helander_HandbookHumanComputerInteraction_1997}
\bibfield{author}{\bibinfo{person}{M.~G. Helander}.} \bibinfo{year}{1997}\natexlab{}.
\newblock \bibinfo{booktitle}{\emph{Handbook of {Human}-{Computer} {Interaction}}}.
\newblock \bibinfo{publisher}{Elsevier}.
\newblock
\showISBNx{978-1-4832-9513-8}
\newblock
\shownote{Google-Books-ID: 6vnSAwAAQBAJ}.


\bibitem[Hershcovis et~al\mbox{.}(2007)]%
        {hershcovis_PredictingWorkplaceAggression_2007}
\bibfield{author}{\bibinfo{person}{M.~Sandy Hershcovis}, \bibinfo{person}{Nick Turner}, \bibinfo{person}{Julian Barling}, \bibinfo{person}{Kara~A. Arnold}, \bibinfo{person}{Kathryne~E. Dupré}, \bibinfo{person}{Michelle Inness}, \bibinfo{person}{Manon~Mireille LeBlanc}, {and} \bibinfo{person}{Niro Sivanathan}.} \bibinfo{year}{2007}\natexlab{}.
\newblock \showarticletitle{Predicting Workplace Aggression: {{A}} Meta-Analysis}.
\newblock \bibinfo{journal}{\emph{Journal of Applied Psychology}} \bibinfo{volume}{92}, \bibinfo{number}{1} (\bibinfo{year}{2007}), \bibinfo{pages}{228--238}.
\newblock
\showISSN{0021-9010}
\urldef\tempurl%
\url{https://doi.org/10.1037/0021-9010.92.1.228}
\showDOI{\tempurl}


\bibitem[Hickok and Maslej(2023)]%
        {hickok_PolicyPrimerRoadmap_2023}
\bibfield{author}{\bibinfo{person}{Merve Hickok} {and} \bibinfo{person}{Nestor Maslej}.} \bibinfo{year}{2023}\natexlab{}.
\newblock \showarticletitle{A policy primer and roadmap on {AI} worker surveillance and productivity scoring tools}.
\newblock \bibinfo{journal}{\emph{Ai and Ethics}} (\bibinfo{date}{March} \bibinfo{year}{2023}), \bibinfo{pages}{1--15}.
\newblock
\showISSN{2730-5953}
\urldef\tempurl%
\url{https://doi.org/10.1007/s43681-023-00275-8}
\showDOI{\tempurl}


\bibitem[Hoffman et~al\mbox{.}(2011)]%
        {hoffman_GreatManGreat_2011}
\bibfield{author}{\bibinfo{person}{Brian~J. Hoffman}, \bibinfo{person}{David~J. Woehr}, \bibinfo{person}{Robyn Maldagen-Youngjohn}, {and} \bibinfo{person}{Brian~D. Lyons}.} \bibinfo{year}{2011}\natexlab{}.
\newblock \showarticletitle{Great man or great myth? {A} quantitative review of the relationship between individual differences and leader effectiveness}.
\newblock \bibinfo{journal}{\emph{Journal of Occupational and Organizational Psychology}} \bibinfo{volume}{84}, \bibinfo{number}{2} (\bibinfo{year}{2011}), \bibinfo{pages}{347--381}.
\newblock
\showISSN{2044-8325}
\urldef\tempurl%
\url{https://doi.org/10.1348/096317909X485207}
\showDOI{\tempurl}
\newblock
\shownote{\_eprint: https://bpspsychub.onlinelibrary.wiley.com/doi/pdf/10.1348/096317909X485207}.


\bibitem[Horvitz(1999)]%
        {horvitz_PrinciplesMixedinitiativeUser_1999}
\bibfield{author}{\bibinfo{person}{Eric Horvitz}.} \bibinfo{year}{1999}\natexlab{}.
\newblock \showarticletitle{Principles of mixed-initiative user interfaces}. In \bibinfo{booktitle}{\emph{Proceedings of the {SIGCHI} conference on {Human} {Factors} in {Computing} {Systems}}} \emph{(\bibinfo{series}{{CHI} '99})}. \bibinfo{publisher}{Association for Computing Machinery}, \bibinfo{address}{New York, NY, USA}, \bibinfo{pages}{159--166}.
\newblock
\showISBNx{978-0-201-48559-2}
\urldef\tempurl%
\url{https://doi.org/10.1145/302979.303030}
\showDOI{\tempurl}


\bibitem[Höhn et~al\mbox{.}(2024)]%
        {hohn_UsingLargeLanguage_2024}
\bibfield{author}{\bibinfo{person}{Sviatlana Höhn}, \bibinfo{person}{Jauwairia Nasir}, \bibinfo{person}{Ali Paikan}, \bibinfo{person}{Pouyan Ziafati}, {and} \bibinfo{person}{Elisabeth André}.} \bibinfo{year}{2024}\natexlab{}.
\newblock \showarticletitle{Using {Large} {Language} {Models} for {Robot}-{Assisted} {Therapeutic} {Role}-{Play}: {Factuality} is not enough!}. In \bibinfo{booktitle}{\emph{Proceedings of the 6th {ACM} {Conference} on {Conversational} {User} {Interfaces}}} \emph{(\bibinfo{series}{{CUI} '24})}. \bibinfo{publisher}{Association for Computing Machinery}, \bibinfo{address}{New York, NY, USA}, \bibinfo{pages}{1--6}.
\newblock
\showISBNx{9798400705113}
\urldef\tempurl%
\url{https://doi.org/10.1145/3640794.3665886}
\showDOI{\tempurl}


\bibitem[Iacovides et~al\mbox{.}({[n.\,d.]})]%
        {iacovides_PlayerStrategiesBreakthroughs_2014}
\bibfield{author}{\bibinfo{person}{Ioanna Iacovides}, \bibinfo{person}{Anna~L. Cox}, \bibinfo{person}{Ara Avakian}, {and} \bibinfo{person}{Thomas Knoll}.} \bibinfo{year}{[n.\,d.]}\natexlab{}.
\newblock \showarticletitle{Player strategies: achieving breakthroughs and progressing in single-player and cooperative games}. In \bibinfo{booktitle}{\emph{Proceedings of the first {ACM} {SIGCHI} annual symposium on Computer-human interaction in play}} (New York, {NY}, {USA}, 2014-10-19) \emph{(\bibinfo{series}{{CHI} {PLAY} '14})}. \bibinfo{publisher}{Association for Computing Machinery}, \bibinfo{pages}{131--140}.
\newblock
\showISBNx{978-1-4503-3014-5}
\urldef\tempurl%
\url{https://doi.org/10.1145/2658537.2658697}
\showDOI{\tempurl}


\bibitem[Iivari and Huisman(2007)]%
        {iivari_RelationshipOrganizationalCulture_2007}
\bibfield{author}{\bibinfo{person}{Juhani Iivari} {and} \bibinfo{person}{Magda Huisman}.} \bibinfo{year}{2007}\natexlab{}.
\newblock \showarticletitle{The {Relationship} between {Organizational} {Culture} and the {Deployment} of {Systems} {Development} {Methodologies}}.
\newblock \bibinfo{journal}{\emph{MIS Quarterly}} \bibinfo{volume}{31}, \bibinfo{number}{1} (\bibinfo{year}{2007}), \bibinfo{pages}{35--58}.
\newblock
\showISSN{0276-7783}
\urldef\tempurl%
\url{https://doi.org/10.2307/25148780}
\showDOI{\tempurl}
\newblock
\shownote{Publisher: Management Information Systems Research Center, University of Minnesota}.


\bibitem[Inkpen et~al\mbox{.}(2023)]%
        {inkpen_AdvancingHumanAIComplementarity_2023}
\bibfield{author}{\bibinfo{person}{Kori Inkpen}, \bibinfo{person}{Shreya Chappidi}, \bibinfo{person}{Keri Mallari}, \bibinfo{person}{Besmira Nushi}, \bibinfo{person}{Divya Ramesh}, \bibinfo{person}{Pietro Michelucci}, \bibinfo{person}{Vani Mandava}, \bibinfo{person}{Libuše~Hannah Vepřek}, {and} \bibinfo{person}{Gabrielle Quinn}.} \bibinfo{year}{2023}\natexlab{}.
\newblock \showarticletitle{Advancing {Human}-{AI} {Complementarity}: {The} {Impact} of {User} {Expertise} and {Algorithmic} {Tuning} on {Joint} {Decision} {Making}}.
\newblock \bibinfo{journal}{\emph{ACM Trans. Comput.-Hum. Interact.}} \bibinfo{volume}{30}, \bibinfo{number}{5} (\bibinfo{date}{Sept.} \bibinfo{year}{2023}), \bibinfo{pages}{71:1--71:29}.
\newblock
\showISSN{1073-0516}
\urldef\tempurl%
\url{https://doi.org/10.1145/3534561}
\showDOI{\tempurl}


\bibitem[Jakesch et~al\mbox{.}(2019)]%
        {jakesch_AIMediatedCommunicationHow_2019}
\bibfield{author}{\bibinfo{person}{Maurice Jakesch}, \bibinfo{person}{Megan French}, \bibinfo{person}{Xiao Ma}, \bibinfo{person}{Jeffrey~T. Hancock}, {and} \bibinfo{person}{Mor Naaman}.} \bibinfo{year}{2019}\natexlab{}.
\newblock \showarticletitle{{AI}-{Mediated} {Communication}: {How} the {Perception} that {Profile} {Text} was {Written} by {AI} {Affects} {Trustworthiness}}. In \bibinfo{booktitle}{\emph{Proceedings of the 2019 {CHI} {Conference} on {Human} {Factors} in {Computing} {Systems}}} \emph{(\bibinfo{series}{{CHI} '19})}. \bibinfo{publisher}{Association for Computing Machinery}, \bibinfo{address}{New York, NY, USA}, \bibinfo{pages}{1--13}.
\newblock
\showISBNx{978-1-4503-5970-2}
\urldef\tempurl%
\url{https://doi.org/10.1145/3290605.3300469}
\showDOI{\tempurl}


\bibitem[Johanson et~al\mbox{.}({[n.\,d.]})]%
        {johanson_HelpingPlayersProgress_2023}
\bibfield{author}{\bibinfo{person}{Colby Johanson}, \bibinfo{person}{Brandon Piller}, \bibinfo{person}{Carl Gutwin}, {and} \bibinfo{person}{Regan~L. Mandryk}.} \bibinfo{year}{[n.\,d.]}\natexlab{}.
\newblock \showarticletitle{If at First You Don’t Succeed: Helping Players Make Progress in Games with Breaks and Checkpoints}.
\newblock   \bibinfo{volume}{7} (\bibinfo{year}{[n.\,d.]}), \bibinfo{pages}{387:342--387:368}.
\newblock
Issue {CHI} {PLAY}.
\urldef\tempurl%
\url{https://doi.org/10.1145/3611033}
\showDOI{\tempurl}


\bibitem[Johnston et~al\mbox{.}(2019)]%
        {johnston2019framework}
\bibfield{author}{\bibinfo{person}{Vivien Johnston}, \bibinfo{person}{Michaela Black}, \bibinfo{person}{Jonathan Wallace}, \bibinfo{person}{Maurice Mulvenna}, {and} \bibinfo{person}{Raymond Bond}.} \bibinfo{year}{2019}\natexlab{}.
\newblock \showarticletitle{A framework for the development of a dynamic adaptive intelligent user interface to enhance the user experience}. In \bibinfo{booktitle}{\emph{Proceedings of the 31st European Conference on Cognitive Ergonomics}}. \bibinfo{pages}{32--35}.
\newblock


\bibitem[Judge et~al\mbox{.}(2002)]%
        {judge_PersonalityLeadershipQualitative_2002}
\bibfield{author}{\bibinfo{person}{Timothy~A. Judge}, \bibinfo{person}{Joyce~E. Bono}, \bibinfo{person}{Remus Ilies}, {and} \bibinfo{person}{Megan~W. Gerhardt}.} \bibinfo{year}{2002}\natexlab{}.
\newblock \showarticletitle{Personality and leadership: {A} qualitative and quantitative review.}
\newblock \bibinfo{journal}{\emph{Journal of Applied Psychology}} \bibinfo{volume}{87}, \bibinfo{number}{4} (\bibinfo{year}{2002}), \bibinfo{pages}{765--780}.
\newblock
\showISSN{1939-1854, 0021-9010}
\urldef\tempurl%
\url{https://doi.org/10.1037/0021-9010.87.4.765}
\showDOI{\tempurl}


\bibitem[Kark(2011)]%
        {kark_GamesManagersPlay_2011}
\bibfield{author}{\bibinfo{person}{Ronit Kark}.} \bibinfo{year}{2011}\natexlab{}.
\newblock \showarticletitle{Games {Managers} {Play}: {Play} as a {Form} of {Leadership} {Development}}.
\newblock \bibinfo{journal}{\emph{Academy of Management Learning \& Education}} \bibinfo{volume}{10}, \bibinfo{number}{3} (\bibinfo{date}{Sept.} \bibinfo{year}{2011}), \bibinfo{pages}{507--527}.
\newblock
\showISSN{1537-260X, 1944-9585}
\urldef\tempurl%
\url{https://doi.org/10.5465/amle.2010.0048}
\showDOI{\tempurl}


\bibitem[Karnieli-Miller(2020)]%
        {karnieli-miller_ReflectivePracticeTeaching_2020}
\bibfield{author}{\bibinfo{person}{Orit Karnieli-Miller}.} \bibinfo{year}{2020}\natexlab{}.
\newblock \showarticletitle{Reflective practice in the teaching of communication skills}.
\newblock \bibinfo{journal}{\emph{Patient Education and Counseling}} \bibinfo{volume}{103}, \bibinfo{number}{10} (\bibinfo{date}{Oct.} \bibinfo{year}{2020}), \bibinfo{pages}{2166--2172}.
\newblock
\showISSN{0738-3991}
\urldef\tempurl%
\url{https://doi.org/10.1016/j.pec.2020.06.021}
\showDOI{\tempurl}


\bibitem[Keyton(2010)]%
        {keyton_CommunicationOrganizationalCulture_2010}
\bibfield{author}{\bibinfo{person}{Joann Keyton}.} \bibinfo{year}{2010}\natexlab{}.
\newblock \bibinfo{booktitle}{\emph{Communication and {Organizational} {Culture}: {A} {Key} to {Understanding} {Work} {Experiences}}}.
\newblock \bibinfo{publisher}{SAGE Publications}.
\newblock
\showISBNx{978-1-4833-6234-2}
\newblock
\shownote{Google-Books-ID: TV12AwAAQBAJ}.


\bibitem[Kimani et~al\mbox{.}(2019)]%
        {kimani_ConversationalAgentSupport_2019}
\bibfield{author}{\bibinfo{person}{Everlyne Kimani}, \bibinfo{person}{Kael Rowan}, \bibinfo{person}{Daniel McDuff}, \bibinfo{person}{Mary Czerwinski}, {and} \bibinfo{person}{Gloria Mark}.} \bibinfo{year}{2019}\natexlab{}.
\newblock \showarticletitle{A {Conversational} {Agent} in {Support} of {Productivity} and {Wellbeing} at {Work}}. In \bibinfo{booktitle}{\emph{2019 8th {International} {Conference} on {Affective} {Computing} and {Intelligent} {Interaction} ({ACII})}}. \bibinfo{pages}{1--7}.
\newblock
\urldef\tempurl%
\url{https://doi.org/10.1109/ACII.2019.8925488}
\showDOI{\tempurl}
\newblock
\shownote{ISSN: 2156-8111}.


\bibitem[Koester(2006)]%
        {koester_InvestigatingWorkplaceDiscourse_2006}
\bibfield{author}{\bibinfo{person}{Almut Koester}.} \bibinfo{year}{2006}\natexlab{}.
\newblock \bibinfo{booktitle}{\emph{Investigating {Workplace} {Discourse}}}.
\newblock \bibinfo{publisher}{Routledge}, \bibinfo{address}{London}.
\newblock
\showISBNx{978-0-203-01574-2}
\urldef\tempurl%
\url{https://doi.org/10.4324/9780203015742}
\showDOI{\tempurl}


\bibitem[Kolb(1984)]%
        {kolb_ExperientialLearningExperience_1984}
\bibfield{author}{\bibinfo{person}{David~A. Kolb}.} \bibinfo{year}{1984}\natexlab{}.
\newblock \bibinfo{booktitle}{\emph{Experiential {Learning}: {Experience} as the {Source} of {Learning} and {Development}}}.
\newblock \bibinfo{publisher}{Prentice-Hall}.
\newblock
\showISBNx{978-0-13-295261-3}
\newblock
\shownote{Google-Books-ID: zXruAAAAMAAJ}.


\bibitem[Kolomaznik et~al\mbox{.}(2024)]%
        {kolomaznik_RoleSocioemotionalAttributes_2024}
\bibfield{author}{\bibinfo{person}{Michal Kolomaznik}, \bibinfo{person}{Vladimir Petrik}, \bibinfo{person}{Michal Slama}, {and} \bibinfo{person}{Vojtech Jurik}.} \bibinfo{year}{2024}\natexlab{}.
\newblock \showarticletitle{The role of socio-emotional attributes in enhancing human-{AI} collaboration}.
\newblock \bibinfo{journal}{\emph{Frontiers in Psychology}}  \bibinfo{volume}{15} (\bibinfo{date}{Oct.} \bibinfo{year}{2024}).
\newblock
\showISSN{1664-1078}
\urldef\tempurl%
\url{https://doi.org/10.3389/fpsyg.2024.1369957}
\showDOI{\tempurl}
\newblock
\shownote{Publisher: Frontiers}.


\bibitem[Kovacevic et~al\mbox{.}(2024)]%
        {kovacevic_ChatbotsAttitudeEnhancing_2024}
\bibfield{author}{\bibinfo{person}{Nikola Kovacevic}, \bibinfo{person}{Tobias Boschung}, \bibinfo{person}{Christian Holz}, \bibinfo{person}{Markus Gross}, {and} \bibinfo{person}{Rafael Wampfler}.} \bibinfo{year}{2024}\natexlab{}.
\newblock \showarticletitle{Chatbots {With} {Attitude}: {Enhancing} {Chatbot} {Interactions} {Through} {Dynamic} {Personality} {Infusion}}. In \bibinfo{booktitle}{\emph{Proceedings of the 6th {ACM} {Conference} on {Conversational} {User} {Interfaces}}} \emph{(\bibinfo{series}{{CUI} '24})}. \bibinfo{publisher}{Association for Computing Machinery}, \bibinfo{address}{New York, NY, USA}, \bibinfo{pages}{1--16}.
\newblock
\showISBNx{9798400705113}
\urldef\tempurl%
\url{https://doi.org/10.1145/3640794.3665543}
\showDOI{\tempurl}


\bibitem[Kurtessis et~al\mbox{.}(2017)]%
        {kurtessis_PerceivedOrganizationalSupport_2017}
\bibfield{author}{\bibinfo{person}{James~N. Kurtessis}, \bibinfo{person}{Robert Eisenberger}, \bibinfo{person}{Michael~T. Ford}, \bibinfo{person}{Louis~C. Buffardi}, \bibinfo{person}{Kathleen~A. Stewart}, {and} \bibinfo{person}{Cory~S. Adis}.} \bibinfo{year}{2017}\natexlab{}.
\newblock \showarticletitle{Perceived {Organizational} {Support}: {A} {Meta}-{Analytic} {Evaluation} of {Organizational} {Support} {Theory}}.
\newblock \bibinfo{journal}{\emph{Journal of Management}} \bibinfo{volume}{43}, \bibinfo{number}{6} (\bibinfo{date}{July} \bibinfo{year}{2017}), \bibinfo{pages}{1854--1884}.
\newblock
\showISSN{0149-2063}
\urldef\tempurl%
\url{https://doi.org/10.1177/0149206315575554}
\showDOI{\tempurl}
\newblock
\shownote{Publisher: SAGE Publications Inc}.


\bibitem[Köchling and Wehner(2020)]%
        {kochling_DiscriminatedAlgorithmSystematic_2020}
\bibfield{author}{\bibinfo{person}{Alina Köchling} {and} \bibinfo{person}{Marius~Claus Wehner}.} \bibinfo{year}{2020}\natexlab{}.
\newblock \showarticletitle{Discriminated by an algorithm: a systematic review of discrimination and fairness by algorithmic decision-making in the context of {HR} recruitment and {HR} development}.
\newblock \bibinfo{journal}{\emph{Business Research}} \bibinfo{volume}{13}, \bibinfo{number}{3} (\bibinfo{date}{Nov.} \bibinfo{year}{2020}), \bibinfo{pages}{795--848}.
\newblock
\showISSN{2198-2627}
\urldef\tempurl%
\url{https://doi.org/10.1007/s40685-020-00134-w}
\showDOI{\tempurl}


\bibitem[Lacerenza et~al\mbox{.}(2017)]%
        {lacerenza_LeadershipTrainingDesign_2017}
\bibfield{author}{\bibinfo{person}{C. Lacerenza}, \bibinfo{person}{Denise~L. Reyes}, \bibinfo{person}{Shannon~L. Marlow}, \bibinfo{person}{Dana~L Joseph}, {and} \bibinfo{person}{E. Salas}.} \bibinfo{year}{2017}\natexlab{}.
\newblock \showarticletitle{Leadership Training Design, Delivery, and Implementation: A Meta-Analysis}.
\newblock \bibinfo{journal}{\emph{Journal of Applied Psychology}}  \bibinfo{volume}{102} (\bibinfo{year}{2017}), \bibinfo{pages}{1686–1718}.
\newblock
\urldef\tempurl%
\url{https://doi.org/10.1037/apl0000241}
\showDOI{\tempurl}


\bibitem[Lackner and Martini(2017)]%
        {lackner_HelpingUniversityStudents_2017}
\bibfield{author}{\bibinfo{person}{Christine Lackner} {and} \bibinfo{person}{Tanya Martini}.} \bibinfo{year}{2017}\natexlab{}.
\newblock \showarticletitle{Helping university students succeed at employment interviews: {The} role of self-reflection in e-portfolios}.
\newblock \bibinfo{journal}{\emph{Teaching and Learning Inquiry}} \bibinfo{volume}{5}, \bibinfo{number}{2} (\bibinfo{date}{Sept.} \bibinfo{year}{2017}), \bibinfo{pages}{3--15}.
\newblock
\showISSN{2167-4787}
\urldef\tempurl%
\url{https://doi.org/10.20343/teachlearninqu.5.2.2}
\showDOI{\tempurl}
\newblock
\shownote{Number: 2}.


\bibitem[Ladousse(1987)]%
        {ladousse_RolePlay_1987}
\bibfield{author}{\bibinfo{person}{Gillian~Porter Ladousse}.} \bibinfo{year}{1987}\natexlab{}.
\newblock \bibinfo{booktitle}{\emph{Role {Play}}}.
\newblock \bibinfo{publisher}{OUP Oxford}.
\newblock
\showISBNx{978-0-19-437095-0}
\newblock
\shownote{Google-Books-ID: oS4STyscmpYC}.


\bibitem[Lewis et~al\mbox{.}(2021)]%
        {lewis_RetrievalAugmentedGenerationKnowledgeIntensive_2021}
\bibfield{author}{\bibinfo{person}{Patrick Lewis}, \bibinfo{person}{Ethan Perez}, \bibinfo{person}{Aleksandra Piktus}, \bibinfo{person}{Fabio Petroni}, \bibinfo{person}{Vladimir Karpukhin}, \bibinfo{person}{Naman Goyal}, \bibinfo{person}{Heinrich Küttler}, \bibinfo{person}{Mike Lewis}, \bibinfo{person}{Wen-tau Yih}, \bibinfo{person}{Tim Rocktäschel}, \bibinfo{person}{Sebastian Riedel}, {and} \bibinfo{person}{Douwe Kiela}.} \bibinfo{year}{2021}\natexlab{}.
\newblock \bibinfo{title}{Retrieval-{Augmented} {Generation} for {Knowledge}-{Intensive} {NLP} {Tasks}}.
\newblock
\newblock
\urldef\tempurl%
\url{https://doi.org/10.48550/arXiv.2005.11401}
\showDOI{\tempurl}
\newblock
\shownote{arXiv:2005.11401 [cs]}.


\bibitem[Licklider(1960)]%
        {licklider_ManComputerSymbiosis_1960}
\bibfield{author}{\bibinfo{person}{J.~C.~R. Licklider}.} \bibinfo{year}{1960}\natexlab{}.
\newblock \bibinfo{title}{Man-{Computer} {Symbiosis}}.
\newblock
\newblock
\urldef\tempurl%
\url{https://groups.csail.mit.edu/medg/people/psz/Licklider.html}
\showURL{%
\tempurl}


\bibitem[Liu et~al\mbox{.}(2018)]%
        {liu_DialogueLearningHuman_2018}
\bibfield{author}{\bibinfo{person}{Bing Liu}, \bibinfo{person}{Gokhan Tur}, \bibinfo{person}{Dilek Hakkani-Tur}, \bibinfo{person}{Pararth Shah}, {and} \bibinfo{person}{Larry Heck}.} \bibinfo{year}{2018}\natexlab{}.
\newblock \bibinfo{title}{Dialogue {Learning} with {Human} {Teaching} and {Feedback} in {End}-to-{End} {Trainable} {Task}-{Oriented} {Dialogue} {Systems}}.
\newblock
\newblock
\urldef\tempurl%
\url{https://doi.org/10.48550/arXiv.1804.06512}
\showDOI{\tempurl}
\newblock
\shownote{arXiv:1804.06512 [cs]}.


\bibitem[Loo(2002)]%
        {loo_JournalingLearningTool_2002}
\bibfield{author}{\bibinfo{person}{Robert Loo}.} \bibinfo{year}{2002}\natexlab{}.
\newblock \showarticletitle{Journaling: {A} {Learning} {Tool} for {Project} {Management} {Training} and {Team}-building}.
\newblock \bibinfo{journal}{\emph{Project Management Journal}} \bibinfo{volume}{33}, \bibinfo{number}{4} (\bibinfo{date}{Dec.} \bibinfo{year}{2002}), \bibinfo{pages}{61--66}.
\newblock
\showISSN{8756-9728}
\urldef\tempurl%
\url{https://doi.org/10.1177/875697280203300407}
\showDOI{\tempurl}
\newblock
\shownote{Publisher: SAGE Publications Inc}.


\bibitem[Magnisalis et~al\mbox{.}(2011)]%
        {magnisalis_AdaptiveIntelligentSystems_2011}
\bibfield{author}{\bibinfo{person}{Ioannis Magnisalis}, \bibinfo{person}{Stavros Demetriadis}, {and} \bibinfo{person}{Anastasios Karakostas}.} \bibinfo{year}{2011}\natexlab{}.
\newblock \showarticletitle{Adaptive and {Intelligent} {Systems} for {Collaborative} {Learning} {Support}: {A} {Review} of the {Field}}.
\newblock \bibinfo{journal}{\emph{IEEE Transactions on Learning Technologies}} \bibinfo{volume}{4}, \bibinfo{number}{1} (\bibinfo{date}{Jan.} \bibinfo{year}{2011}), \bibinfo{pages}{5--20}.
\newblock
\showISSN{1939-1382}
\urldef\tempurl%
\url{https://doi.org/10.1109/TLT.2011.2}
\showDOI{\tempurl}
\newblock
\shownote{Conference Name: IEEE Transactions on Learning Technologies}.


\bibitem[Mahmud et~al\mbox{.}(2023)]%
        {mahmud_StudyHumanAI_2023}
\bibfield{author}{\bibinfo{person}{Bahar Mahmud}, \bibinfo{person}{Guan Hong}, {and} \bibinfo{person}{Bernard Fong}.} \bibinfo{year}{2023}\natexlab{}.
\newblock \showarticletitle{A {Study} of {Human}–{AI} {Symbiosis} for {Creative} {Work}: {Recent} {Developments} and {Future} {Directions} in {Deep} {Learning}}.
\newblock \bibinfo{journal}{\emph{ACM Trans. Multimedia Comput. Commun. Appl.}} \bibinfo{volume}{20}, \bibinfo{number}{2} (\bibinfo{date}{Sept.} \bibinfo{year}{2023}), \bibinfo{pages}{47:1--47:21}.
\newblock
\showISSN{1551-6857}
\urldef\tempurl%
\url{https://doi.org/10.1145/3542698}
\showDOI{\tempurl}


\bibitem[Maxwell(1993)]%
        {maxwell_DevelopingLeaderYou_1993}
\bibfield{author}{\bibinfo{person}{John~C. Maxwell}.} \bibinfo{year}{1993}\natexlab{}.
\newblock \bibinfo{booktitle}{\emph{Developing the {{Leader Within You}}}}.
\newblock \bibinfo{publisher}{Thomas Nelson Inc}.
\newblock
\showISBNx{978-0-8407-6744-8}
\showeprint[googlebooks]{tBdBo0dfX88C}


\bibitem[Maxwell(2007a)]%
        {maxwell_21IndispensableQualities_2007}
\bibfield{author}{\bibinfo{person}{John~C. Maxwell}.} \bibinfo{year}{2007}\natexlab{a}.
\newblock \bibinfo{booktitle}{\emph{The 21 {{Indispensable Qualities}} of a {{Leader}}: {{Becoming}} the {{Person Others Will Want}} to {{Follow}}}}.
\newblock \bibinfo{publisher}{HarperCollins Leadership}.
\newblock
\showISBNx{978-1-4185-0823-4}
\showeprint[googlebooks]{HaO2qbfBMGEC}


\bibitem[Maxwell(2007b)]%
        {maxwell_21IrrefutableLaws_2007}
\bibfield{author}{\bibinfo{person}{John~C. Maxwell}.} \bibinfo{year}{2007}\natexlab{b}.
\newblock \bibinfo{booktitle}{\emph{The 21 {{Irrefutable Laws}} of {{Leadership}}: {{Follow Them}} and {{People Will Follow You}}}}.
\newblock \bibinfo{publisher}{HarperCollins Leadership}.
\newblock
\showISBNx{978-1-4185-0826-5}


\bibitem[McCall(2004)]%
        {mccall_LeadershipDevelopmentExperience_2004}
\bibfield{author}{\bibinfo{person}{Morgan~W. McCall}.} \bibinfo{year}{2004}\natexlab{}.
\newblock \showarticletitle{Leadership {Development} through {Experience}}.
\newblock \bibinfo{journal}{\emph{The Academy of Management Executive (1993-2005)}} \bibinfo{volume}{18}, \bibinfo{number}{3} (\bibinfo{year}{2004}), \bibinfo{pages}{127--130}.
\newblock
\showISSN{1079-5545}
\urldef\tempurl%
\url{https://www.jstor.org/stable/4166101}
\showURL{%
\tempurl}
\newblock
\shownote{Publisher: Academy of Management}.


\bibitem[McDermott and Roth(1978)]%
        {mcdermott_SocialOrganizationBehavior_1978}
\bibfield{author}{\bibinfo{person}{R.~P. McDermott} {and} \bibinfo{person}{David~R. Roth}.} \bibinfo{year}{1978}\natexlab{}.
\newblock \showarticletitle{The {Social} {Organization} of {Behavior}: {Interactional} {Approaches}}.
\newblock \bibinfo{journal}{\emph{Annual Review of Anthropology}} \bibinfo{volume}{7}, \bibinfo{number}{1} (\bibinfo{date}{Oct.} \bibinfo{year}{1978}), \bibinfo{pages}{321--345}.
\newblock
\showISSN{0084-6570, 1545-4290}
\urldef\tempurl%
\url{https://doi.org/10.1146/annurev.an.07.100178.001541}
\showDOI{\tempurl}


\bibitem[McGovern et~al\mbox{.}(2020)]%
        {mcgovern_ApplicationVirtualReality_2020}
\bibfield{author}{\bibinfo{person}{Enda McGovern}, \bibinfo{person}{Gerardo Moreira}, {and} \bibinfo{person}{Cuauhtemoc Luna-Nevarez}.} \bibinfo{year}{2020}\natexlab{}.
\newblock \showarticletitle{An application of virtual reality in education: {Can} this technology enhance the quality of students’ learning experience?}
\newblock \bibinfo{journal}{\emph{Journal of Education for Business}} \bibinfo{volume}{95}, \bibinfo{number}{7} (\bibinfo{date}{Oct.} \bibinfo{year}{2020}), \bibinfo{pages}{490--496}.
\newblock
\showISSN{0883-2323}
\urldef\tempurl%
\url{https://doi.org/10.1080/08832323.2019.1703096}
\showDOI{\tempurl}
\newblock
\shownote{Publisher: Routledge \_eprint: https://doi.org/10.1080/08832323.2019.1703096}.


\bibitem[Mclaren et~al\mbox{.}(2010)]%
        {mclaren_SupportingCollaborativeLearning_2010}
\bibfield{author}{\bibinfo{person}{Bruce~M. Mclaren}, \bibinfo{person}{Oliver Scheuer}, {and} \bibinfo{person}{Jan Mikšátko}.} \bibinfo{year}{2010}\natexlab{}.
\newblock \showarticletitle{Supporting {Collaborative} {Learning} and {E}-{Discussions} {Using} {Artificial} {Intelligence} {Techniques}}.
\newblock \bibinfo{journal}{\emph{International Journal of Artificial Intelligence in Education}} \bibinfo{volume}{20}, \bibinfo{number}{1} (\bibinfo{date}{Jan.} \bibinfo{year}{2010}), \bibinfo{pages}{1--46}.
\newblock
\showISSN{1560-4292}
\urldef\tempurl%
\url{https://doi.org/10.3233/JAI-2010-0001}
\showDOI{\tempurl}
\newblock
\shownote{Publisher: IOS Press}.


\bibitem[Mena-Guacas et~al\mbox{.}(2023)]%
        {mena-guacas_CollaborativeLearningSkill_2023}
\bibfield{author}{\bibinfo{person}{Andres~F. Mena-Guacas}, \bibinfo{person}{Jairo Alonso~Urueña Rodríguez}, \bibinfo{person}{David Mauricio~Santana Trujillo}, \bibinfo{person}{José Gómez-Galán}, {and} \bibinfo{person}{Eloy López-Meneses}.} \bibinfo{year}{2023}\natexlab{}.
\newblock \showarticletitle{Collaborative learning and skill development for educational growth of artificial intelligence: {A} systematic review}.
\newblock \bibinfo{journal}{\emph{Contemporary Educational Technology}} \bibinfo{volume}{15}, \bibinfo{number}{3} (\bibinfo{date}{July} \bibinfo{year}{2023}), \bibinfo{pages}{ep428}.
\newblock
\showISSN{1309-517X}
\urldef\tempurl%
\url{https://doi.org/10.30935/cedtech/13123}
\showDOI{\tempurl}
\newblock
\shownote{Publisher: Bastas}.


\bibitem[Metzger et~al\mbox{.}(2017)]%
        {metzger_HowMachinesAre_2017}
\bibfield{author}{\bibinfo{person}{Dirk Metzger}, \bibinfo{person}{Christina Niemöller}, \bibinfo{person}{Benjamin Wingert}, \bibinfo{person}{Tobias Schultze}, \bibinfo{person}{Matthias Bues}, {and} \bibinfo{person}{Oliver Thomas}.} \bibinfo{year}{2017}\natexlab{}.
\newblock \showarticletitle{How {{Machines}} Are {{Serviced}} - {{Design}} of a {{Virtual Reality-based Training System}} for {{Technical Customer Services}}}.
\newblock \bibinfo{journal}{\emph{Wirtschaftsinformatik 2017 Proceedings}} (\bibinfo{year}{2017}).
\newblock
\urldef\tempurl%
\url{https://aisel.aisnet.org/wi2017/track06/paper/3}
\showURL{%
\tempurl}


\bibitem[Moldoveanu and Narayandas(2019)]%
        {moldoveanu_FutureLeadershipDevelopment_2019}
\bibfield{author}{\bibinfo{person}{Mihnea Moldoveanu} {and} \bibinfo{person}{Das Narayandas}.} \bibinfo{year}{2019}\natexlab{}.
\newblock \showarticletitle{The {Future} of {Leadership} {Development}}.
\newblock \bibinfo{journal}{\emph{Harvard business review}}  \bibinfo{volume}{March April 2019} (\bibinfo{date}{March} \bibinfo{year}{2019}), \bibinfo{pages}{40}.
\newblock


\bibitem[Muller and Kogan(2012)]%
        {muller_GroundedTheoryMethod_2012}
\bibfield{author}{\bibinfo{person}{Michael~J. Muller} {and} \bibinfo{person}{Sandra Kogan}.} \bibinfo{year}{2012}\natexlab{}.
\newblock \showarticletitle{Grounded {Theory} {Method} in {Human}-{Computer} {Interaction} and {Computer}-{Supported} {Cooperative} {Work}}.
\newblock In \bibinfo{booktitle}{\emph{Human {Computer} {Interaction} {Handbook}} (\bibinfo{edition}{3} ed.)}. \bibinfo{publisher}{CRC Press}.
\newblock
\showISBNx{978-0-429-10397-1}
\newblock
\shownote{Num Pages: 21}.


\bibitem[Neves and Eisenberger(2012)]%
        {neves_ManagementCommunicationEmployee_2012}
\bibfield{author}{\bibinfo{person}{Pedro Neves} {and} \bibinfo{person}{Robert Eisenberger}.} \bibinfo{year}{2012}\natexlab{}.
\newblock \showarticletitle{Management {Communication} and {Employee} {Performance}: {The} {Contribution} of {Perceived} {Organizational} {Support}}.
\newblock \bibinfo{journal}{\emph{Human Performance}} \bibinfo{volume}{25}, \bibinfo{number}{5} (\bibinfo{date}{Nov.} \bibinfo{year}{2012}), \bibinfo{pages}{452--464}.
\newblock
\showISSN{0895-9285}
\urldef\tempurl%
\url{https://doi.org/10.1080/08959285.2012.721834}
\showDOI{\tempurl}
\newblock
\shownote{Publisher: Routledge \_eprint: https://doi.org/10.1080/08959285.2012.721834}.


\bibitem[Ng et~al\mbox{.}(2023)]%
        {ng_RealTimeHybridLanguage_2023}
\bibfield{author}{\bibinfo{person}{Han~Wei Ng}, \bibinfo{person}{Aiden Koh}, \bibinfo{person}{Anthea Foong}, {and} \bibinfo{person}{Jeremy Ong}.} \bibinfo{year}{2023}\natexlab{}.
\newblock \showarticletitle{Real-{Time} {Hybrid} {Language} {Model} for {Virtual} {Patient} {Conversations}}. In \bibinfo{booktitle}{\emph{Artificial {Intelligence} in {Education}}}, \bibfield{editor}{\bibinfo{person}{Ning Wang}, \bibinfo{person}{Genaro Rebolledo-Mendez}, \bibinfo{person}{Noboru Matsuda}, \bibinfo{person}{Olga~C. Santos}, {and} \bibinfo{person}{Vania Dimitrova}} (Eds.). \bibinfo{publisher}{Springer Nature Switzerland}, \bibinfo{address}{Cham}, \bibinfo{pages}{780--785}.
\newblock
\showISBNx{978-3-031-36272-9}
\urldef\tempurl%
\url{https://doi.org/10.1007/978-3-031-36272-9_71}
\showDOI{\tempurl}


\bibitem[Northouse(2021)]%
        {northouse_LeadershipTheoryPractice_2021}
\bibfield{author}{\bibinfo{person}{Peter~G. Northouse}.} \bibinfo{year}{2021}\natexlab{}.
\newblock \bibinfo{booktitle}{\emph{Leadership: {Theory} and {Practice}}}.
\newblock \bibinfo{publisher}{SAGE Publications}.
\newblock
\showISBNx{978-1-07-183447-3}
\newblock
\shownote{Google-Books-ID: 6qYLEAAAQBAJ}.


\bibitem[Ortiz and Harrell(2018)]%
        {ortiz_EnablingCriticalSelfReflection_2018}
\bibfield{author}{\bibinfo{person}{Pablo Ortiz} {and} \bibinfo{person}{D.~Fox Harrell}.} \bibinfo{year}{2018}\natexlab{}.
\newblock \showarticletitle{Enabling {Critical} {Self}-{Reflection} through {Roleplay} with {Chimeria}: {Grayscale}}. In \bibinfo{booktitle}{\emph{Proceedings of the 2018 {Annual} {Symposium} on {Computer}-{Human} {Interaction} in {Play}}} \emph{(\bibinfo{series}{{CHI} {PLAY} '18})}. \bibinfo{publisher}{Association for Computing Machinery}, \bibinfo{address}{New York, NY, USA}, \bibinfo{pages}{353--364}.
\newblock
\showISBNx{978-1-4503-5624-4}
\urldef\tempurl%
\url{https://doi.org/10.1145/3242671.3242687}
\showDOI{\tempurl}


\bibitem[Othlinghaus-Wulhorst and Hoppe({[n.\,d.]})]%
        {othlinghaus-wulhorst_TechnicalConceptualFramework_2020}
\bibfield{author}{\bibinfo{person}{Julia Othlinghaus-Wulhorst} {and} \bibinfo{person}{H.~Ulrich Hoppe}.} \bibinfo{year}{[n.\,d.]}\natexlab{}.
\newblock \showarticletitle{A Technical and Conceptual Framework for Serious Role-Playing Games in the Area of Social Skill Training}. In \bibinfo{booktitle}{\emph{Frontiers in Computer Science}} (2020-07-31), Vol.~\bibinfo{volume}{2}. \bibinfo{pages}{28}.
\newblock
\urldef\tempurl%
\url{https://doi.org/10.3389/fcomp.2020.00028}
\showDOI{\tempurl}
\newblock
\shownote{{ISSN}: 2624-9898 Journal Abbreviation: Front. Comput. Sci.}.


\bibitem[Pang et~al\mbox{.}(2025)]%
        {pang_UnderstandingLLMificationCHI_2025}
\bibfield{author}{\bibinfo{person}{Rock~Yuren Pang}, \bibinfo{person}{Hope Schroeder}, \bibinfo{person}{Kynnedy~Simone Smith}, \bibinfo{person}{Solon Barocas}, \bibinfo{person}{Ziang Xiao}, \bibinfo{person}{Emily Tseng}, {and} \bibinfo{person}{Danielle Bragg}.} \bibinfo{year}{2025}\natexlab{}.
\newblock \bibinfo{title}{Understanding the {LLM}-ification of {CHI}: {Unpacking} the {Impact} of {LLMs} at {CHI} through a {Systematic} {Literature} {Review}}.
\newblock
\newblock
\urldef\tempurl%
\url{https://doi.org/10.48550/arXiv.2501.12557}
\showDOI{\tempurl}
\newblock
\shownote{arXiv:2501.12557 [cs]}.


\bibitem[Pangh et~al\mbox{.}(2019)]%
        {pangh_EffectReflectionNursePatient_2019}
\bibfield{author}{\bibinfo{person}{Bahman Pangh}, \bibinfo{person}{Leila Jouybari}, \bibinfo{person}{Mohamad~Ali Vakili}, \bibinfo{person}{Akram Sanagoo}, {and} \bibinfo{person}{Aysheh Torik}.} \bibinfo{year}{2019}\natexlab{}.
\newblock \showarticletitle{The {Effect} of {Reflection} on {Nurse}-{Patient} {Communication} {Skills} in {Emergency} {Medical} {Centers}}.
\newblock \bibinfo{journal}{\emph{Journal of Caring Sciences}} \bibinfo{volume}{8}, \bibinfo{number}{2} (\bibinfo{date}{June} \bibinfo{year}{2019}), \bibinfo{pages}{75--81}.
\newblock
\urldef\tempurl%
\url{https://doi.org/10.15171/jcs.2019.011}
\showDOI{\tempurl}
\newblock
\shownote{Num Pages: 75-81 Place: Tabriz, Iran Publisher: Tabriz University of Medical Sciences Section: Original Article}.


\bibitem[Park and Choi(2023)]%
        {park_AudiLensConfigurableLLMGenerated_2023}
\bibfield{author}{\bibinfo{person}{Jeongeon Park} {and} \bibinfo{person}{DaEun Choi}.} \bibinfo{year}{2023}\natexlab{}.
\newblock \showarticletitle{{AudiLens}: {Configurable} {LLM}-{Generated} {Audiences} for {Public} {Speech} {Practice}}. In \bibinfo{booktitle}{\emph{Adjunct {Proceedings} of the 36th {Annual} {ACM} {Symposium} on {User} {Interface} {Software} and {Technology}}} \emph{(\bibinfo{series}{{UIST} '23 {Adjunct}})}. \bibinfo{publisher}{Association for Computing Machinery}, \bibinfo{address}{New York, NY, USA}, \bibinfo{pages}{1--3}.
\newblock
\showISBNx{9798400700965}
\urldef\tempurl%
\url{https://doi.org/10.1145/3586182.3625114}
\showDOI{\tempurl}


\bibitem[Park et~al\mbox{.}(2023)]%
        {park_GenerativeAgentsInteractive_2023a}
\bibfield{author}{\bibinfo{person}{Joon~Sung Park}, \bibinfo{person}{Joseph O'Brien}, \bibinfo{person}{Carrie~Jun Cai}, \bibinfo{person}{Meredith~Ringel Morris}, \bibinfo{person}{Percy Liang}, {and} \bibinfo{person}{Michael~S. Bernstein}.} \bibinfo{year}{2023}\natexlab{}.
\newblock \showarticletitle{Generative {{Agents}}: {{Interactive Simulacra}} of {{Human Behavior}}}. In \bibinfo{booktitle}{\emph{Proceedings of the 36th {{Annual ACM Symposium}} on {{User Interface Software}} and {{Technology}}}} (San Francisco CA USA, 2023-10-29). \bibinfo{publisher}{ACM}, \bibinfo{pages}{1--22}.
\newblock
\showISBNx{9798400701320}
\urldef\tempurl%
\url{https://doi.org/10.1145/3586183.3606763}
\showDOI{\tempurl}


\bibitem[Park et~al\mbox{.}(2022)]%
        {park_SocialSimulacraCreating_2022}
\bibfield{author}{\bibinfo{person}{Joon~Sung Park}, \bibinfo{person}{Lindsay Popowski}, \bibinfo{person}{Carrie~J. Cai}, \bibinfo{person}{Meredith~Ringel Morris}, \bibinfo{person}{Percy Liang}, {and} \bibinfo{person}{Michael~S. Bernstein}.} \bibinfo{year}{2022}\natexlab{}.
\newblock \bibinfo{title}{Social {Simulacra}: {Creating} {Populated} {Prototypes} for {Social} {Computing} {Systems}}.
\newblock
\newblock
\urldef\tempurl%
\url{https://doi.org/10.48550/arXiv.2208.04024}
\showDOI{\tempurl}
\newblock
\shownote{arXiv:2208.04024 [cs]}.


\bibitem[Plaza-del Arco et~al\mbox{.}(2024)]%
        {plaza-del-arco_AngryMenSad_2024}
\bibfield{author}{\bibinfo{person}{Flor~Miriam Plaza-del Arco}, \bibinfo{person}{Amanda~Cercas Curry}, \bibinfo{person}{Alba Curry}, \bibinfo{person}{Gavin Abercrombie}, {and} \bibinfo{person}{Dirk Hovy}.} \bibinfo{year}{2024}\natexlab{}.
\newblock \bibinfo{title}{Angry {Men}, {Sad} {Women}: {Large} {Language} {Models} {Reflect} {Gendered} {Stereotypes} in {Emotion} {Attribution}}.
\newblock
\newblock
\urldef\tempurl%
\url{https://doi.org/10.48550/arXiv.2403.03121}
\showDOI{\tempurl}
\newblock
\shownote{arXiv:2403.03121 [cs]}.


\bibitem[Raffel et~al\mbox{.}(2019)]%
        {raffel_ExploringLimitsTransfer_2019}
\bibfield{author}{\bibinfo{person}{Colin Raffel}, \bibinfo{person}{Noam~M. Shazeer}, \bibinfo{person}{Adam Roberts}, \bibinfo{person}{Katherine Lee}, \bibinfo{person}{Sharan Narang}, \bibinfo{person}{Michael Matena}, \bibinfo{person}{Yanqi Zhou}, \bibinfo{person}{Wei Li}, {and} \bibinfo{person}{Peter~J. Liu}.} \bibinfo{year}{2019}\natexlab{}.
\newblock \showarticletitle{Exploring the Limits of Transfer Learning with a Unified Text-to-Text Transformer}.
\newblock \bibinfo{journal}{\emph{J. Mach. Learn. Res.}}  \bibinfo{volume}{21} (\bibinfo{year}{2019}), \bibinfo{pages}{140:1--140:67}.
\newblock
\urldef\tempurl%
\url{https://api.semanticscholar.org/CorpusID:204838007}
\showURL{%
\tempurl}


\bibitem[Raisch and Krakowski(2021)]%
        {raisch_ArtificialIntelligenceManagement_2021}
\bibfield{author}{\bibinfo{person}{Sebastian Raisch} {and} \bibinfo{person}{Sebastian Krakowski}.} \bibinfo{year}{2021}\natexlab{}.
\newblock \showarticletitle{Artificial {Intelligence} and {Management}: {The} {Automation}–{Augmentation} {Paradox}}.
\newblock \bibinfo{journal}{\emph{Academy of Management Review}} \bibinfo{volume}{46}, \bibinfo{number}{1} (\bibinfo{date}{Jan.} \bibinfo{year}{2021}), \bibinfo{pages}{192--210}.
\newblock
\showISSN{0363-7425}
\urldef\tempurl%
\url{https://doi.org/10.5465/amr.2018.0072}
\showDOI{\tempurl}
\newblock
\shownote{Publisher: Academy of Management}.


\bibitem[Rashkin et~al\mbox{.}(2019)]%
        {rashkin_EmpatheticOpendomainConversation_2019}
\bibfield{author}{\bibinfo{person}{Hannah Rashkin}, \bibinfo{person}{Eric~Michael Smith}, \bibinfo{person}{Margaret Li}, {and} \bibinfo{person}{Y-Lan Boureau}.} \bibinfo{year}{2019}\natexlab{}.
\newblock \showarticletitle{Towards {Empathetic} {Open}-domain {Conversation} {Models}: {A} {New} {Benchmark} and {Dataset}}. In \bibinfo{booktitle}{\emph{Proceedings of the 57th {Annual} {Meeting} of the {Association} for {Computational} {Linguistics}}}. \bibinfo{publisher}{Association for Computational Linguistics}, \bibinfo{address}{Florence, Italy}, \bibinfo{pages}{5370--5381}.
\newblock
\urldef\tempurl%
\url{https://doi.org/10.18653/v1/P19-1534}
\showDOI{\tempurl}


\bibitem[Retkowsky et~al\mbox{.}(2024)]%
        {retkowsky_ManagingChatGPTempoweredWorkforce_2024}
\bibfield{author}{\bibinfo{person}{Jana Retkowsky}, \bibinfo{person}{Ella Hafermalz}, {and} \bibinfo{person}{Marleen Huysman}.} \bibinfo{year}{2024}\natexlab{}.
\newblock \showarticletitle{Managing a {ChatGPT}-empowered workforce: {Understanding} its affordances and side effects}.
\newblock \bibinfo{journal}{\emph{Business Horizons}} (\bibinfo{date}{April} \bibinfo{year}{2024}).
\newblock
\showISSN{0007-6813}
\urldef\tempurl%
\url{https://doi.org/10.1016/j.bushor.2024.04.009}
\showDOI{\tempurl}


\bibitem[Rost(1993)]%
        {rost_LeadershipTwentyFirstCentury_1993}
\bibfield{author}{\bibinfo{person}{Joseph Rost}.} \bibinfo{year}{1993}\natexlab{}.
\newblock \bibinfo{booktitle}{\emph{Leadership for the {Twenty}-{First} {Century}}}.
\newblock \bibinfo{publisher}{Bloomsbury Academic}.
\newblock
\showISBNx{978-0-275-94610-4}


\bibitem[Saliba and Ostojic(2014)]%
        {saliba_PersonalityParticipationWho_2014}
\bibfield{author}{\bibinfo{person}{Anthony Saliba} {and} \bibinfo{person}{Peter Ostojic}.} \bibinfo{year}{2014}\natexlab{}.
\newblock \showarticletitle{Personality and {Participation}: {Who} {Volunteers} to {Participate} in {Studies}}.
\newblock \bibinfo{journal}{\emph{Psychology}} \bibinfo{volume}{5}, \bibinfo{number}{3} (\bibinfo{date}{March} \bibinfo{year}{2014}), \bibinfo{pages}{230--243}.
\newblock
\urldef\tempurl%
\url{https://doi.org/10.4236/psych.2014.53034}
\showDOI{\tempurl}
\newblock
\shownote{Number: 3 Publisher: Scientific Research Publishing}.


\bibitem[Sap et~al\mbox{.}(2019)]%
        {sap_RiskRacialBias_2019}
\bibfield{author}{\bibinfo{person}{Maarten Sap}, \bibinfo{person}{Dallas Card}, \bibinfo{person}{Saadia Gabriel}, \bibinfo{person}{Yejin Choi}, {and} \bibinfo{person}{Noah~A. Smith}.} \bibinfo{year}{2019}\natexlab{}.
\newblock \showarticletitle{The {Risk} of {Racial} {Bias} in {Hate} {Speech} {Detection}}. In \bibinfo{booktitle}{\emph{Proceedings of the 57th {Annual} {Meeting} of the {Association} for {Computational} {Linguistics}}}, \bibfield{editor}{\bibinfo{person}{Anna Korhonen}, \bibinfo{person}{David Traum}, {and} \bibinfo{person}{Lluís Màrquez}} (Eds.). \bibinfo{publisher}{Association for Computational Linguistics}, \bibinfo{address}{Florence, Italy}, \bibinfo{pages}{1668--1678}.
\newblock
\urldef\tempurl%
\url{https://doi.org/10.18653/v1/P19-1163}
\showDOI{\tempurl}


\bibitem[Scandura({[n.\,d.]})]%
        {scandura_MentorshipCareerMobility_1992}
\bibfield{author}{\bibinfo{person}{Terri~A. Scandura}.} \bibinfo{year}{[n.\,d.]}\natexlab{}.
\newblock \showarticletitle{Mentorship and Career Mobility: {{An}} Empirical Investigation}.
\newblock  \bibinfo{volume}{13}, \bibinfo{number}{2} (\bibinfo{year}{[n.\,d.]}), \bibinfo{pages}{169--174}.
\newblock
\showISSN{1099-1379}
\urldef\tempurl%
\url{https://doi.org/10.1002/job.4030130206}
\showDOI{\tempurl}


\bibitem[Scandura and Williams({[n.\,d.]})]%
        {scandura_MentoringTransformationalLeadership_2004}
\bibfield{author}{\bibinfo{person}{Terri~A. Scandura} {and} \bibinfo{person}{Ethlyn~A. Williams}.} \bibinfo{year}{[n.\,d.]}\natexlab{}.
\newblock \showarticletitle{Mentoring and Transformational Leadership: {{The}} Role of Supervisory Career Mentoring}.
\newblock  \bibinfo{volume}{65}, \bibinfo{number}{3} (\bibinfo{year}{[n.\,d.]}), \bibinfo{pages}{448--468}.
\newblock
\showISSN{0001-8791}
\urldef\tempurl%
\url{https://doi.org/10.1016/j.jvb.2003.10.003}
\showDOI{\tempurl}


\bibitem[Schön(1983)]%
        {schon_ReflectivePractitionerHow_1983}
\bibfield{author}{\bibinfo{person}{Donald~A. Schön}.} \bibinfo{year}{1983}\natexlab{}.
\newblock \bibinfo{booktitle}{\emph{The {Reflective} {Practitioner}: {How} {Professionals} {Think} in {Action}}}.
\newblock \bibinfo{publisher}{Ashgate}.
\newblock
\showISBNx{978-1-85742-319-8}
\newblock
\shownote{Google-Books-ID: E85qAAAAMAAJ}.


\bibitem[Serban et~al\mbox{.}(2016)]%
        {serban_BuildingEndToEndDialogue_2016}
\bibfield{author}{\bibinfo{person}{Iulian Serban}, \bibinfo{person}{Alessandro Sordoni}, \bibinfo{person}{Yoshua Bengio}, \bibinfo{person}{Aaron Courville}, {and} \bibinfo{person}{Joelle Pineau}.} \bibinfo{year}{2016}\natexlab{}.
\newblock \showarticletitle{Building {End}-{To}-{End} {Dialogue} {Systems} {Using} {Generative} {Hierarchical} {Neural} {Network} {Models}}.
\newblock \bibinfo{journal}{\emph{Proceedings of the AAAI Conference on Artificial Intelligence}} \bibinfo{volume}{30}, \bibinfo{number}{1} (\bibinfo{date}{March} \bibinfo{year}{2016}).
\newblock
\showISSN{2374-3468, 2159-5399}
\urldef\tempurl%
\url{https://doi.org/10.1609/aaai.v30i1.9883}
\showDOI{\tempurl}


\bibitem[Shaikh et~al\mbox{.}({[n.\,d.]})]%
        {shaikh_RehearsalSimulatingConflict_2024}
\bibfield{author}{\bibinfo{person}{Omar Shaikh}, \bibinfo{person}{Valentino~Emil Chai}, \bibinfo{person}{Michele Gelfand}, \bibinfo{person}{Diyi Yang}, {and} \bibinfo{person}{Michael~S. Bernstein}.} \bibinfo{year}{[n.\,d.]}\natexlab{}.
\newblock \showarticletitle{Rehearsal: {{Simulating Conflict}} to {{Teach Conflict Resolution}}}. In \bibinfo{booktitle}{\emph{Proceedings of the {{CHI Conference}} on {{Human Factors}} in {{Computing Systems}}}} (New York, NY, USA, 2024-05-11) \emph{(\bibinfo{series}{{{CHI}} '24})}. \bibinfo{publisher}{Association for Computing Machinery}, \bibinfo{pages}{1--20}.
\newblock
\showISBNx{9798400703300}
\urldef\tempurl%
\url{https://doi.org/10.1145/3613904.3642159}
\showDOI{\tempurl}


\bibitem[Shneiderman(2020)]%
        {shneiderman_HumanCenteredArtificialIntelligence_2020}
\bibfield{author}{\bibinfo{person}{Ben Shneiderman}.} \bibinfo{year}{2020}\natexlab{}.
\newblock \bibinfo{title}{Human-{Centered} {Artificial} {Intelligence}: {Reliable}, {Safe} \& {Trustworthy}}.
\newblock
\newblock
\urldef\tempurl%
\url{https://doi.org/10.48550/arXiv.2002.04087}
\showDOI{\tempurl}
\newblock
\shownote{arXiv:2002.04087 [cs]}.


\bibitem[St-Hilaire and Gilbert(2019)]%
        {st-hilaire_WhatLeadersNeed_2019}
\bibfield{author}{\bibinfo{person}{France St-Hilaire} {and} \bibinfo{person}{Marie-Hélène Gilbert}.} \bibinfo{year}{2019}\natexlab{}.
\newblock \showarticletitle{What do leaders need to know about managers’ mental health?}
\newblock \bibinfo{journal}{\emph{Organizational Dynamics}} \bibinfo{volume}{48}, \bibinfo{number}{3} (\bibinfo{date}{July} \bibinfo{year}{2019}), \bibinfo{pages}{85--92}.
\newblock
\showISSN{00902616}
\urldef\tempurl%
\url{https://doi.org/10.1016/j.orgdyn.2018.11.002}
\showDOI{\tempurl}


\bibitem[Sun et~al\mbox{.}(2024)]%
        {sun_BuildingBetterAI_2024}
\bibfield{author}{\bibinfo{person}{Guangzhi Sun}, \bibinfo{person}{Xiao Zhan}, {and} \bibinfo{person}{Jose Such}.} \bibinfo{year}{2024}\natexlab{}.
\newblock \showarticletitle{Building {Better} {AI} {Agents}: {A} {Provocation} on the {Utilisation} of {Persona} in {LLM}-based {Conversational} {Agents}}. In \bibinfo{booktitle}{\emph{Proceedings of the 6th {ACM} {Conference} on {Conversational} {User} {Interfaces}}} \emph{(\bibinfo{series}{{CUI} '24})}. \bibinfo{publisher}{Association for Computing Machinery}, \bibinfo{address}{New York, NY, USA}, \bibinfo{pages}{1--6}.
\newblock
\showISBNx{9798400705113}
\urldef\tempurl%
\url{https://doi.org/10.1145/3640794.3665887}
\showDOI{\tempurl}


\bibitem[Tabak and Lebron(2017)]%
        {tabak_LearningDoingLeadership_2017}
\bibfield{author}{\bibinfo{person}{Filiz Tabak} {and} \bibinfo{person}{Mariana Lebron}.} \bibinfo{year}{2017}\natexlab{}.
\newblock \showarticletitle{Learning by {Doing} in {Leadership} {Education}: {Experiencing} {Followership} and {Effective} {Leadership} {Communication} {Through} {Role}-{Play}}.
\newblock \bibinfo{journal}{\emph{Journal of Leadership Education}} \bibinfo{volume}{16}, \bibinfo{number}{2} (\bibinfo{date}{April} \bibinfo{year}{2017}), \bibinfo{pages}{199--212}.
\newblock
\showISSN{1552-9045}
\urldef\tempurl%
\url{https://doi.org/10.12806/V16/I2/A1}
\showDOI{\tempurl}


\bibitem[Tepper({[n.\,d.]})]%
        {tepper_ConsequencesAbusiveSupervision_2000}
\bibfield{author}{\bibinfo{person}{Bennett~J. Tepper}.} \bibinfo{year}{[n.\,d.]}\natexlab{}.
\newblock \showarticletitle{Consequences of {{Abusive Supervision}}}.
\newblock  \bibinfo{volume}{43}, \bibinfo{number}{2} (\bibinfo{year}{[n.\,d.]}), \bibinfo{pages}{178--190}.
\newblock
\showISSN{00014273}
\urldef\tempurl%
\url{https://doi.org/10.5465/1556375}
\showDOI{\tempurl}


\bibitem[Thiel et~al\mbox{.}(2022)]%
        {thiel2022monitoring}
\bibfield{author}{\bibinfo{person}{Chase Thiel}, \bibinfo{person}{Juliet~M. Bonner}, \bibinfo{person}{Jon Bush}, \bibinfo{person}{David~T. Welsh}, {and} \bibinfo{person}{Nitya Garud}.} \bibinfo{year}{2022}\natexlab{}.
\newblock \showarticletitle{Monitoring employees makes them more likely to break rules}.
\newblock \bibinfo{journal}{\emph{Harvard Business Review}} (\bibinfo{year}{2022}).
\newblock
\urldef\tempurl%
\url{https://hbr.org/2022/06/monitoring-employees-makes-them-more-likely-to-break-rules}
\showURL{%
\tempurl}


\bibitem[Wang et~al\mbox{.}(2023)]%
        {wang_EmotionalIntelligenceLarge_2023}
\bibfield{author}{\bibinfo{person}{Xuena Wang}, \bibinfo{person}{Xueting Li}, \bibinfo{person}{Zi Yin}, \bibinfo{person}{Yue Wu}, {and} \bibinfo{person}{Liu Jia}.} \bibinfo{year}{2023}\natexlab{}.
\newblock \bibinfo{title}{Emotional {Intelligence} of {Large} {Language} {Models}}.
\newblock
\newblock
\urldef\tempurl%
\url{https://doi.org/10.48550/arXiv.2307.09042}
\showDOI{\tempurl}
\newblock
\shownote{arXiv:2307.09042 [cs]}.


\bibitem[Wen et~al\mbox{.}(2017)]%
        {wen_NetworkbasedEndtoEndTrainable_2017}
\bibfield{author}{\bibinfo{person}{Tsung-Hsien Wen}, \bibinfo{person}{David Vandyke}, \bibinfo{person}{Nikola Mrksic}, \bibinfo{person}{Milica Gasic}, \bibinfo{person}{Lina~M. Rojas-Barahona}, \bibinfo{person}{Pei-Hao Su}, \bibinfo{person}{Stefan Ultes}, {and} \bibinfo{person}{Steve Young}.} \bibinfo{year}{2017}\natexlab{}.
\newblock \bibinfo{title}{A {Network}-based {End}-to-{End} {Trainable} {Task}-oriented {Dialogue} {System}}.
\newblock
\newblock
\urldef\tempurl%
\url{https://doi.org/10.48550/arXiv.1604.04562}
\showDOI{\tempurl}
\newblock
\shownote{arXiv:1604.04562 [cs, stat]}.


\bibitem[Wintersberger et~al\mbox{.}(2022)]%
        {wintersberger2022designing}
\bibfield{author}{\bibinfo{person}{Philipp Wintersberger}, \bibinfo{person}{Niels Van~Berkel}, \bibinfo{person}{Nadia Fereydooni}, \bibinfo{person}{Benjamin Tag}, \bibinfo{person}{Elena~L Glassman}, \bibinfo{person}{Daniel Buschek}, \bibinfo{person}{Ann Blandford}, {and} \bibinfo{person}{Florian Michahelles}.} \bibinfo{year}{2022}\natexlab{}.
\newblock \showarticletitle{Designing for continuous interaction with artificial intelligence systems}. In \bibinfo{booktitle}{\emph{CHI Conference on Human Factors in Computing Systems Extended Abstracts}}. \bibinfo{pages}{1--4}.
\newblock


\bibitem[Wu et~al\mbox{.}({[n.\,d.]})]%
        {wu_LargeLanguageModels_2023a}
\bibfield{author}{\bibinfo{person}{Ning Wu}, \bibinfo{person}{Ming Gong}, \bibinfo{person}{Linjun Shou}, \bibinfo{person}{Shining Liang}, {and} \bibinfo{person}{Daxin Jiang}.} \bibinfo{year}{[n.\,d.]}\natexlab{}.
\newblock \bibinfo{booktitle}{\emph{Large {{Language Models}} Are {{Diverse Role-Players}} for {{Summarization Evaluation}}}}.
\newblock
\showeprint[arxiv]{2303.15078}~[cs]
\urldef\tempurl%
\url{http://arxiv.org/abs/2303.15078}
\showURL{%
\tempurl}


\bibitem[Wu et~al\mbox{.}(2022)]%
        {wu_AIChainsTransparent_2022}
\bibfield{author}{\bibinfo{person}{Tongshuang Wu}, \bibinfo{person}{Michael Terry}, {and} \bibinfo{person}{Carrie~Jun Cai}.} \bibinfo{year}{2022}\natexlab{}.
\newblock \showarticletitle{{{AI Chains}}: {{Transparent}} and {{Controllable Human-AI Interaction}} by {{Chaining Large Language Model Prompts}}}. In \bibinfo{booktitle}{\emph{Proceedings of the 2022 {{CHI Conference}} on {{Human Factors}} in {{Computing Systems}}}} (New York, NY, USA, 2022-04-29) \emph{(\bibinfo{series}{{{CHI}} '22})}. \bibinfo{publisher}{Association for Computing Machinery}.
\newblock
\urldef\tempurl%
\url{https://doi.org/10.1145/3491102.3517582}
\showDOI{\tempurl}


\bibitem[Yablonski(2020)]%
        {yablonski_LawsUXUsing_2020}
\bibfield{author}{\bibinfo{person}{Jon Yablonski}.} \bibinfo{year}{2020}\natexlab{}.
\newblock \bibinfo{booktitle}{\emph{Laws of {UX}: {Using} {Psychology} to {Design} {Better} {Products} \& {Services}}}.
\newblock \bibinfo{publisher}{"O'Reilly Media, Inc."}.
\newblock
\showISBNx{978-1-4920-5528-0}
\newblock
\shownote{Google-Books-ID: BuneDwAAQBAJ}.


\bibitem[Zambrana et~al\mbox{.}(2015)]%
        {zambrana_DontLeaveUs_2015}
\bibfield{author}{\bibinfo{person}{Ruth~Enid Zambrana}, \bibinfo{person}{Rashawn Ray}, \bibinfo{person}{Michelle~M. Espino}, \bibinfo{person}{Corinne Castro}, \bibinfo{person}{Beth Douthirt~Cohen}, {and} \bibinfo{person}{Jennifer Eliason}.} \bibinfo{year}{2015}\natexlab{}.
\newblock \showarticletitle{“{Don}’t {Leave} {Us} {Behind}”: {The} {Importance} of {Mentoring} for {Underrepresented} {Minority} {Faculty}}.
\newblock \bibinfo{journal}{\emph{American Educational Research Journal}} \bibinfo{volume}{52}, \bibinfo{number}{1} (\bibinfo{date}{Feb.} \bibinfo{year}{2015}), \bibinfo{pages}{40--72}.
\newblock
\showISSN{0002-8312}
\urldef\tempurl%
\url{https://doi.org/10.3102/0002831214563063}
\showDOI{\tempurl}
\newblock
\shownote{Publisher: American Educational Research Association}.


\bibitem[Zhang and Zhang(2019)]%
        {zhang_ThinkaloudProtocols_2019}
\bibfield{author}{\bibinfo{person}{Lawrence~Jun Zhang} {and} \bibinfo{person}{Donglan Zhang}.} \bibinfo{year}{2019}\natexlab{}.
\newblock \showarticletitle{Think-aloud protocols}.
\newblock In \bibinfo{booktitle}{\emph{The {Routledge} {Handbook} of {Research} {Methods} in {Applied} {Linguistics}}}. \bibinfo{publisher}{Routledge}.
\newblock
\showISBNx{978-0-367-82447-1}
\newblock
\shownote{Num Pages: 10}.


\bibitem[Zhang et~al\mbox{.}(2021)]%
        {zhang_IdealHumanExpectations_2021}
\bibfield{author}{\bibinfo{person}{Rui Zhang}, \bibinfo{person}{Nathan~J. McNeese}, \bibinfo{person}{Guo Freeman}, {and} \bibinfo{person}{Geoff Musick}.} \bibinfo{year}{2021}\natexlab{}.
\newblock \showarticletitle{"{An} {Ideal} {Human}": {Expectations} of {AI} {Teammates} in {Human}-{AI} {Teaming}}.
\newblock \bibinfo{journal}{\emph{Proc. ACM Hum.-Comput. Interact.}} \bibinfo{volume}{4}, \bibinfo{number}{CSCW3} (\bibinfo{date}{Jan.} \bibinfo{year}{2021}), \bibinfo{pages}{246:1--246:25}.
\newblock
\urldef\tempurl%
\url{https://doi.org/10.1145/3432945}
\showDOI{\tempurl}


\bibitem[Zhou et~al\mbox{.}(2021)]%
        {zhou_VirtualRealityReflection_2021}
\bibfield{author}{\bibinfo{person}{Hangyu Zhou}, \bibinfo{person}{Yuichiro Fujimoto}, \bibinfo{person}{Masayuki Kanbara}, {and} \bibinfo{person}{Hirokazu Kato}.} \bibinfo{year}{2021}\natexlab{}.
\newblock \showarticletitle{Virtual {Reality} as a {Reflection} {Technique} for {Public} {Speaking} {Training}}.
\newblock \bibinfo{journal}{\emph{Applied Sciences}} \bibinfo{volume}{11}, \bibinfo{number}{9} (\bibinfo{date}{Jan.} \bibinfo{year}{2021}), \bibinfo{pages}{3988}.
\newblock
\showISSN{2076-3417}
\urldef\tempurl%
\url{https://doi.org/10.3390/app11093988}
\showDOI{\tempurl}
\newblock
\shownote{Number: 9 Publisher: Multidisciplinary Digital Publishing Institute}.


\end{thebibliography}
